\documentclass[aps,prd,nofootinbib,twocolumn]{revtex4-1}
\usepackage{amsmath}
\usepackage[scientific-notation=true]{siunitx}
\usepackage{graphicx}
\usepackage{hyperref}
\usepackage{placeins}
\usepackage{bm}
\usepackage{enumitem}

\let\vec\bm
\newcommand{\uvec}[1]{\bm{\hat{#1}}}
\newcommand{\approptoinn}[2]{\mathrel{\vcenter{
			\offinterlineskip\halign{\hfil$##$\cr
				#1\propto\cr\noalign{\kern2pt}#1\sim\cr\noalign{\kern-2pt}}}}}

\begin{document}
\title{Tidal evolution of dark matter annihilation rates in subhalos}
\author{M. Sten Delos}
\email{Electronic address: delos@unc.edu}
\affiliation{Department of Physics and Astronomy, University of North Carolina at Chapel Hill, Phillips Hall CB3255, Chapel Hill, North Carolina 27599, USA}

\begin{abstract}
Dark matter halos grow by hierarchical clustering as they merge together to produce ever larger structures.  During these merger processes, the smaller halo can potentially survive as a subhalo of the larger halo, so a galaxy-scale halo today likely possesses a rich abundance of substructure.  This substructure can greatly boost the rate of dark matter annihilation within the host halo, but the precise magnitude of this boost is clouded by uncertainty about the survival prospects of these subhalos.  In particular, tidal forces gradually strip material from the subhalos, reducing their annihilation signals and potentially destroying them.  In this work, we use high-resolution idealized $N$-body simulations to develop and tune a model that can predict the impact of this tidal evolution on the annihilation rates within subhalos.  This model predicts the time evolution of a subhalo's annihilation rate as a function of three physically motivated parameters of the host-subhalo system: the energy injected into subhalo particles per orbit about the host, the ratio of stretching to compressive tidal forces, and the radial distribution of tidal heating within the subhalo.  Our model will improve the accuracy of predictions of the magnitude and morphology of annihilation signals from dark matter substructure.  Additionally, our parametrization can describe the time evolution of other subhalo properties, so it has implications for understanding aspects of subhalo tidal evolution beyond the annihilation rate.
\end{abstract}

                              
\maketitle

\section{Introduction}

Despite overwhelming evidence for the existence of dark matter (e.g., Refs.~\cite{zwicky1933rotverschiebung,clowe2004weak,2016planck}), its microphysical details remain unknown.  Numerous models have been considered, but none have been experimentally confirmed (see Refs.~\cite{bertone2018new,feng2010dark,bertone2010particle,bertone2005particle,bergstrom2000non} for reviews).  However, to explain the present abundance of dark matter, a large class of models, including the popular weakly interacting massive particle \cite{jungman1996supersymmetric}, propose that dark matter was pair produced from the thermal plasma in the hot early universe.  In this scenario, the dark matter can annihilate back into standard-model particles today, leading to prospects for the detection of high-energy gamma rays or other annihilation products (e.g., Ref.~\cite{strigari2007precise}).

The rate of dark matter annihilation scales as the square of the dark matter density, so it is strongly sensitive to the spatial distribution of the dark matter.  At galactic scales and above, this spatial distribution is well understood.  Initially overdense patches in the Universe collapse into gravitationally bound dark matter halos, which thereafter merge to produce successively larger structures.  Numerical simulations demonstrate that the spherically averaged mass distributions of the resulting dark matter halos are well described by the NFW density profile \cite{navarro1996structure,navarro1997universal},
\begin{equation}\label{NFW}
\rho(r) = \frac{\rho_s}{(r/r_s)(1+r/r_s)^{2}},
\end{equation}
which has scale parameters $r_s$ and $\rho_s$.  Baryonic effects may subsequently alter this density profile (e.g., Ref.~\cite{brooks2014re}).

However, at subgalactic scales, the spatial distribution of the dark matter is less clear.  As a halo is built up through hierarchical merging, it accretes smaller halos that survive as subhalos within the larger host.  These subhalos gradually lose mass due to the influence of tidal forces from the host (e.g., Refs.~\cite{binney1987galactic,mo2010galaxy}), surviving until either they are completely stripped or dynamical friction causes them to sink into the host's center \cite{chandrasekhar1943dynamical}.  However, for sufficiently small subhalos dynamical friction is inefficient \cite{mo2010galaxy}.  Moreover, numerous analyses have found that if the subhalos possess divergent central density, as in the NFW profile, then tidal forces may never fully strip them \cite{goerdt2007survival,penarrubia2010impact,errani2019can,berezinsky2008remnants}.

Thus, a galactic-scale dark matter halo is likely to possess a multitude of subhalos.  This substructure can significantly boost the rate of dark matter annihilation, depending on the scale of the smallest halos and when they form \cite{berezinsky2003small,diemand2005earth,pieri2008dark,berezinsky2008remnants,ishiyama2010gamma,anderhalden2013density,*anderhalden2013erratum,ishiyama2014hierarchical,sanchez2014flattening,anderson2016search,gao2011will,springel2008prospects,stref2017modeling,stref2019remnants,bartels2015boosting,hiroshima2018modeling} (see Ref.~\cite{ando2019halo} for a recent review).  When the smallest halos are microhalos of roughly earth mass, annihilation rates may be boosted by a factor of about $10$ \cite{stref2017modeling,hiroshima2018modeling} relative to those expected in the absence of substructure, assuming that these halos arise from primordial density fluctuations comparable to the large-scale fluctuations inferred from the cosmic microwave background (e.g., Ref.~\cite{2016planck}).  Moreover, this boost can be raised by orders of magnitude by cosmological scenarios that amplify small-scale density fluctuations and thereby lead to earlier and more abundant microhalo formation.  Such scenarios include a period of domination by a heavy species \cite{erickcek2011reheating,barenboim2014structure,fan2014nonthermal,erickcek2015dark}
or a fast-rolling scalar field \cite{redmond2018growth} prior to nucleosynthesis, along with a variety of inflationary models \cite{silk1987double,salopek1989designing,starobinsky1992,ivanov1994inflation,randall1996supernatural,stewart1997flattening,adams1997multiple,starobinsky1998beyond,covi1999running,martin2000nonvacuum,chung2000probing,martin2001trans,joy2008new,barnaby2009particle,barnaby2010features,ben2010cosmic,gong2011waterfall,lyth2011contribution,bugaev2011curvature,barnaby2011large,achucarro2011features,cespedes2012importance,barnaby2012gauge}.  In these cases, the high density within these microhalos causes them to completely dominate any dark matter annihilation signal (e.g., Refs.~\cite{bringmann2012improved,delos2018density,blanco2019annihilation}).

Unfortunately, all estimates of the substructure's boost to annihilation rates are subject to uncertainties about the impact of tidal effects on subhalos.  Cosmological simulations cannot resolve subhalos that are much smaller than the host, and those that are resolved are prone to artificial destruction \cite{van2017dissecting,van2017disruption,van2018dark}.  Numerous semianalytic models have been developed to describe the dynamical evolution of subhalos; Refs.~\cite{taylor2001dynamics,penarrubia2005effects,van2005mass,zentner2005physics,kampakoglou2006tidal,gan2010improved,pullen2014nonlinear,jiang2016statistics} model a subhalo's loss of mass due to tidal stripping, and Refs.~\cite{hayashi2003structural,penarrubia2010impact} predict the impact of this mass loss on the halo's density profile.  However, these models are typically tuned to the results of cosmological simulations, so their predictions are affected by the artificial subhalo disruption occurring therein.  They cannot fully reproduce the results of idealized simulations \cite{van2017disruption,ogiya2019dash}.

Meanwhile, calculations of the dark matter annihilation rate in the substructure have employed a number of different treatments of tidal evolution.  Some, such as Refs.~\cite{bartels2015boosting,hiroshima2018modeling}, employ a combination of the models above to predict the time evolution of subhalo density profiles.  Others, such as Refs.~\cite{springel2008prospects,erickcek2015dark,stref2017modeling,stref2019remnants}, employ simpler models, often either truncating subhalos at a characteristic tidal radius or formulating a destruction condition for subhalos and assuming the survivors are unaltered.  Still others, such as Refs.~\cite{bringmann2012improved,anderhalden2013density,*anderhalden2013erratum,ishiyama2014hierarchical,sanchez2014flattening,anderson2016search,delos2018density,blanco2019annihilation} neglect the tidal disruption of the substructure altogether.

Our work is motivated by this context.  Since cosmological simulations cannot resolve the smallest substructures, we follow Refs.~\cite{hayashi2003structural,kazantzidis2004density,read2006tidal,penarrubia2010impact,van2018dark,ogiya2019dash} in using idealized simulations of an $N$-body subhalo inside an analytic galactic potential.  However, unlike these works, we focus on understanding the impact of tides on the subhalo's annihilation rate, a goal that requires significantly better resolution than has been attained in previous studies.  Moreover, previous works have focused on understanding the evolution of subhalos of scales resolvable in cosmological simulations, such as halos associated with dwarf galaxies within a galactic halo.  Accordingly, they probe only the subhalo properties and orbits that are found in such simulations.  For instance, Ref.~\cite{ogiya2019dash} only studies subhalos orbiting above the host's scale radius.  In contrast, we seek to probe the full range of subhalos down to the smallest microhalos, which span a far broader range of properties and orbits.

Using the results of 52 high-resolution $N$-body simulations, we develop a physically motivated model that can predict the time evolution of a subhalo's annihilation rate due to tidal effects.  In the process, we isolate three physical variables that determine this evolution:
\begin{enumerate}[label={(\arabic*)}]
	\item The energy injected by tidal forces into subhalo particles over the course of each orbit about the host, in units of the particle's binding energy to the subhalo;
	\item The ratio of stretching (radial) tidal forces to compressive (tangential) tidal forces;
	\item The range of radii in the subhalo across which material is heated by tidal forces, which is set by the shape of the subhalo's orbit.
\end{enumerate}
This model predicts the suppression of a subhalo's annihilation rate, characterized by its $J$ factor\footnote{We assume the dark matter annihilation cross section is velocity independent in the nonrelativistic limit.}
\begin{equation}\label{J}
J\equiv\int \rho^2 \mathrm{d}V,
\end{equation}
as a function of its orbit about the host.  To assist the application of our model, we supply convenient fitting functions.

This article is organized as follows.  In Sec.~\ref{sec:sim}, we detail how we carry out our $N$-body simulations.  Section~\ref{sec:trends} qualitatively discusses the trajectory of a subhalo's $J$ factor, interpreting simulation trends physically and motivating our model.  In Sec.~\ref{sec:model}, we develop our predictive model for the evolution of a subhalo's $J$ factor, and Sec.~\ref{sec:disc} summarizes the model and discusses limitations and extensions.  In Sec.~\ref{sec:compare}, we compare the model's predictions to those of previous semianalytic models.  Section~\ref{sec:conclusion} concludes, after which we supply a variety of appendixes.  Appendix \ref{sec:sim_app} supplies further details about our simulations, while Appendix~\ref{sec:size} quantifies the range of subhalo sizes over which the results of these simulations are applicable.  Appendix~\ref{sec:fits} presents fitting formulas and other computational details that aid in applying our model.  Appendix~\ref{sec:dash} tests our model against a publicly available simulation library \cite{ogiya2019dash}.  Finally, in Appendix~\ref{sec:rmax}, we observe that our model can be adapted to describe the evolution of subhalo properties beyond the $J$ factor.

\section{Simulations}\label{sec:sim}

Owing to the difference in scales between a host and its smallest subhalos, the computational challenge in simulating subhalo evolution in a cosmological context is formidable.  A number of previous works have addressed this problem by simulating an $N$-body subhalo inside an analytic host potential \cite{hayashi2003structural,kazantzidis2004density,read2006tidal,penarrubia2010impact,van2018dark,ogiya2019dash}; our approach is similar but differs in one key step.  Instead of placing a subhalo in orbit about the host potential, we subject the subhalo directly to the time-dependent tidal force field experienced by an analytic orbit about the host.  This procedure minimizes the impact of numerical precision errors that can result from differences in scale between the subhalo's orbital and internal dynamics.  In this section, we detail that procedure and present qualitative results.

We assume that both the host and the subhalo possess the NFW density profile given by Eq.~(\ref{NFW}).  While there is evidence that many galactic halos possess constant-density cores instead of the NFW profile's cusp \cite{moore1994evidence}, at least some galactic halos appear to be cuspy \cite{read2018case}.  Additionally, while microhalos are expected to form with $\rho\propto r^{-3/2}$ inner profiles \cite{ishiyama2010gamma,anderhalden2013density,*anderhalden2013erratum,ishiyama2014hierarchical,polisensky2015fingerprints,ogiya2017sets,delos2018ultracompact,delos2018density,angulo2017earth,delos2019predicting}, it is likely that mergers will drive their inner cusps toward the $\rho\propto r^{-1}$ of the NFW profile \cite{ogiya2016dynamical,angulo2017earth,delos2019predicting}.

To model the host's tidal field we begin with an analytically computed orbit, described by the time-dependent vector $\vec R(t)$ pointing from host center to subhalo center.  The tidal acceleration at position $\vec r$ relative to the subhalo center is\footnote{We experimented with using the full tidal force $\vec F_\mathrm{tidal}(\vec r) = \vec F(\vec R + \vec r) - \vec F(\vec R)$, but because $r\ll R$ in our simulations, it offers no advantage; moreover, it is less numerically stable due to the subtraction of two close numbers.}
\begin{equation}\label{tidalforce}
\vec F_\mathrm{tidal}(\vec r) = -\frac{\mathrm{d}F}{\mathrm{d}R} (\vec r \cdot \uvec R)\uvec R - F(R)\frac{\vec r-(\vec r\cdot\uvec R)\uvec R}{R}
\end{equation}
at linear order in $r/R$, where $F(R)$ is the force profile of the host, $R=|\vec R|$, and $\uvec{R}=\vec R/R$.  We modified the \textsc{Gadget-2} $N$-body simulation code \cite{springel2001gadget,springel2005cosmological} to include this tidal acceleration.

We prepare the initial $N$-body subhalo with an NFW profile by drawing particles from an isotropic distribution function computed using the fitting form in Ref.~\cite{widrow2000distribution}.  Additionally, we sample the subhalo's central region at increased resolution; particles whose orbital pericenters are below $r_s/3$, where $r_s$ is the subhalo scale radius, have $1/64$ the mass and $64$ times the number density of the other particles.  Appendix~\ref{sec:sim_app} demonstrates that there is no significant relaxation associated with the use of particles of different masses.  We cut off the density profile at $r = 500 r_s$; subhalo particles this far out are stripped immediately, so as long as the cutoff radius is much larger than $r_s$, the precise choice makes no difference.\footnote{The natural place to cut off the density profile would be where the density reaches that of the subhalo's background: the host.  However, tidal forces automatically truncate a subhalo's density profile at roughly the radius where its average density equals that of the host [see, e.g., Eq.~(\ref{rhot})], so it is not necessary to tune a cutoff radius by hand.}  We represent the subhalo using a total of $8\times 10^6$ particles, and roughly 70\% of them, carrying roughly 4\% of the total mass, are high-resolution particles.  All of our subhalos have $r_s\simeq 10^{-6}R_s$, where $R_s$ is the scale radius of the host, but as we will soon discuss, the precise choice of $r_s$ has no impact on dynamics.

For our simulations, we consider a variety of orbits about the host.  An orbit in a spherically symmetric potential is characterized by two parameters: energy $E$ and angular momentum $L$ or, equivalently, a scale parameter and a shape parameter.  For convenience, we use the circular orbit radius\footnote{Note that $R_c$ is roughly the time-averaged radius; for a power-law potential $\phi(R)\propto R^n$, $R_c=\langle R^n\rangle^{1/n}$.  See Appendix~\ref{sec:fits} for a more precise relationship for NFW profiles.} $R_c$, defined as the radius of the circular orbit with energy $E$, and the ``circularity'' $\eta=L/L_c$, where $L_c$ is the angular momentum of the circular orbit with the same energy.  In each simulation the subhalo begins at its orbital apocenter.

\begin{figure}[t]
	\includegraphics[width=.49\columnwidth]{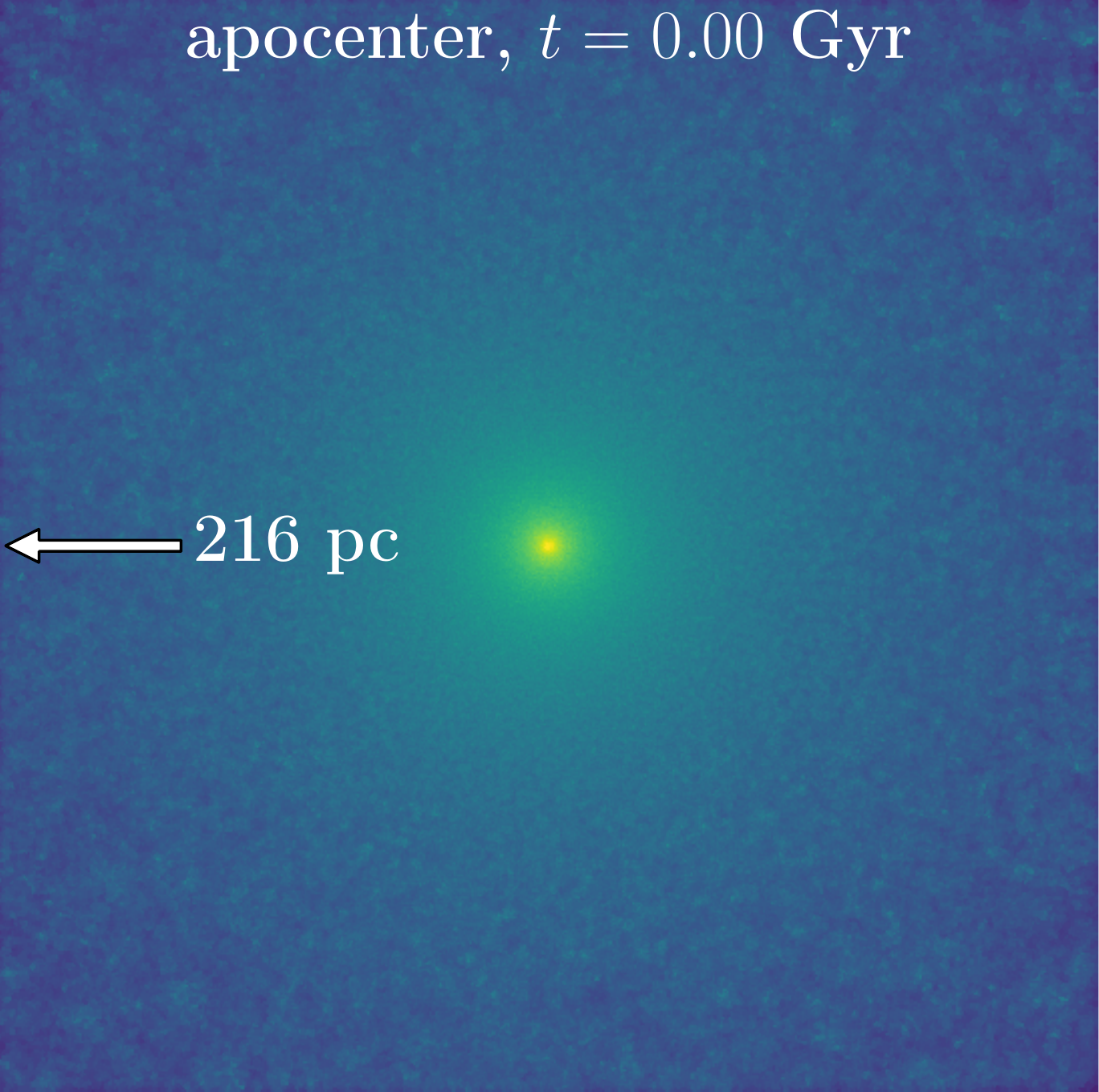} \hspace{-2mm}
	\includegraphics[width=.49\columnwidth]{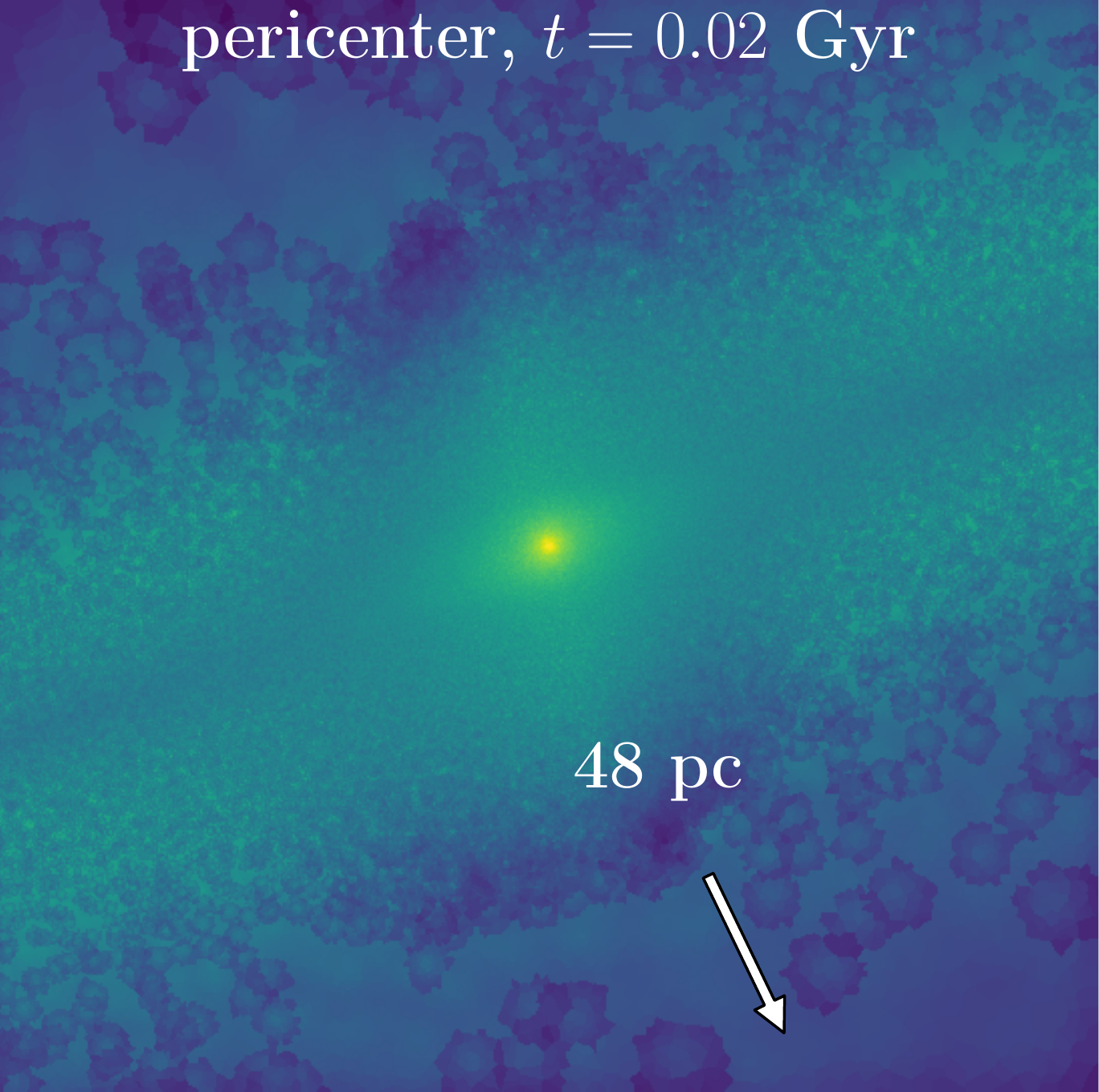} \\
	\vspace{-3mm}
	\includegraphics[width=.49\columnwidth]{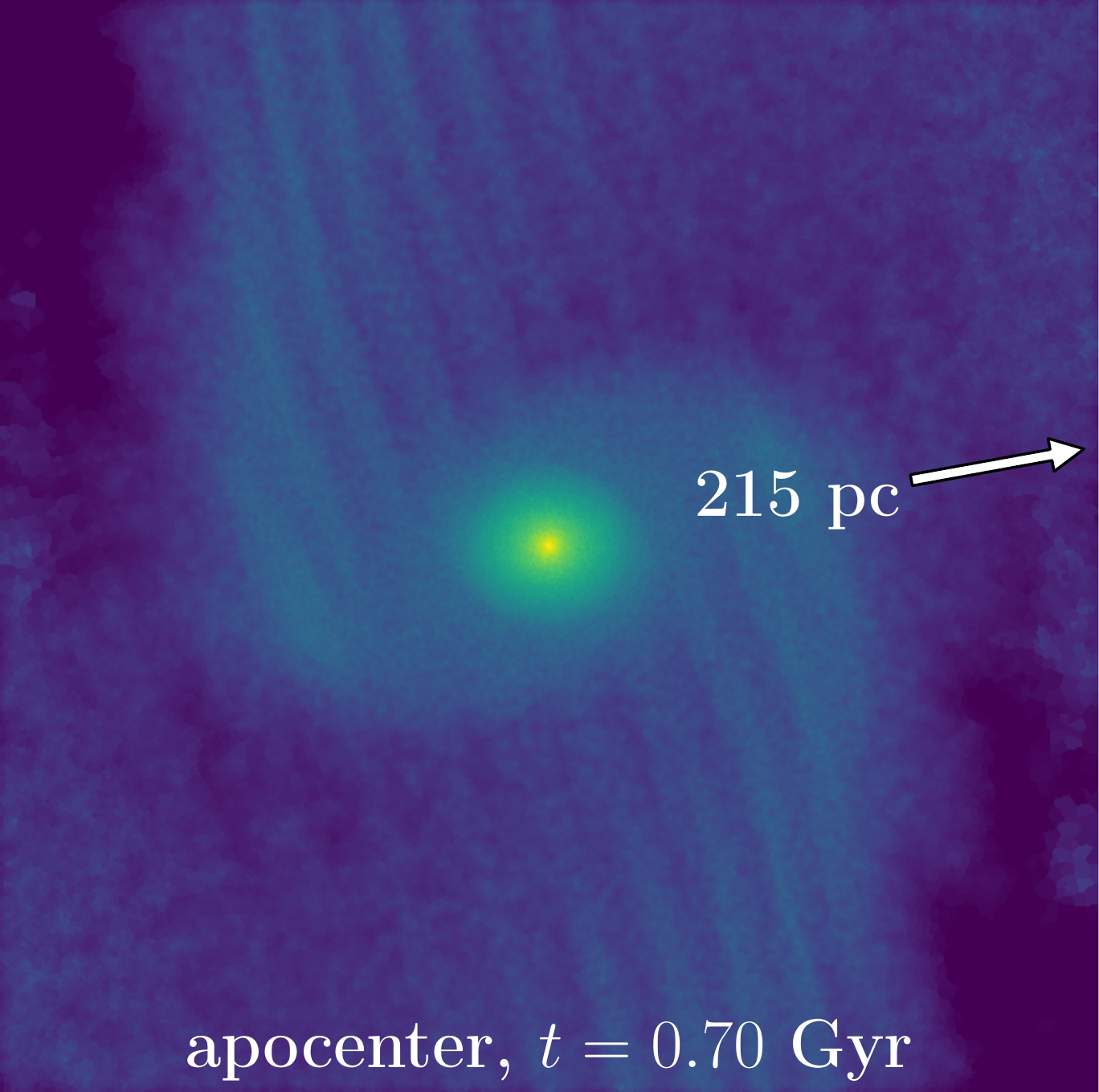} \hspace{-2mm}
	\includegraphics[width=.49\columnwidth]{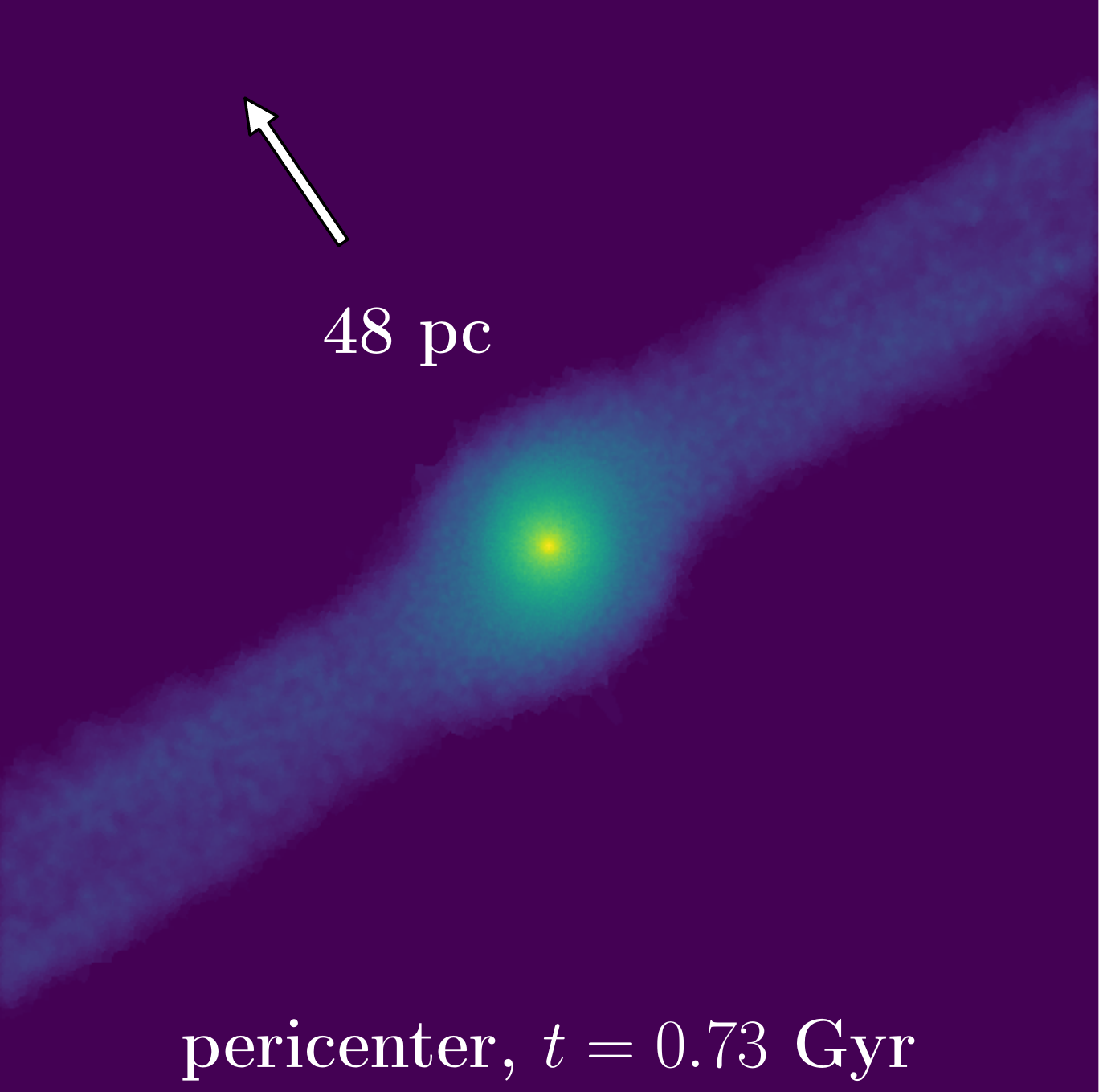}
	\caption{\label{fig:tidalsims} The projected density field of a microhalo simulated in orbit about a galactic halo.  The width of each frame is $0.03$ pc, and the arrow indicates the direction and distance to the host center.  The density is computed using a $k$-nearest-neighbor density estimate with $k=50$ and is plotted with a logarithmic color scale (lighter is denser).}
\end{figure}

\begin{figure}[t]
	\includegraphics[width=\columnwidth]{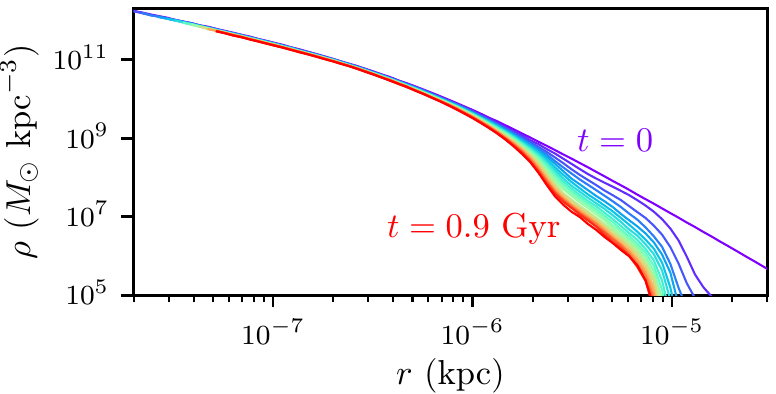}
	\caption{\label{fig:tidalsims_r}The density profile evolution of the halo depicted in Fig.~\ref{fig:tidalsims}.  91\% of the initial mass of the microhalo is stripped by $t=0.9$ Gyr, but the central density profile is largely unaffected.}
\end{figure}

Figure~\ref{fig:tidalsims} illustrates a simulation executed through this arrangement.  The host has scale radius $R_s=0.8$ kpc and scale density $P_s=\num{5e7}$ $M_\odot/\text{kpc}^3$, while the $N$-body subhalo is initially a microhalo with scale radius $r_s = \SI{6e-7}{kpc}$ that has $\rho_s/P_s=1285$ times the scale density of the host.  The subhalo orbit has $R_c=0.15$ kpc and $\eta=0.5$.  The simulation runs through 18 orbits about the host, and Fig.~\ref{fig:tidalsims_r} plots the density profile of the subhalo at each apocenter.  Consistently with the results of other works, such as Refs.~\cite{goerdt2007survival,penarrubia2010impact,errani2019can,berezinsky2008remnants}, we find that this subhalo's central cusp is highly resistant to disruption by the host's tidal forces.  91\% of the subhalo's mass is stripped by simulation termination, but its central density profile is largely unscathed.


Our goal in this work is to understand how the annihilation signal decays due to tidal effects.  For this purpose, we consider the $J$ factor, Eq.~(\ref{J}), integrated over the subhalo mass distribution.  This $J$ factor is the factor in the annihilation rate that depends on mass distribution, and Appendix~\ref{sec:sim_app} discusses our procedure to extract it from the simulations.  We also show in Appendix~\ref{sec:sim_app} that the resulting $J$-factor trajectories are converged with respect to simulation parameters.

Finally, we conclude this section by discussing the applicability of our simulation results.  The linearized tidal force in Eq.~(\ref{tidalforce}) is valid for $r\ll R$, and in Appendix~\ref{sec:size} we show that it yields accurate results in simulations as long as
\begin{equation}\label{size}
r_s\lesssim 0.1 R_c.
\end{equation}
Our simulation results are only applicable if this condition is satisfied.  Additionally, since we do not simulate the host halo's dynamics, we cannot account for dynamical friction.  Reference~\cite{ogiya2019dash} found that dynamical friction\footnote{Specifically, Ref.~\cite{ogiya2019dash} studied dynamical self-friction, or the dynamical friction that results from the subhalo's own tidal tail.  This friction can be considerably more efficient than that resulting from the host's material alone \cite{fujii2006dynamical,fellhauer2007influence}.} has minimal impact on a subhalo's mass evolution for host-to-subhalo mass ratios $M/m\gtrsim 100$.  However, since subhalos relevant to dark matter annihilation may orbit over significantly longer timescales than considered in Ref.~\cite{ogiya2019dash}, it is also useful to have an analytic estimate for when dynamical friction can be neglected.  It follows from the analysis in Ref.~\cite{mo2010galaxy} that for a subhalo that accretes onto a host at redshift $z$, dynamical friction can be neglected as long as the host-to-subhalo mass ratio $M/m$ satisfies
\begin{equation}\label{friction}
\frac{M/m}{\ln(M/m)} \gtrsim 10(1+z)^{3/2}.
\end{equation}
Our results may be considered applicable as long as Eqs. (\ref{size}) and~(\ref{friction}) are satisfied, but we remark that if one is satisfied, then the other likely is too.

\section{Trends in the tidal evolution}\label{sec:trends}

In this section, we explore trends in the evolution of $J$ as a function of system and time and attempt to explain them physically.  Our goal is to find the $J$ factor as a function of time $t$, orbital parameters $R_c$ and $\eta$, subhalo parameters $r_s$ and $\rho_s$, and host parameters $R_s$ and $P_s$.  The dimensionality of this space is large, but some immediate simplifications are evident:
\begin{enumerate}[label={(\arabic*)}]
	\item As long as the subhalo is much smaller than its orbit, or $r_s\ll R_c$, the value of $r_s$ has no impact on dynamics.\footnote{For fixed density, the internal velocities of particles in a subhalo are proportional to its radius.  Since the tidal acceleration in Eq.~(\ref{tidalforce}) is also proportional to the radius, fractional velocity changes induced by tidal forces are independent of the subhalo's radius.}  All of our simulated subhalos have $r_s\simeq 10^{-6}R_s$, which leads to $r_s\ll R_c$ for all orbits we consider.
	\item If instead of time $t$ we use the orbit count $n=t/T$, where $T$ is the orbital period, then the overall density scale has no impact on dynamics, and only the ratio $\rho_s/P_s$ enters.
	\item The overall size of the host-subhalo system is irrelevant, so only the ratio $R_c/R_s$ affects dynamics.
\end{enumerate}
We have verified that all of these simplifications are borne out in our simulations.  Hence, if $J_\mathrm{init}$ is the initial $J$ factor, then $J/J_\mathrm{init}$ is now a function of time $n=t/T$ and just three system parameters: $\rho_s/P_s$, $R_c/R_s$, and $\eta$.  Notably, the tidal evolution is independent of the subhalo's mass, a property also noted in prior works (e.g., Ref.~\cite{ogiya2019dash}).

\subsection{Trends in the simulations}

\begin{figure}[t]
	\centering
	\includegraphics[width=\columnwidth]{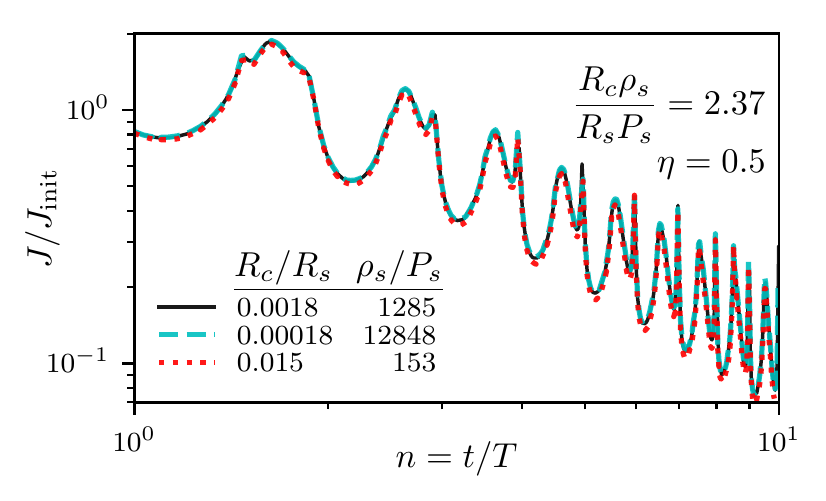}
	\caption{\label{fig:x} Trajectories of the $J$ factor for different systems with the same $\tilde x=\frac{R_c \rho_s}{R_s P_s}$ and $\eta$.  Scaled to the orbital period, these systems all have the same trajectory.}
\end{figure}

We first inspect the results of selected simulations in order to find trends in the behavior of $J$.  As a further simplification, we focus on the $R_c\ll R_s$ regime.  The host potential is self-similar in this regime, reducing the tidal evolution problem in two additional ways:
\begin{enumerate}[label={(\arabic*)}]
	\item The orbital radius $R_c$ is degenerate with properties of the host and subhalo.  For instance, reducing the orbital radius is equivalent to making the host denser.
	\item Orbits with the same circularity $\eta$ have the same shape; they are rescaled versions of one another.
\end{enumerate}
The first simplification further reduces the parameter space so that in this self-similar regime, there are only two parameters,\footnote{We reserve $x$ (without the tilde) for later use as a modified version of $\tilde x$.}
\begin{equation}\label{tildex}
\tilde x\equiv \frac{R_c \rho_s}{R_s P_s}
\end{equation}
and $\eta$.  Figure~\ref{fig:x} shows the success of this parameter reduction; different systems with the same $\tilde x$ and $\eta$ follow precisely the same $J(n)$ trajectories.  Meanwhile, the second simplification allows us to isolate the impact of these two parameters; we can vary the ``reduced orbital radius'' $\tilde x$ without altering the orbit's shape.

\begin{figure}[t]
	\centering
	\includegraphics[width=\columnwidth]{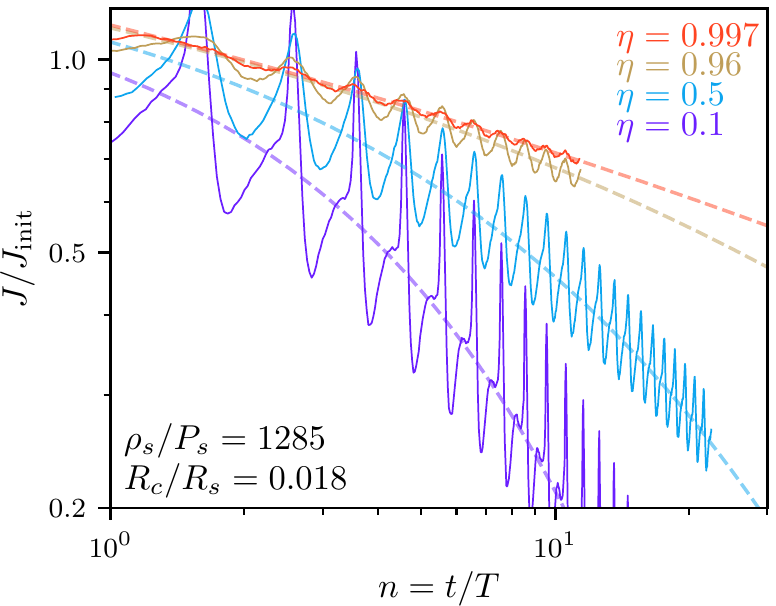}
	\caption{\label{fig:Jt} Trajectories of the $J$ factor for different orbital shapes.  The $J$ factor oscillates with the orbital period; the dashed lines show fits using Eq.~(\ref{dlnJ}) (for $|\mathrm{d}\ln J/\mathrm{d}\ln t|<1$).}
\end{figure}

We first investigate the impact of orbit shape.  Figure~\ref{fig:Jt} shows the trajectory of the $J$ factor for several values of $\eta$.  We see immediately that the $J$ factor oscillates with the orbital period with a larger amplitude for more eccentric orbits.  This trend is explained by noting that all tidal forces are compressive in the self-similar regime,\footnote{Tangential tidal forces are always compressive, while radial tidal forces are negligible when $R_c\ll R_s$; see Sec.~\ref{sec:b} for further discussion.} so the subhalo becomes most compact near the orbital pericenter.  The subhalo's $J$ factor, being proportional to its mass-weighted average density [e.g., Eq.~(\ref{Jsum})], is maximized at this point.  The precise appearance of these oscillations can be complicated because a subhalo's response to these tidal forces is delayed; for instance, double peaks in Fig.~\ref{fig:x} arise because the subhalo and its unbound tidal stream are maximally compressed at different times.  However, these oscillations are relatively unimportant.  If subhalos are at random points in their orbits, then the $J$ factor averaged over an orbital period suffices to predict the aggregate signal from a population of subhalos.  We discuss this point further in Sec.~\ref{sec:disc}.

\begin{figure}[t]
	\centering
	\includegraphics[width=\columnwidth]{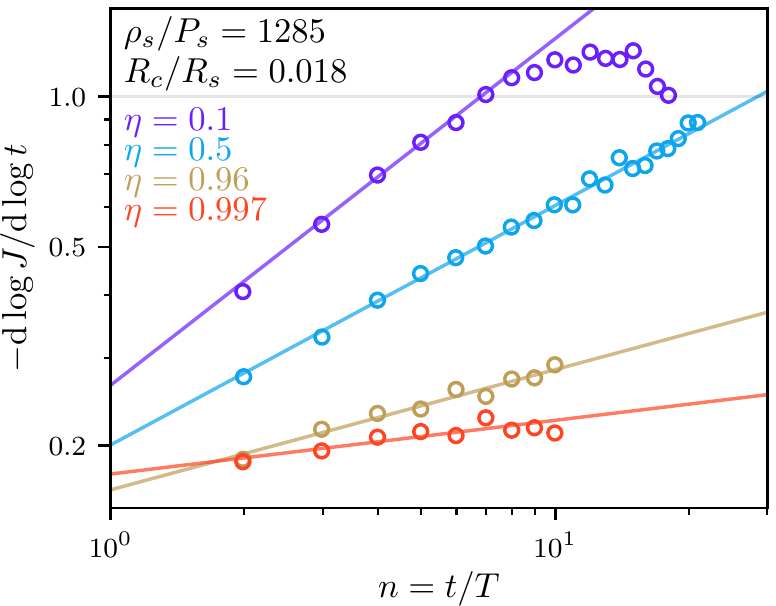}
	\caption{\label{fig:Jtd} The impact of orbit shape on the $J$-factor trajectory.  This figure plots the logarithmic slope of the orbital period-averaged trajectory to make the trends clearer.  Notably, the slope runs more rapidly for more eccentric orbits.  The points show the simulation results, while the lines correspond to fits using Eq.~(\ref{dlnJ}) (for $|\mathrm{d}\ln J/\mathrm{d}\ln t|<1$).}
\end{figure}

More interesting trends arise in the broader time evolution.  Figure~\ref{fig:Jtd} plots the running power-law index $\mathrm{d}\ln J/\mathrm{d}\ln t$ of the $J$ factor with time, and the equation
\begin{equation}\label{dlnJ}
\frac{\mathrm{d}\ln J}{\mathrm{d}\ln t} = -b n^{1-c}
\end{equation}
describes the evolution of this index reasonably well as long as $|\mathrm{d}\ln J/\mathrm{d}\ln t| <\mathcal{O}(1)$.  Here, $b>0$ and $c>0$ are constant parameters, and $c$ is smaller for more eccentric orbits.  Figures \ref{fig:Jt} and~\ref{fig:Jtd} also show fits to the $J$-factor trajectories using this form, which determines $J(n)$ up to a constant multiple.  The exponent in Eq.~(\ref{dlnJ}) is so defined because it leads to the more evocative expression
\begin{equation}\label{dJ}
\frac{1}{J}\frac{\mathrm{d}J}{\mathrm{d}n}=-bn^{-c}.
\end{equation}
If $c=0$, this equation tells us that the $J$ factor would decay by the same factor $\mathrm{e}^{-b}$ over each orbit.  The parameter $c$, when $c>0$, accommodates some physical process by which tidal effects lose efficiency over time.

\begin{figure}[t]
	\centering
	\includegraphics[width=\columnwidth]{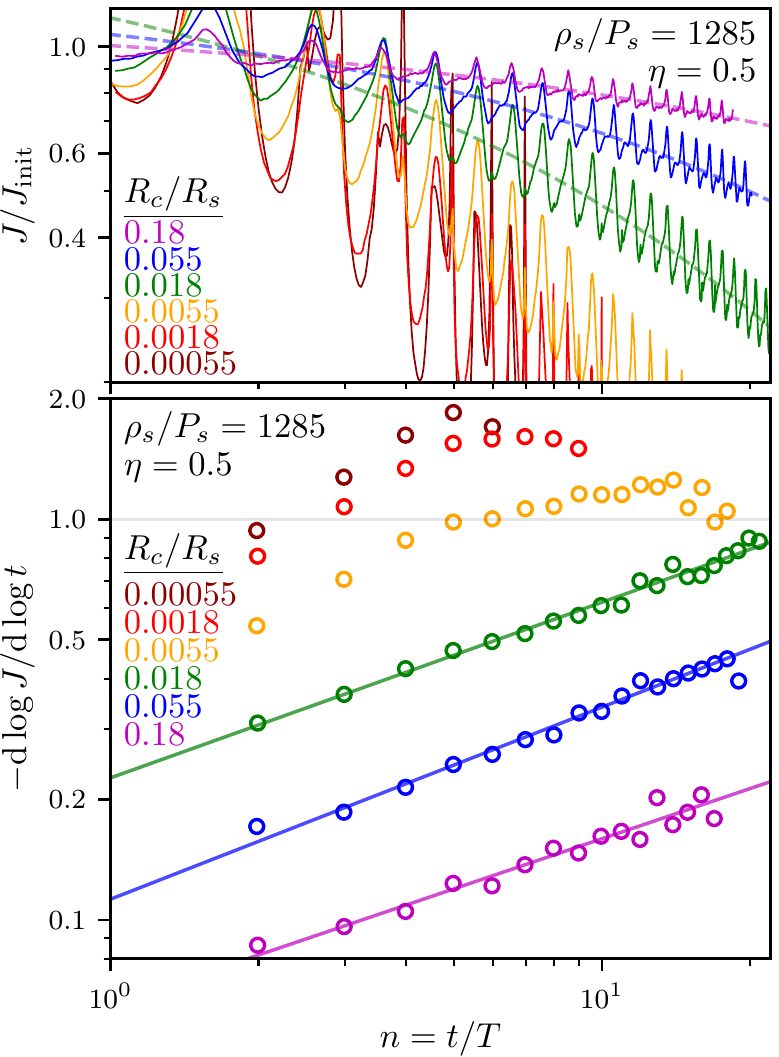}
	\caption{\label{fig:Jtdr} The impact of the orbital radius on the $J$-factor trajectory.  Top: The $J$-factor trajectory as in Fig.~\ref{fig:Jt} (solid lines); the dashed lines show fits using Eq.~(\ref{dlnJ}).  Bottom: The logarithmic slope of the orbital period-averaged trajectory, as in Fig.~\ref{fig:Jtd}; the points show the simulation results, while the lines correspond to the fits.  We only fit the trajectories that do not pass $|\mathrm{d}\ln J/\mathrm{d}\ln t|=1$.}
\end{figure}

We show the impact of the orbital radius in Fig.~\ref{fig:Jtdr}, which plots the trajectories of $J$ and $\mathrm{d}\ln J/\mathrm{d}\ln t$ for a variety of reduced orbital radii $\tilde x$ ranging from $0.7$ to $230$.  Evidently, $\tilde x$ affects the initial decay rate of the $J$ factor, described by the parameter $b$ in Eq.~(\ref{dlnJ}), without altering the rate at which the decay slows over time.  This figure also shows more clearly that there is a steepness limit to the decay of the $J$ factor:
\begin{equation}\label{dlnJ_}
\frac{\mathrm{d}\ln J}{\mathrm{d}\ln t} \simeq -\min\!\left\{b n^{1-c},B\right\},
\end{equation}
where $B\sim \mathcal{O}(1)$.

\subsection{Physical interpretation}\label{sec:toy}

Behavior similar to that of Eq.~(\ref{dlnJ_}) can be reproduced in a toy model.  Suppose the subhalo has potential ${\phi(r)\propto r^\gamma}$ up to an additive constant; for instance, an NFW profile would have $\gamma=1$ for ${r\ll r_s}$.  Now discretize time, perhaps as a count of orbits, letting $r_n$ be the radius of the subhalo's outer boundary at time $t_n$.  Any material outside $r_n$ at time $t_n$ is free, fixing the additive constant in the potential such that $\phi_n(r)\propto r^\gamma-r_n^\gamma$ (where $\phi\geq 0$ implies freedom).  Now suppose that at each time step, particles in the subhalo experience an injection of energy ${\Delta E\propto r^\alpha}$ due to tidal forces, and any radius with $\Delta E + \phi > 0$ no longer belongs to the halo.  For instance, if the time step is constant, then $\alpha=2$ since the energy injection is proportional to the square of the tidal force, which is in turn proportional to the radius.  This tidal heating rule leads to the evolution equation $\phi_n(r_{n+1})+\Delta E(r_{n+1})=0$, or 
\begin{equation}\label{toy}
r_{n+1}^\gamma-r_n^\gamma+f r_{n+1}^\alpha = 0,
\end{equation}
where $f$ incorporates all of the proportionality constants.

For simplicity, we may assume $r_0=1$, absorbing its dimensionful value into $f$.  For $\alpha>\gamma$, Eq.~(\ref{toy}) obeys
\begin{equation}\label{toysol}
\frac{\Delta \ln r}{\Delta \ln n} \simeq
\begin{cases} 
-fn/\gamma, & fn\ll 1, \\
-1/(\alpha-\gamma), & fn\gg 1,
\end{cases}
\end{equation}
where $\Delta$ denotes the discrete difference across time steps.  If $J\propto r^\beta$, then
\begin{equation}\label{toysolJ}
\frac{\Delta \ln J}{\Delta \ln n} \simeq -\min\left\{b n,B\right\}
\end{equation}
with $b=\beta f/\gamma$ and $B=\beta/(\alpha-\gamma)$.  The quantity $fn$ can be understood as (up to factors of order unity) the ratio of the tidal energy injection to the subhalo's internal energy.  The two separate regimes arise physically because when $fn\ll 1$, the total radius change $|r_n-r_0|\ll r_0$.  Since the radius does not change appreciably, the efficiency of tidal heating does not change, so $r$ and $J$ drop by the same fraction in each orbit.  However, when $fn\gg 1$, $|r_n-r_0|\sim r_0$.  In this case, the radius is decreasing significantly, which implies that the density at the halo's shrinking outer boundary is increasing and hence that the halo is becoming more difficult to strip.

This toy model has reproduced Eq.~(\ref{dlnJ}) with ${c=0}$, successfully explaining the apparent upper limit in $|\mathrm{d}\ln J/\mathrm{d}\ln t|$.  We remark, however, that there is another, completely different, physical reason to expect an upper limit in $|\mathrm{d}\ln J/\mathrm{d}\ln t|$: an unbound tidal stream grows in length $L$ as $L\propto t$.  Hence, its volume grows as ${V\propto t}$, so its $J$ factor drops as $J\propto M^2/V\propto t^{-1}$.  Once a subhalo has been stripped to the point that its own $J$ factor is dwarfed by that of its tidal stream, the $J$ factor of the subhalo remnant decays as ${J\propto t^{-1}}$.  The combination of these two processes---the increasing density of the subhalo as its radius drops and the $J$ factor of its tidal stream---may explain the behavior in Figs. \ref{fig:Jtd} and~\ref{fig:Jtdr} wherein $|\mathrm{d}\ln J/\mathrm{d}\ln t|$ initially shallows toward some value larger than $1$ before subsequently returning back to $1$.  Note, however, that the precise evolution of the $J$ factor in the $|\mathrm{d}\ln J/\mathrm{d}\ln t|\sim 1$ regime is of little consequence.  By this point, the subhalo has already lost most of its $J$ factor and contributes little to annihilation signals.

The physical explanation for the $c>0$ behavior observed in the simulations remains unclear.  However, it is necessarily connected to how the shape of the subhalo's density profile changes in response to tidal effects (see Fig.~\ref{fig:tidalsims_r}), which the toy model does not account for.  In a more complete picture, the rate $\mathrm{d}J/\mathrm{d}n$ of tidal evolution should be sensitive only to the instantaneous host-subhalo system with no explicit dependence on the time $n$.  Hence, it should be possible to replace the factor $n^{-c}$ in Eq.~(\ref{dJ}) with a function of the subhalo's density profile (and other properties of the system).  However, in the $|\mathrm{d}\ln J/\mathrm{d}\ln t|<1$ regime, the total change in $J$ is much smaller than $J$ itself, and if we neglect changes in the shape of the density profile, then any parameter of the density profile (e.g., $\rho_s$ or $r_s$) must experience a similarly small change.  Since the factor $n^{-c}$ can change by an order of magnitude in the same regime, it is not possible, except in a very contrived way, to replace this factor with a function of the density profile.

Thus, the $c>0$ behavior must follow from changes in the density profile's shape.  As a result of these changes, the density profile picks up new parameters that can potentially vary wildly without significantly altering $J$, and the factor $n^{-c}$ can be replaced with a function of those parameters.  For instance, by introducing a new parameter $q$, we can write
\begin{equation}\label{dJq}
\frac{1}{J}\frac{\mathrm{d}J}{\mathrm{d}n}=-b q,
\ \ \ \ 
\frac{1}{q}\frac{\mathrm{d}q}{\mathrm{d}n}=-c q^{1/c}.
\end{equation}
This system no longer has explicit time dependence, but if $q=1$ when $n=1$, then it is equivalent to Eq.~(\ref{dJ}).

\section{Modeling the tidal evolution}\label{sec:model}

\begin{figure}[t]
	\centering
	\includegraphics[width=\columnwidth]{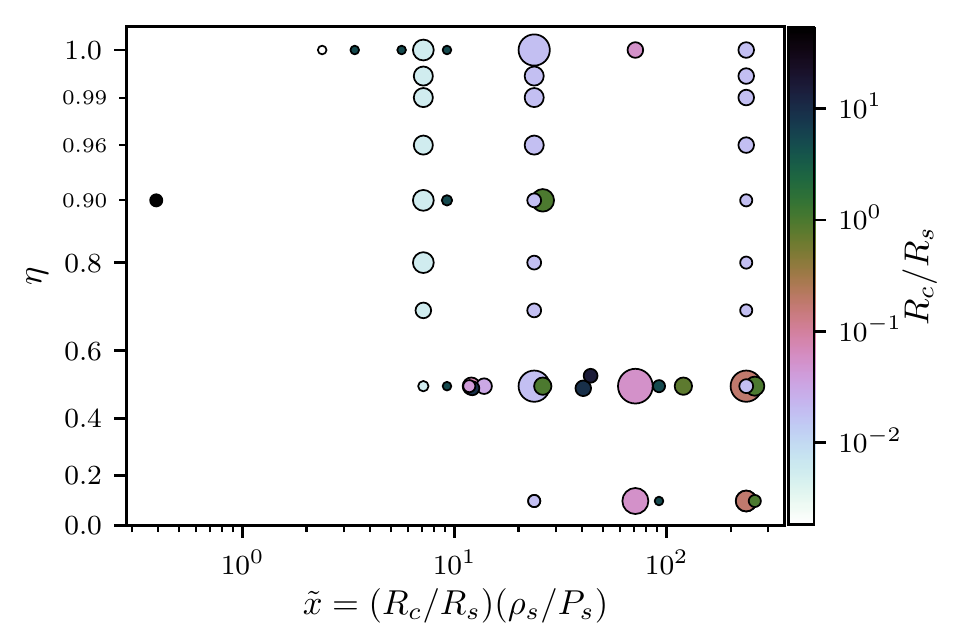}\\
	\includegraphics[width=\columnwidth]{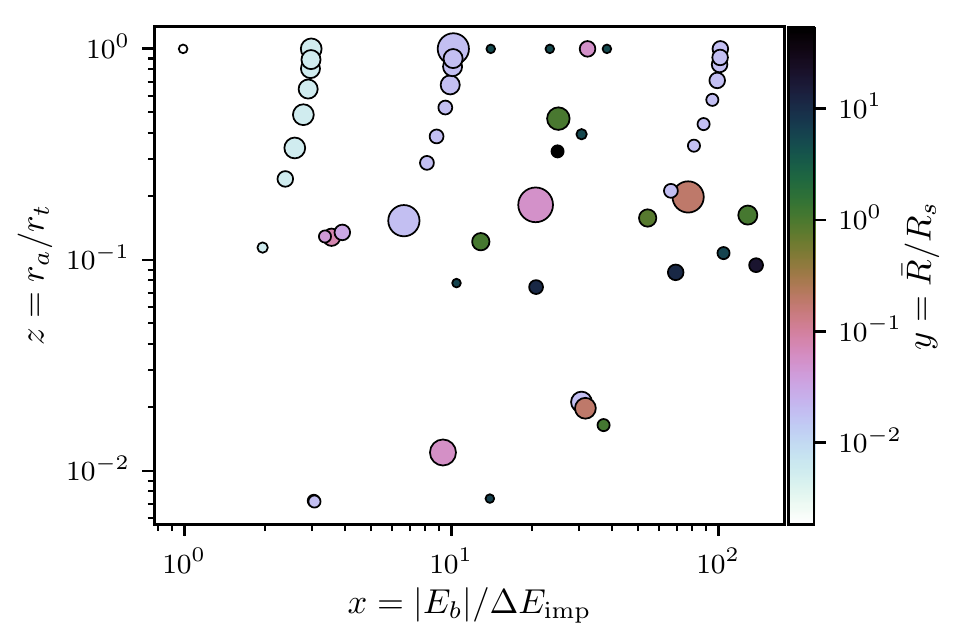}
	\caption{\label{fig:sims} These figures summarize the 52 simulations we use to tune our model.  Top: The simulations distributed in the host-subhalo system parameters.  Bottom: The simulations distributed in the reduced parameters $x$, $y$, and $z$ (see the text).  Simulations with $x\lesssim 1$ are not included in this sample because they lead to $|\mathrm{d}\ln J/\mathrm{d}\ln t| \geq 1$ too quickly.  The radius of each marker is proportional to the number of orbital periods, which ranges from 5 to 20.}
\end{figure}

Motivated by the results of the previous section, we seek a model of the form
\begin{equation}\label{model}
\ln \frac{J}{J_\mathrm{init}} = b\left[a-\frac{1}{1-c}\left(n^{1-c}-1\right)\right]
\end{equation}
for the case where $|\mathrm{d}\ln J/\mathrm{d}\ln t|<1$.  The parameters $b$ and $c$ follow immediately from Eq.~(\ref{dlnJ}), and we have inserted another parameter $a$ to fix the overall normalization.  Our goal is now to relate $a$, $b$, and $c$ to the parameters of the host-subhalo system.  For this purpose, we use the results of 52 idealized $N$-body simulations that we carried out as described in Sec.~\ref{sec:sim}.  The parameter space covered by these simulations is depicted in Fig.~\ref{fig:sims}.

\begin{figure}[t]
	\centering
	\includegraphics[width=\columnwidth]{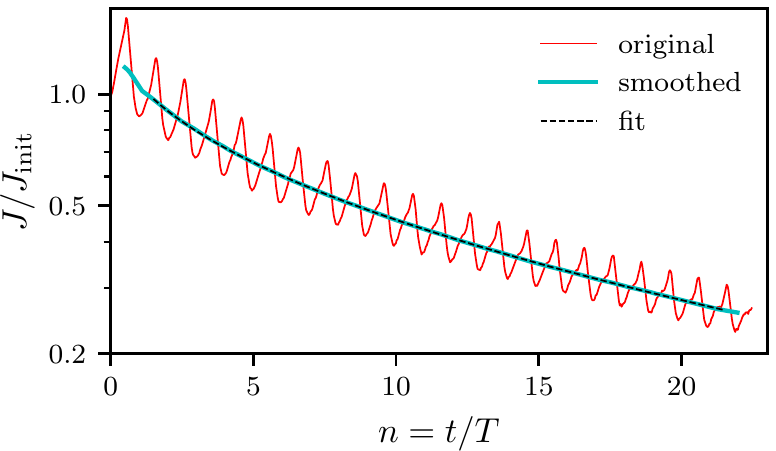}
	\caption{\label{fig:smooth} Demonstration of the fitting procedure for $J$-factor trajectories.  The trajectory (thin oscillating line) obtained from the simulation is smoothed (thick line) using a top-hat filter with width equal to the orbital period.  Equation~(\ref{model}) is fit (dashed line) to the smoothed trajectory.}
\end{figure}

For each simulation we obtain the trajectory of the subhalo's $J$ factor, stopping if ${|\mathrm{d}\ln J/\mathrm{d}\ln t|\geq 1}$ or otherwise at an arbitrarily chosen simulation termination time.  As we discussed in the previous section, the evolutionary behavior changes markedly when ${|\mathrm{d}\ln J/\mathrm{d}\ln t|\geq 1}$, but precise predictions in this regime are unnecessary.  Next, we convolve the $J$-factor trajectory in log space with a top-hat filter of width equal to the radial (apocenter-to-apocenter) orbit period.\footnote{For circular orbits the radial orbit period is ill defined, and we substitute its limit as the orbit approaches circular as obtained using the fitting form in Appendix~\ref{sec:fits}.}  This step suppresses the influence of the periodic oscillatory behavior observed in Sec.~\ref{sec:trends}; we are effectively finding the moving logarithmic average of $J$ over this period.  Finally, we fit Eq.~(\ref{model}) to this smoothed trajectory of $J$, but we only employ times after the end of the first radial period (so the first point is at $n=1.5$, whose corresponding $J$ factor averages from $n=1$ to $n=2$).  This restriction is intended to remove the influence of any transient effects associated with suddenly turning on the tidal field.  Additionally, in case the smoothing procedure fails to fully suppress periodic effects, we minimize any resulting bias by ending the fit at an integer number of orbits (so for instance, we might end at $n=15.5$, corresponding to the average $J$ from $n=15$ to $n=16$).  The number of radial orbits fit through this procedure is represented in Fig.~\ref{fig:sims} as the marker size; this number is a proxy for how much information that simulation provides.\footnote{When performing fits, we weight a simulation spanning $n$ orbits by $\sqrt n$.}  Figure~\ref{fig:smooth} illustrates the procedure; the smoothing filter suppresses oscillations quite effectively.

\subsection{Parameter $b$: The initial $J$-factor decay rate}\label{sec:b}

\begin{figure}[t]
	\centering
	\includegraphics[width=\columnwidth]{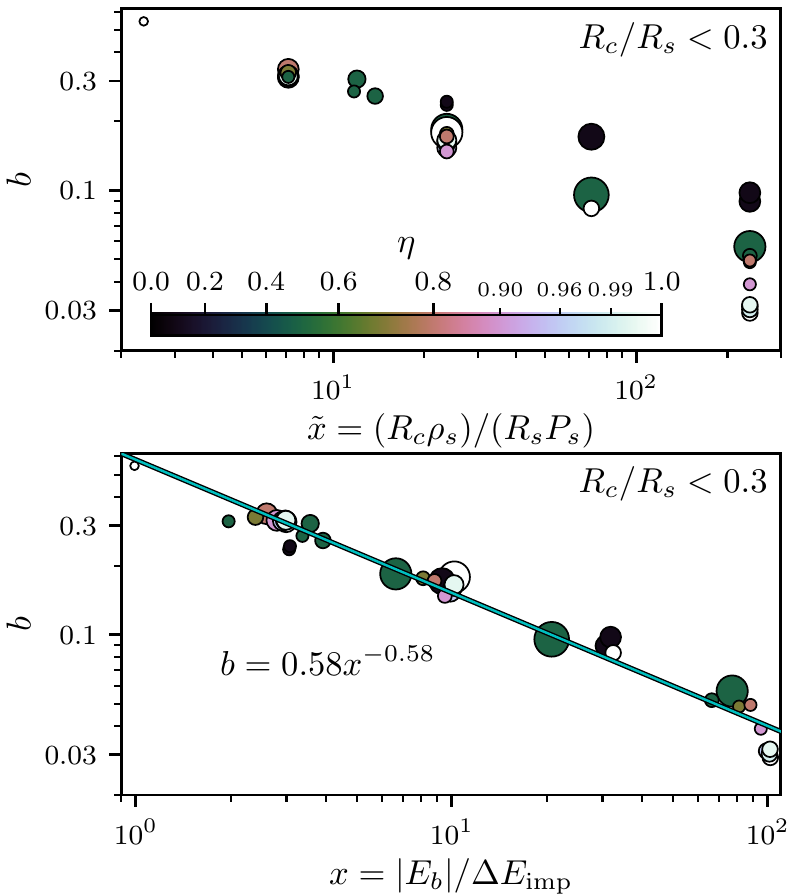}
	\caption{\label{fig:xvsx} The dependence of the trajectory parameter $b$ on system parameters in the self-similar regime ($R_c/R_s\ll 1$).  Top: There is a trend between $b$ and $\tilde x$, but it is polluted by residual sensitivity to the orbit-shape parameter $\eta$.  Bottom: $b$ is a power law in $x$ with little residual sensitivity to the orbit shape, and the best fit is plotted as a solid line.  The color scale is the same for both panels.  Each marker is a simulation, and the marker radius is proportional to the number of orbital periods, which ranges from 7 to 20 for this sample.}
\end{figure}

From Fig.~\ref{fig:Jtdr} we anticipate that $b$ should depend strongly on the orbital radius.  For simplicity we first study the self-similar regime, $R_c\ll R_s$.  The upper panel of Fig.~\ref{fig:xvsx} plots $b$ against the reduced orbital radius $\tilde x$ for the 36 of our simulations that satisfy $R_c/R_s < 0.3$.  While $b$ is strongly sensitive to $\tilde x$, there is also significant sensitivity to the orbital shape, parametrized by $\eta$.  However, it turns out that we can eliminate the shape dependence of $b$ by defining the reduced orbital radius more carefully.

We first remark that $\tilde x \sim |E_b|/\Delta E$, where $E_b$ is the binding energy of a particle at the subhalo's initial scale radius $r_s$ and $\Delta E$ is the energy injected into that particle by tidal forces over the subhalo's orbital period.  To see this, observe that the particle's binding energy is
\begin{equation}\label{Eb}
E_b = -4\pi (\ln 2) G \rho_s r_s^2
\end{equation}
(per mass).  Meanwhile, the tidal acceleration on this particle is roughly $(r_s/R)F$, where $R$ is the subhalo's orbital radius and $F$ is the host force (per mass) at radius $R$.  In the self-similar regime, $F\sim G P_s R_s$.  The total velocity injected into the particle is $\Delta v\sim F T$, where $T$ is the subhalo orbital period.  Since $T\sim \sqrt{R/F}$, the energy injection (per mass) is $\Delta E \sim (\Delta v)^2 \sim G P_s R_s r_s^2/R$, and since $R\sim R_c$, this leads to $|E_b|/\Delta E\sim \tilde x$.

With this motivation, we define
\begin{equation}\label{x}
x\equiv|E_b|/\Delta E_\mathrm{imp}
\end{equation}
as a more exact version of $\tilde x$.  Here, $\Delta E_\mathrm{imp}$ is the energy injection per orbit on a particle at $r_s$ computed using the impulse approximation as in Ref.~\cite{gnedin1999tidal}.  In this approximation, the subhalo particle is treated as stationary while Eq.~(\ref{tidalforce}) is integrated to find the velocity (and hence energy) injection.  We supply a fitting formula for $\Delta E_\mathrm{imp}$ in Appendix~\ref{sec:fits} for convenience.  As intuition, $x$ is of order the ratio
\begin{equation}\label{xsim}
x\sim\rho_s/\bar P(<\!R_c)
\end{equation}
between the subhalo's density and the average host density within the subhalo's orbital radius, a connection that follows from the observation that ${\Delta E_\mathrm{imp}/r_s^2 \sim F(R_c)/R_c}$ (see Appendix~\ref{sec:fits}).  In the bottom panel of Fig.~\ref{fig:xvsx} we plot $b$ against $x$ for the self-similar regime.  Evidently, our definition of $x$ captures most or all of the sensitivity of the parameter $b$ to the orbit shape, and
\begin{equation}\label{b_}
b=b_0 x^{-b_1}, \ \text{if}\ R_c\ll R_s,
\end{equation}
with $b_0=0.58$ and $b_1=0.58$.  This success is remarkable; the approximation that subhalo particles are stationary during the application of tidal forces can only be valid for highly eccentric orbits, and yet the impulsive energy calculation accurately predicts the tidal evolution for more circular orbits as well.

\begin{figure}[t]
	\centering
	\includegraphics[width=\columnwidth]{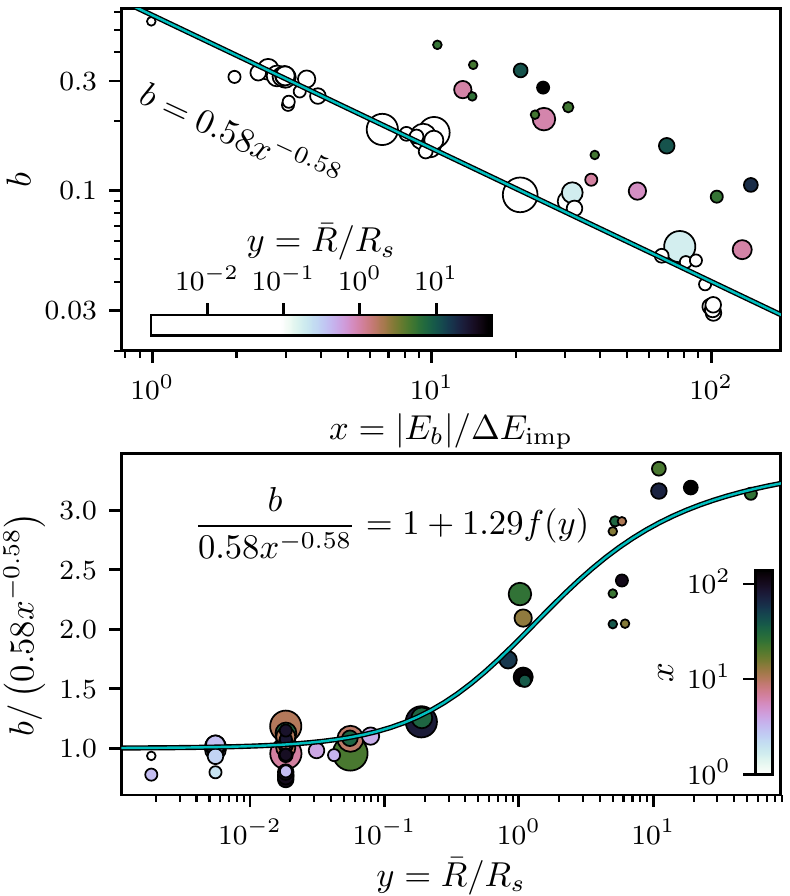}
	\caption{\label{fig:b} The dependence of $b$ on $x$ and $y$.  Top: At each radius $y=\bar R/R_s$, $b$ appears to follow a similar power law in $x$ with a different normalization.  The solid line is duplicated from Fig.~\ref{fig:xvsx}.  Bottom: Scaling of the normalization of $b$ with radius $y$.  The best fit is plotted, with $f(y)$ defined in Eq.~(\ref{asymf}).  Each marker is a simulation, and the marker radius is proportional to the number of orbital periods, which ranges from 5 to 20.}
\end{figure}

To complete our understanding of the parameter $b$ we must move beyond the self-similar regime.  In the upper panel of Fig.~\ref{fig:b}, we plot $b$ against $x$ for all of our simulated subhalos.  The color scale indicates the time-averaged orbital radius $\bar R$ in units of $R_s$.  It appears that the effect of leaving the self-similar regime is to alter the normalization of $b$ while keeping the power-law sensitivity to $x$ unchanged.  In particular, we may write
\begin{equation}\label{b}
b=b_0 x^{-b_1}\left[1+b_2 f(y)\right],
\end{equation}
for some function $f(y)$ and parameter $b_2$, where we define
\begin{equation}\label{y}
y\equiv\bar R/R_s.
\end{equation}
For convenience, we supply a fitting formula for $\bar R$ in Appendix~\ref{sec:fits}.  While we could use the circular orbit radius $R_c$ instead, we favor $\bar R$ because its physical significance is clearer.

To define the function $f(y)$, we consider the physical impact of leaving the self-similar regime.  The magnitudes of the tidal forces are altered, but this effect should be accounted for by the definition of $x$.  However, the directions of the tidal forces also change.  In particular, Eq.~(\ref{tidalforce}) implies that there are stretching tidal forces proportional to $\mathrm{d}F/\mathrm{d}R$ along the radial axis from the host and compressive tidal forces proportional to $F/R$ along the perpendicular directions.  When the host force profile $F(R)$ is self-similar, the ratio between the stretching and compressive forces is fixed.\footnote{In fact, for an NFW profile, $\mathrm{d}F/\mathrm{d}R=0$ when $R\ll R_s$.}  Beyond the self-similar regime, however, the ratio between these forces can change.  With this motivation, we define $f(R/R_s)\equiv (\mathrm{d}F/\mathrm{d}R)/(F/R)$ as this ratio.  For the NFW profile this definition implies that
\begin{equation}\label{asymf}
f(y) = \frac{2\ln(1+y)-y(2+3y)/(1+y)^2}{\ln(1+y)-y/(1+y)}.
\end{equation}
In the bottom panel of Fig.~\ref{fig:b}, we plot $b/(b_0 x^{-b_1})$ against $y$ for the purpose of tuning the parameter $b_2$ in Eq.~(\ref{b}).  We find that this equation\footnote{The force-ratio argument motivates any expression of the form $\left[1+b_2 f(y)^\alpha\right]^\beta$, but we assume for simplicity that $\alpha=\beta=1$.} works reasonably well, and we obtain $b_2=1.29$.  The introduction of stretching tidal forces increases the efficiency of tidal effects, which is reflected as an increase in the decay rate $b$ of the $J$ factor.

\subsection{Parameter $a$: The $J$-factor normalization}\label{sec:a}

\begin{figure}[t]
	\centering
	\includegraphics[width=\columnwidth]{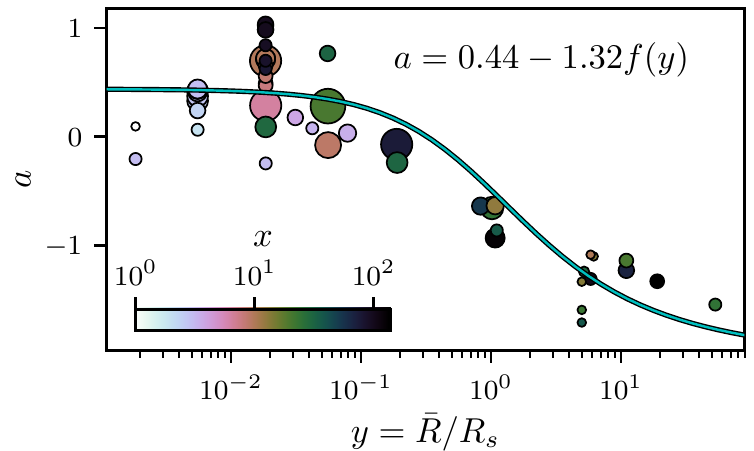}
	\caption{\label{fig:a} The dependence of the trajectory parameter $a$ on the orbital radius parameter $y$.  The best-fitting curve is plotted as a solid line using the definition of $f(y)$ in Eq.~(\ref{asymf}).  Each marker is a simulation, and the marker radius is proportional to the number of orbital periods, which ranges from 5 to 20.}
\end{figure}

We next handle the overall normalization of $J/J_\mathrm{init}$.  According to Eq.~(\ref{model}), the $J$ factor changes by the factor $\mathrm{e}^{ab}$ after the first orbit, which is sensitive to a second parameter: $a$.  Because of the way we defined this parameter, it turns that $a$ is almost wholly sensitive to $y=\bar R/R_s$ alone.  Figure~\ref{fig:a} plots $a$ against $y$ for all of our simulations, and we find that with only moderate scatter,
\begin{equation}\label{a}
a = a_0-a_1 f(y)
\end{equation}
with $a_0=0.44$ and $a_1=1.32$.

In some sense, the parameter $a$ describes the initial behavior of the subhalo as it equilibrates---to the extent that this is possible---into the tidal field generated by the host.  When $y\ll 1$ all nonzero tidal forces are compressive, so the $J$ factor is initially slightly boosted ($a>0$).  However, when $y\gtrsim 1$, the stretching tidal forces cause the $J$ factor to be initially suppressed ($a<0$).  Note that the trajectory given by Eq.~(\ref{model}) is only valid after this equilibration takes place, so it is not valid for $n<1$.

\subsection{Parameter $c$: The loss of tidal efficiency}\label{sec:c}

Finally, we address the parameter $c$ that characterizes the drop in the efficiency of tidal effects over time.  As we found in Sec.~\ref{sec:trends}, $c$ is sensitive to the orbit shape; more eccentric orbits yield smaller values of $c$ while more circular orbits yield larger values.  In the self-similar regime we could write $c$ as a function of $\eta$, since $\eta$ completely describes the orbit shape.  However, beyond this regime, orbits with the same $\eta$ could have different shapes.  Thus, to accurately describe the sensitivity of the parameter $c$ to the host-subhalo system, it is necessary to find the correct orbit-shape parametrization.

\begin{figure}[t]
	\centering
	\includegraphics[width=\columnwidth]{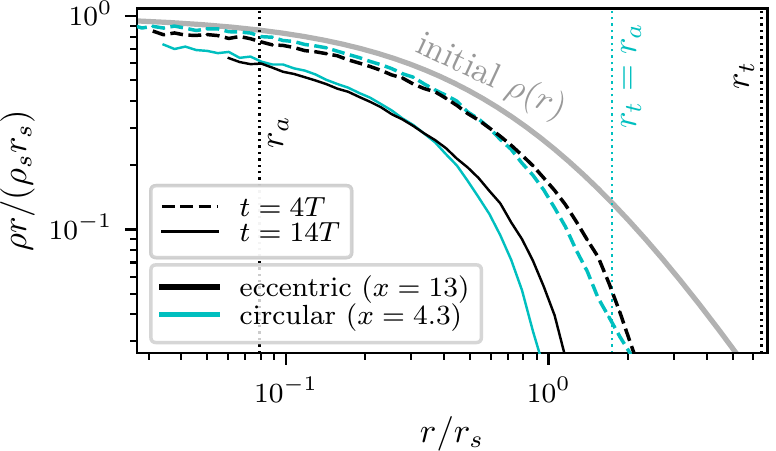}
	\caption{\label{fig:radial_shape} The influence of the shape of a subhalo's orbit on its density profile after tidal evolution.  One subhalo is on a highly eccentric orbit ($\eta=0.1$) while the other its on a circular orbit; we plot the density profiles after 4 and 14 orbits.  The key difference is that the circular orbit strips material primarily from the outskirts, while the eccentric orbit strips more material from the interior.  This difference can be understood in terms of the adiabatic shielding radius $r_a$ and its comparison to the tidal radius $r_t$ (see the text), shown as dotted lines for both orbits.  The subhalos are chosen to yield similar density profiles and do not have the same energy parameter $x$.}
\end{figure}

We argued in Sec.~\ref{sec:trends} that the loss of tidal efficiency encoded in $c$ is related to changes in the shape of the subhalo density profile.  The connection to the orbital shape is that circular orbits tidally heat material more predominantly in the outskirts of the subhalo, while eccentric orbits can alter the density profile further inward.  This tendency is illustrated in Fig.~\ref{fig:radial_shape}, which depicts the tidally altered density profiles of two subhalos with different orbit shapes.  The subhalo on the circular orbit loses more material from its outskirts, while the subhalo on the eccentric orbit loses more material from its interior.

Differences in the radii at which material is heated can be understood in terms of adiabatic shielding (e.g., Refs.~\cite{spitzer1987dynamical,weinberg1994adiabatic,weinberg1994adiabatic2,gnedin1999self}).  Deep within the subhalo, the internal dynamical timescale is much shorter than the timescale over which the external tidal field changes.  In this case, the conservation of adiabatic invariants prevents any energy injection by tidal forces; these radii are adiabatically shielded.  Meanwhile, adiabatic shielding is connected to the shape of the subhalo's orbit.  The timescale over which tidal forces change is related to the timescale of the pericenter passage, which can be very short for highly eccentric orbits.

Up to factors of order unity, the subhalo's internal dynamical timescale is $t_\mathrm{dyn} \sim (G m(r)/r^3)^{-1/2}$ at radius $r$, where $m(r)$ is the subhalo mass profile \cite{binney1987galactic}.  Meanwhile, the pericenter passage timescale is $t_p\sim R_p/V_p$, where $R_p$ and $V_p$ are the radius and velocity at the pericenter, respectively.  To make precise the connection between the orbit shape and the radii at which tidal heating is efficient, we define the adiabatic shielding radius $r_a$ as the radius at which $t_\mathrm{dyn}=t_p$.  This definition motivates a characteristic density scale
\begin{equation}\label{rhoa}
\rho_a\equiv \frac{V_p^2}{G R_p^2} = \eta^2\frac{M(R_c)R_c}{R_p^4},
\end{equation}
so that $r_a$ is the radius at which $m(r_a)/r_a^3=\rho_a$.  Note that we used the definitions of the circular orbit radius $R_c$ and orbit circularity $\eta$ to eliminate $V_p$ from Eq.~(\ref{rhoa}); $M(R)$ is the host mass profile at radius $R$.

To quantify changes in the shape of the subhalo density profile, we can compare the radius $r_a$ below which material is shielded to the tidal radius $r_t$ above which all material is stripped.  The tidal radius is the radius above which the tidal force from the host exceeds the gravitational force from the subhalo.  There are several definitions of the tidal radius in the literature, but they are all related to the expression \cite{klypin1999galaxies,klypin2015halo} $r_t=R[m(r_t)/M(R)]^{1/3}$ by (possibly nonconstant) factors of order unity (see, e.g., Ref.~\cite{van2017disruption}).  The tidal radius is only well defined for circular orbits, but it is common to apply the concept to eccentric orbits as well \cite{van2017disruption}.  In particular, if we seek the radius above which all material is stripped, we can define the tidal radius $r_t$ using the orbital apocenter radius $R_a$.  In this case, there is a characteristic density scale
\begin{equation}\label{rhot}
\rho_t\equiv M(R_a)/R_a^3,
\end{equation}
and $r_t$ is the solution to $m(r_t)/r_t^3=\rho_t$.

\begin{figure}[t]
	\centering
	\includegraphics[width=\columnwidth]{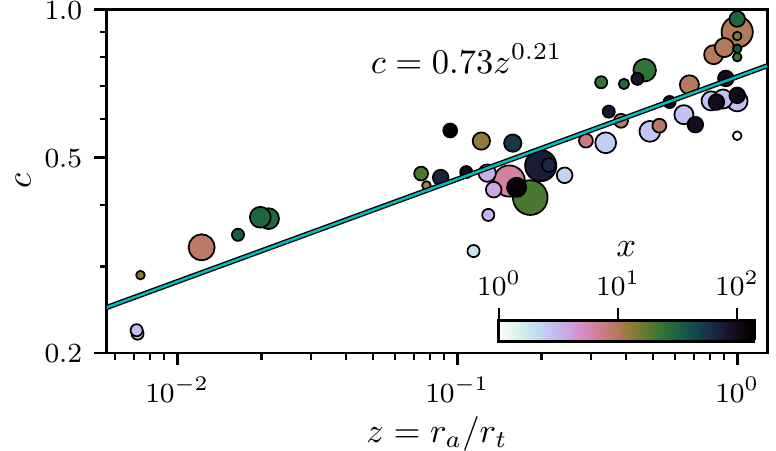}
	\caption{\label{fig:c} The dependence of the trajectory parameter $c$ on the system parameter $z=r_a/r_t$.  With moderate scatter, $c$ follows a power law in $z$, shown as the solid line.  Each marker is a simulation, and the marker radius is proportional to the number of orbital periods, which ranges from 5 to 20.}
\end{figure}

Anticipating that the drop in the efficiency of tidal effects encoded in the parameter $c$ is a consequence of changes to the shape of the subhalo density profile, we may hypothesize that $c$ is sensitive to the ratio
\begin{equation}\label{z}
z\equiv r_a/r_t,
\end{equation}
which ranges from 0 for radial orbits to 1 for circular orbits.  For simplicity, in defining $z$ we employ the subhalo's initial NFW mass profile [see Eq.~(\ref{nfwmass})].  Figure~\ref{fig:c} shows the relationship between $c$ and $z$; there is some scatter, but the trend is that\footnote{We argued in Sec.~\ref{sec:toy} that $c>0$ is connected to changes in the density profile's shape.  In this light, Eq.~(\ref{c}) implies that in the limit $z=0$ where the tidal energy injection is completely impulsive, the shape of the density profile does not change over successive orbits.  This notion is consistent with the results of Ref.~\cite{delos2019evolution}, which found that impulsive point-object encounters yield a universal density profile.}
\begin{equation}\label{c}
c = c_0 z^{c_1}
\end{equation}
with $c_0=0.73$ and $c_1=0.21$.  Note that $z$ is not solely a function of the subhalo's orbit.  Because it depends on the subhalo mass profile $m(r)$, it is also sensitive to the density ratio $\rho_s/P_s$.  We explored using $\rho_t/\rho_a$, a purely orbital parameter, instead of $r_a/r_t$.  This parameter exhibited a similar power-law relationship with $c$, but it left significant residual sensitivity to the parameter $x$, which is related to $\rho_s/P_s$.  Using $z=r_a/r_t$ mostly eliminates that sensitivity.

\section{Model summary and discussion}\label{sec:disc}

In the last section, we developed a model for the evolution of a subhalo's $J$ factor due to tidal effects as a function of parameters of the host-subhalo system.  As long as the $J$ factor decays slower than $|\mathrm{d}\ln J/\mathrm{d}\ln t|=1$, its trajectory is well fit by the expression
\begin{equation}\tag{\ref{model}}
\ln \frac{J}{J_\mathrm{init}} = b\left[a-\frac{1}{1-c}\left(n^{1-c}-1\right)\right].
\end{equation}
Here, $n=t/T$ is the number of subhalo orbits, and $a$, $b$, and $c$ are parameters that depend on the host-subhalo reduced system parameters $x$, $y$, and $z$ through
\begin{align}
\tag{\ref{a}}
a &= a_0-a_1 f(y),
\\
\tag{\ref{b}}
b &= b_0 x^{-b_1}\left[1+b_2 f(y)\right],
\\
\tag{\ref{c}}
c &= c_0 z^{c_1}
\end{align}
with ${a_0=0.44}$, ${a_1=1.32}$, ${b_0=0.58}$, ${b_1=0.58}$, ${b_2=1.29}$, ${c_0=0.73}$, and ${c_1=0.21}$.  In Appendix~\ref{sec:fits}, we detail how to compute $x$, $y$, $z$, and $T$ from the subhalo parameters $r_s$ and $\rho_s$, host parameters $R_s$ and $P_s$, and orbital parameters $R_c$ and $\eta$.

Equation~(\ref{model}) applies only when ${|\mathrm{d}\ln J/\mathrm{d}\ln t|<1}$.  The $J$ factor's precise behavior when $|\mathrm{d}\ln J/\mathrm{d}\ln t|\gtrsim 1$ is of little consequence, as the subhalos in this regime contribute only minimally to aggregate annihilation signals.  Nevertheless, it is useful to have an approximate treatment in this regime.  As we discussed in Sec.~\ref{sec:trends}, when $|\mathrm{d}\ln J/\mathrm{d}\ln t|\geq 1$ it is a reasonable approximation to enforce $-\mathrm{d}\ln J/\mathrm{d}\ln t=1$, i.e., $J\propto n^{-1}$.  We define 
\begin{equation}\label{n1}
n_1=b^{1/(c-1)}
\end{equation}
as the orbit count at which $-\mathrm{d}\ln J/\mathrm{d}\ln t=1$.  Additionally, when $b>1$ (so $n_1<1$), we cannot expect our treatment of the normalization of $J$ (Sec.~\ref{sec:a}) to be accurate.  To handle these issues, we can write
\begin{equation}\label{model1}
\frac{J}{J_\mathrm{init}}=
\begin{cases} 
\exp\!\left\{b\left[a-\frac{1}{1-c}\left(n^{1-c}\!-\!1\right)\right]\right\},
&\!\!\!\text{if } n\leq n_1, b<1,  \\
\exp\!\left\{b\left[a-\frac{1}{1-c}\left(\frac{1}{b}-1\right)\right]\right\}\frac{n_1}{n},
&\!\!\!\text{if } n> n_1, b<1,
\\
\mathrm{e}^a/n,
&\!\!\!\text{if } b\geq 1,
\end{cases}
\end{equation}
where the last case follows from continuity considerations.  We further note that this equation is valid only when $n\geq 1$ (see Sec.~\ref{sec:a}).

One can now use our model to understand the emission from a host halo due to dark matter annihilation in subhalos.  In particular, one can sample subhalos from an orbital distribution in $R_c$ and $\eta$ (e.g., Refs.~\cite{tormen1997rise,zentner2005physics,khochfar2006orbital,wetzel2011orbits,jiang2015orbital,van2017dissecting}).  Accounting for tidal evolution, each subhalo's contribution to the dark matter annihilation signal is then scaled by the orbit-dependent function given by Eq.~(\ref{model1}).  In Ref.~\cite{delos2019gamma} (in preparation), we will use this model to study the annihilation signature arising from the extreme-density microhalos that result from certain early universe scenarios.  In this case, the orbital distribution of subhalos is the same as that of particles, and one may employ the host halo's distribution function (e.g., Ref.~\cite{widrow2000distribution}) to sample subhalo orbits.

Our model does not include the periodic oscillations in the $J$ factor observed in Sec.~\ref{sec:trends}.  These oscillations do not affect the overall annihilation rate in subhalos, but they still introduce a systematic biasing effect where subhalos at smaller radii have larger $J$, and this effect can alter the morphology of an annihilation signal.  However, we remark that these oscillations only have a significant amplitude in the $R_c\ll R_s$ regime, when all tidal forces are compressive, and at small $x\lesssim 10$.  Because of these restrictions, we anticipate that their impact is minor.  However, in forthcoming work \cite{delos2019gamma} we will quantify the impact of these oscillations.

We also address another potential limitation to our model.  The differential equation driving it, Eq.~(\ref{dJ}), has explicit time dependence in the factor $n^{-c}$, so the resulting tidal evolution is not completely determined by the system's instantaneous state.  Physically, we view $n^{-c}$ as a proxy for unknown physical variables [e.g., Eq.~(\ref{dJq})], and as long as the host halo's density profile and the subhalo's orbit are static, this formulation poses no difficulty.  Since halos grow from the inside outward, subsequent accretion is not expected to significantly alter the density profile of a host halo at the radii of already-present subhalos, so the host halo is generally expected to remain static.  Moreover, if dynamical friction is negligible [see Eq.~(\ref{friction})], the subhalo's orbit is also static.  However, there is a scenario where a subhalo's host is expected to change dramatically.  If the host is itself a satellite of a larger host halo, then the subhalo may be tidally stripped from its host, becoming itself a satellite of the superhost.  In this scenario, it is not obvious how to continue the subhalo's tidal evolution.

If the initial host-subhalo system yields trajectory parameters $a^\prime$, $b^\prime$, and $c^\prime$ and the new host-subhalo system yields parameters $a$, $b$, and $c$, then a self-consistent way to treat this problem is to substitute the factor $n^{-c}$ in Eq.~(\ref{dJ}) with $(n+n^{\prime c^\prime/c})^{-c}$ and integrate the resulting expression.  This treatment follows from the assumption that the parameter $q=n^{-c}$ in Eq.~(\ref{dJq}) is a function of the subhalo alone.  Additionally, the $J$ factor should be rescaled by $\mathrm{e}^{ba-b^\prime a^\prime}$, a consideration motivated by the discussion in Sec.~\ref{sec:a}.  However, it turns out that while this treatment works reasonably well for a portion of the $a^\prime$, $b^\prime$, $c^\prime$, $a$, $b$, $c$ parameter space, it does not accurately predict every scenario; the parameter $q$ in Eq.~(\ref{dJq}) is not a function of the subhalo alone.  We leave a detailed investigation of this problem to future work.

As another caveat, the long-term accuracy of the trajectory in Eq.~(\ref{model}) relies on the assumption that the efficiency of tidal effects follows precisely the power law $n^{-c}$, as described by Eq.~(\ref{dJ}).  While such a power law is a natural assumption [e.g, Eq.~(\ref{dJq})] and is borne out in our simulations, it does not have a direct physical motivation; tidal heating models considered in Sec.~\ref{sec:toy} and elsewhere \cite{pullen2014nonlinear} can only reproduce $c=0$.  Without such motivation it is unclear that this power-law behavior should extend beyond the $n=20$ orbits of our longest simulations.  Also, the system parameter $z=r_a/r_t$ that sets the power-law index $c$ is defined based on two concepts that are not themselves entirely well defined: the adiabatic shielding radius $r_a$ and the tidal radius $r_t$.  Moreover, since our model does not predict the larger evolution of the subhalo density profile, we use the subhalo's initial density profile to define $r_a$ and $r_t$ even though the density profile quickly begins to change.  For these reasons, we anticipate that it is possible to find a better-motivated parameter to replace $z=r_a/r_t$.

Nevertheless, this model describes the results of our simulations with remarkable success.  As further validation, we consider the library of idealized subhalo simulations, called DASH (for dynamical aspects of subhaloes), published by Ref.~\cite{ogiya2019dash}.  These simulations have a lower resolution than ours, but because of the extraordinary volume of this library, it still supplies a valuable test for our model.  In Appendix~\ref{sec:dash} we verify that modulo substantial scatter and certain systematic effects associated with their lower resolution, the DASH simulations are consistent with our model.

\section{Comparison to previous work}\label{sec:compare}

Numerous prior works have endeavored to model the impact of tidal effects on a subhalo's dynamical evolution
\cite{taylor2001dynamics,hayashi2003structural,penarrubia2005effects,van2005mass,zentner2005physics,kampakoglou2006tidal,gan2010improved,penarrubia2010impact,pullen2014nonlinear,jiang2016statistics}.  In this section, we explore how our results compare to those of previous studies.  Motivated primarily by simulations, our model is based on the notion that a subhalo's $J$-factor evolution is determined by
\begin{equation}\tag{\ref{dJ}}
\frac{1}{J}\frac{\mathrm{d}J}{\mathrm{d}n}=-bn^{-c},
\end{equation}
where $b$ and $c$ are functions of the host-subhalo system and $n$ counts the number of orbits.  In contrast to our focus on the $J$ factor, previous works have largely focused on the evolution of a subhalo's total bound mass $m_\mathrm{bound}$ and of its maximum circular velocity $v_\mathrm{max}$ and corresponding radius $r_\mathrm{max}$.  However, the general form of our model is not specific to the $J$ factor, and we show in Appendix~\ref{sec:rmax} that it can also describe the evolution of $v_\mathrm{max}$, $r_\mathrm{max}$, and $m_\mathrm{bound}$.

Despite the broad applicability of our model suggested by Appendix~\ref{sec:rmax}, no prior work (to our knowledge) has proposed tidal evolution of the form given in Eq.~(\ref{dJ}).  Broadly, prior models of tidal evolution fall into two main categories, although a given work may employ more than one:
\begin{enumerate}[label={(\arabic*)}]
	\item Tidal stripping models, where material outside the characteristic tidal radius [e.g., Eq.~(\ref{rhot})] is assumed to be stripped over some time period;
	\item Tidal heating models, where energy injected by tidal forces heats subhalo material, causing it to rise and possibly become freed from the subhalo.
\end{enumerate}
We found in Sec.~\ref{sec:b} that the parameter $b$ in Eq.~(\ref{dJ}), which characterizes the rate of tidal evolution, is tightly sensitive to the energy injected by tidal forces (see Fig.~\ref{fig:xvsx}).  Additionally, we observed in Sec.~\ref{sec:trends} that the tidal evolution in our simulations closely resembles that predicted by a toy model of tidal heating.  Consequently, we anticipate that of the two classes of models, tidal heating models should yield results most similar to those of our model.  We will first compare the results of our model to those of the tidal heating model developed by Ref.~\cite{pullen2014nonlinear}, hereafter P14.

However, prior treatments of dark matter annihilation within subhalos predominantly treat the impact of the host halo's tidal forces using models based on tidal stripping \cite{bartels2015boosting,hiroshima2018modeling,ando2019halo,springel2008prospects,erickcek2015dark,stref2017modeling,stref2019remnants}.  Tidal stripping models cannot prescribe how to change a subhalo's density profile below the tidal radius, but it is possible to apply a simulation-tuned prescription for how the density profile responds to mass loss \cite{hayashi2003structural,penarrubia2010impact}.  We will subsequently compare the results of our model to those of a tidal stripping model developed by Ref.~\cite{jiang2016statistics} (hereafter J16), using the prescription of Ref.~\cite{penarrubia2010impact} (hereafter P10) to predict the subhalo's density profile.  This pair of models has been employed by Refs.~\cite{bartels2015boosting,hiroshima2018modeling} to predict dark matter annihilation rates in subhalos.

\subsection{Comparison to a tidal heating model}

\begin{figure}[t]
	\centering
	\includegraphics[width=\columnwidth]{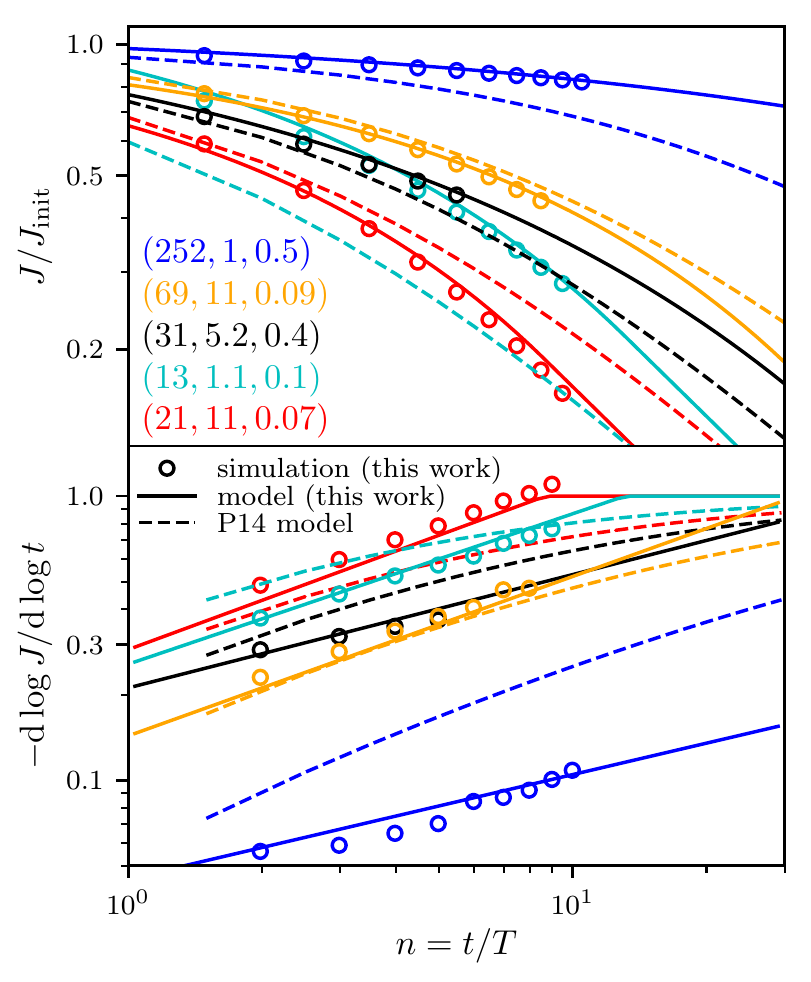}
	\caption{\label{fig:SAp} A comparison between our tidal evolution model (solid lines) and the analytic tidal heating model developed in Ref.~\cite{pullen2014nonlinear} (P14, dashed lines).  This figure shows the $J$-factor trajectory and its logarithmic derivative for different host-subhalo parameters $(x,y,z)$, listed on the figure.  We also show our simulation results (as circles) for these parameters.  The P14 predictions exhibit the correct trends, but they are only reasonably accurate for a small range of host-subhalo system parameters.}
\end{figure}

We first compare our model's predictions to those of the analytic tidal heating model given in P14.  In the tidal heating picture, energy injected by tidal forces causes subhalo material to move to higher radii, and P14 employed the assumption of virial equilibrium to predict this change in radius and consequently the subhalo's new density profile.  We follow the prescription in P14\footnote{For simplicity, we compute the energy injection $\Delta E$ directly using the impulse approximation (Appendix~\ref{sec:fits}), neglecting additional corrections suggested in P14; these corrections will not qualitatively alter the results.} to compute the evolution of a subhalo's density profile, subsequently integrating it to obtain the $J$ factor.  Figure~\ref{fig:SAp} shows a sample of the resulting $J$-factor trajectories, and we compare those trajectories to our model's predictions and to the results of our simulations.  Generally, we find that for a model constructed from first principles, the P14 model is remarkably accurate.  However, it does not fully capture the sensitivity of tidal evolution to system parameters, a matter we explore next.

The quantity $Q=\Delta E/r^2$ employed by P14 is related to host-subhalo system parameters by $Q/(G \rho_s) = 4\pi(\ln 2) N/x$, where $N$ is the number of orbits over which the energy injection is taken and $x$ is the system parameter (see Sec.~\ref{sec:model}).  For our comparison, we take $N=1$ and iterate the calculation, assuming the halo revirializes during each orbit.  Since the density profile evolution in P14 is only sensitive to the ratio $Q/(G \rho_s)$, we see immediately that this model's predictions are sensitive only to the system parameter $x$ and are insensitive to $y$ and $z$.  Additionally, the $J$-factor evolution predicted by P14 turns out to be only sensitive to $x$ in the combination $n/x$, where $n=t/T$ is the number of orbits, so every system follows the same trajectory rescaled in time.  In this respect, the P14 model is similar to the toy model we explored in Sec.~\ref{sec:toy}, which was only sensitive to the combination $fn$ of system parameters $f$ and orbit count $n$.  In fact, the P14 model approximately obeys the toy model solution Eq.~(\ref{toysolJ}) with $b\simeq 3.2/x$ and $B=1$, but it can potentially transition between the $n/x\ll 1$ and $n/x\gg 1$ regimes extremely slowly, and all behavior seen in Fig.~\ref{fig:SAp} is in the intermediate regime.

The combination of its single time-rescaled trajectory and its insensitivity to $y$ and $z$ leaves the P14 model unable to accurately predict tidal evolution in the full host-subhalo parameter space.  We see evidence of this deficiency in Fig.~\ref{fig:SAp}, but we further note that we did not plot any subhalos in the $y\ll 1$ regime.  In this regime, the P14 model dramatically overestimates the impact of tidal stripping since it does not account for the directions of tidal forces, which are encapsulated in the parameter $y$.  While the P14 model yields reasonably accurate predictions over a small range of host-subhalo system parameters, our model can accurately predict the evolution of a much broader variety of systems.

\subsection{Comparison to a tidal stripping model}

Finally, we compare our model to a semianalytic model of tidal stripping that has been employed in previous calculations of annihilation rates in the substructure \cite{bartels2015boosting,hiroshima2018modeling}.  This semianalytic model uses the tidal stripping model in J16 to characterize a subhalo's mass loss, subsequently using the results of P10 to connect this mass loss to the subhalo's density profile and hence annihilation signal.  In J16, the rate of mass loss for subhalos of mass $m$ inside a host halo of mass $M$, averaged over subhalo orbits, is modeled using
\begin{equation}\label{J16}
\frac{\mathrm{d}m}{\mathrm{d}t} = -\mathcal{A} \frac{m}{t_\mathrm{dyn}}\left(\frac{m}{M}\right)^\zeta.
\end{equation}
Here, $\mathcal{A}$ and $\zeta$ are simulation-tuned parameters and $t_\mathrm{dyn}$ is the host's dynamical timescale at its virial radius (e.g., Ref.~\cite{binney1987galactic}).  For this comparison we adopt J16's central values $\mathcal{A}=0.86$ and $\zeta=0.07$.

In P10, it is shown that the subhalo's maximum circular velocity $v_\mathrm{max}$ and the radius $r_\mathrm{max}$ at which it is attained are related to the fraction $m/m_\mathrm{acc}$ of the subhalo's mass that remains gravitationally bound, where $m_\mathrm{acc}$ is the subhalo's virial mass at accretion.  We confirm in Appendix~\ref{sec:rmax} that these relations are reasonably accurate if $y>1$ and the subhalos have concentration $r_\mathrm{vir}/r_s\simeq 20$ at accretion.  If we assume that subhalos possess NFW profiles, then the mass fraction $m/m_\mathrm{acc}$ predicted by J16 thereby determines each subhalo's $J$ factor.

\begin{figure}[t]
	\centering
	\includegraphics[width=\columnwidth]{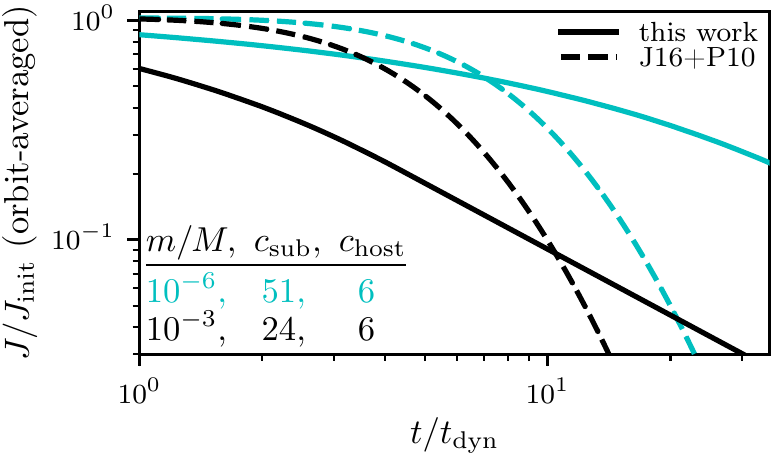}
	\caption{\label{fig:SA} A comparison between our tidal evolution model and the semianalytic model developed in Refs. \cite{jiang2016statistics} (J16) and~\cite{penarrubia2010impact} (P10).  This figure shows the orbit-averaged $J$-factor trajectory of subhalos of mass $m$ and concentration $c_\mathrm{sub}$ within a host halo of mass $M$ and concentration $c_\mathrm{host}$.  Compared to our model, the semianalytic model underestimates the impact of tidal stripping early on while overestimating its impact at late times.  Since the plotted trajectories are averaged over subhalo orbits, we cannot plot simulation results for comparison; nevertheless, Fig.~\ref{fig:SAp} illustrates that our model matches simulation results reasonably well.}
\end{figure}

To compare our model, we employ the same subhalo orbital distribution considered in J16, which is drawn from Ref.~\cite{zentner2005physics}.  The circular orbit radius $R_c$ is taken to be uniformly distributed between $0.6R_\mathrm{vir}$ and $R_\mathrm{vir}$, where $R_\mathrm{vir}$ is the host's virial radius.  Meanwhile, the circularity $\eta$ is distributed proportionally to $\sin\pi\eta$, and we assume that the distributions of $R_c$ and $\eta$ are independent.  By drawing subhalo orbits from this distribution, we are able to compute the orbit-averaged\footnote{Specifically, we take the median $J/J_\mathrm{init}$ at each time, but using the mean or the logarithmic mean instead does not significantly alter the results.} value of $J/J_\mathrm{init}$ using our model given in Eq.~(\ref{model1}).  In Fig.~\ref{fig:SA}, we plot the resulting orbit-averaged $J$-factor trajectories along with those predicted by the semianalytic model of J16 and P10.  We consider two different host-subhalo systems, listed on the figure, and since the semianalytic model is sensitive to the total virial masses of the host and subhalo, we employ the concentration parameter $c\equiv r_\mathrm{vir}/r_s$ to describe these systems; $c_\mathrm{host}$ is the host halo's concentration, while $c_\mathrm{sub}$ is the subhalo's concentration when it is accreted.

Compared to our model, Fig.~\ref{fig:SA} shows that the semianalytic model underestimates the impact of tidal forces early on while overestimating their impact at late times.  These discrepancies arise from several sources.  As we show in Appendix~\ref{sec:rmax}, a subhalo's $J$ factor after tidal evolution is about 30\% smaller than what would be predicted from its parameters $r_\mathrm{max}$ and $v_\mathrm{max}$ assuming an NFW profile.  However, this source of error is relatively minor.  The main differences arise from the model in J16 given by Eq.~(\ref{J16}).  Since this model does not account for the subhalo's density profile, it takes too long to strip the subhalo's weakly bound outskirts (beyond $r_s$) that contribute little to annihilation rates.\footnote{For the two cases shown in Fig.~\ref{fig:SA}, it takes, respectively, 6 and 3 dynamical times for the J16 model to bring the subhalo's bound mass below its initial $m_\mathrm{max}$, the mass enclosed within the radius $r_\mathrm{max}$ at which the maximum circular velocity is attained.  As we find in Appendix~\ref{sec:rmax} (see Fig.~\ref{fig:Mbd}), the drop in the $J$ factor is minimal above this mass threshold.}  This behavior partially explains why the semianalytic model underestimates the early impact of tidal effects.  Meanwhile, for small $\zeta \ll 1$, Eq.~(\ref{J16}) describes nearly exponential decay, analogous to our model, Eq.~(\ref{dJ}), with $c=0$.  Without the braking behavior contributed by $c>0$ (and attributed to changes in the shape of the subhalo's density profile; see Sec.~\ref{sec:trends}) along with the limiting $|\mathrm{d}\ln J/\mathrm{d}\ln t|\sim 1$ behavior, the semianalytic model overestimates the impact of tidal effects at late times.  For these reasons, our simulation-tuned model supplies significantly more accurate predictions of subhalo annihilation rates.

\section{Conclusion}\label{sec:conclusion}

In this work, we used 52 idealized $N$-body simulations to develop a model that can predict the impact of a host halo's tidal forces on the rates of dark matter annihilation within its subhalos.  Our model is given by Eq.~(\ref{model1}) and summarized in Sec.~\ref{sec:disc}, and it predicts the evolution of the subhalo's $J$ factor, the factor in the annihilation rate that depends on mass distribution, as a function of the subhalo's orbit and other properties of the host-subhalo system.  These properties are distilled into three physically motivated variables $x$, $y$, and $z$ that characterize the energy injected by tidal forces, the ratio of stretching to compressive tidal forces, and the radial distribution of tidally heated material, respectively.  Appendix~\ref{sec:fits} details how to compute these variables from standard properties of the host-subhalo system.

Our model is based on the notion that for sufficiently small changes in $J$, the $J$ factor evolves according to
\begin{equation}\tag{\ref{dJ}}
\frac{1}{J}\frac{\mathrm{d}J}{\mathrm{d}n}=-bn^{-c},
\end{equation}
where $n=t/T$ is the time in units of the subhalo's orbital period and $b$ and $c$ are parameters that depend on the system.  If $c=0$, Eq.~(\ref{dJ}) states that the subhalo loses a fixed fraction $\mathrm{e}^{-b}$ of its $J$ factor in each orbit.  The parameter $c\geq 0$ is motivated by simulation results and adds a braking mechanism to the $J$ factor's decay.  To our knowledge, a model of this form has not previously been put forward, even though we find that it can also describe other structural properties of the subhalo.  We also find that our model predicts significantly different $J$-factor trajectories than prior semianalytic models.  We further validate our model by testing it against the publicly available DASH library of subhalo simulations \cite{ogiya2019dash}, finding reasonable agreement.

Our model has limitations.  As presented, it is restricted to host-subhalo systems in which both halos possess NFW density profiles.  The NFW profile (possibly with minor corrections; e.g., Ref.~\cite{navarro2010diversity}) arises generically in dark matter simulations of halos built by hierarchical clustering \cite{navarro1996structure,navarro1997universal}.  However, the smallest subhalos, forming by direct collapse, exhibit steeper density profiles \cite{ishiyama2010gamma,anderhalden2013density,*anderhalden2013erratum,ishiyama2014hierarchical,polisensky2015fingerprints,angulo2017earth,ogiya2017sets,delos2018ultracompact,delos2018density,delos2019predicting}.  Additionally, the density profiles of many galactic halos (but not all \cite{read2018case}) are inferred to be shallower than the NFW profile, an observation that may be explained by baryonic effects or unknown dark matter properties (see Refs.~\cite{brooks2014re,bullock2017small} for reviews).  Despite being developed using NFW profiles, we anticipate that the physical manner in which we defined the model parameters $x$, $y$, and $z$ implies that our model can be adapted to accommodate different host or subhalo density profiles.

Also, our model only accounts for tidal forces from the host halo.  Subhalos can also be disrupted by encounters with other subhalos, but the results of Ref.~\cite{van2017disruption} suggest that this effect is subdominant.  More importantly, subhalos can be affected by baryonic content residing within the host, such as stars (e.g., Refs.~\cite{berezinsky2003small,berezinsky2006destruction,goerdt2007survival,zhao2007tidal,green2007mini,schneider2010impact,ishiyama2010gamma,angus2007cold,berezinsky2014small,delos2019evolution}) or a disk (e.g., Refs.~\cite{taylor2001dynamics,angus2007cold,schneider2010impact,d2010substructure,berezinsky2014small,zhu2016baryonic,errani2016effect,garrison2017not,stref2017modeling,kelley2018phat,hutten2019gamma}).  These effects are not included in our model and must be accounted for separately.  However, we remark that many of the dwarf spheroidal galaxies, already some of the most promising targets for dark matter annihilation searches \cite{strigari2007precise}, have such little baryonic content (e.g., Ref.~\cite{mcconnachie2012observed}) that it may be possible to neglect the influence of this content on their subhalos.

Despite these limitations, we anticipate that our model will prove useful in understanding the annihilation signals of dark matter substructure.  In a subsequent paper \cite{delos2019gamma}, we will explore the consequences of our model by using it to study microhalo-dominated annihilation signals in nearby dwarf galaxies.  Such signals are expected to arise from certain cosmological scenarios, such as an early matter-dominated era prior to nucleosynthesis, and our model enables precise characterization of the magnitude and morphology of these signals.

\begin{acknowledgments}
The simulations for this work were carried out on the Killdevil and Dogwood computing clusters at the University of North Carolina at Chapel Hill.  This work was funded by NASA through the Fermi
Guest Investigator Cycle 10 Award No. 80NSSC17K0751 (PI A. Erickcek).  The author thanks Adrienne Erickcek and Tim Linden for helpful discussions.  Key figures in this work employ the cube-helix color scheme developed by Ref.~\cite{green2011colour}.
\end{acknowledgments}

\appendix

\section{Simulation details}\label{sec:sim_app}

\subsection{High- and low-resolution particles}

\begin{figure}[t]
	\centering
	\includegraphics[width=\columnwidth]{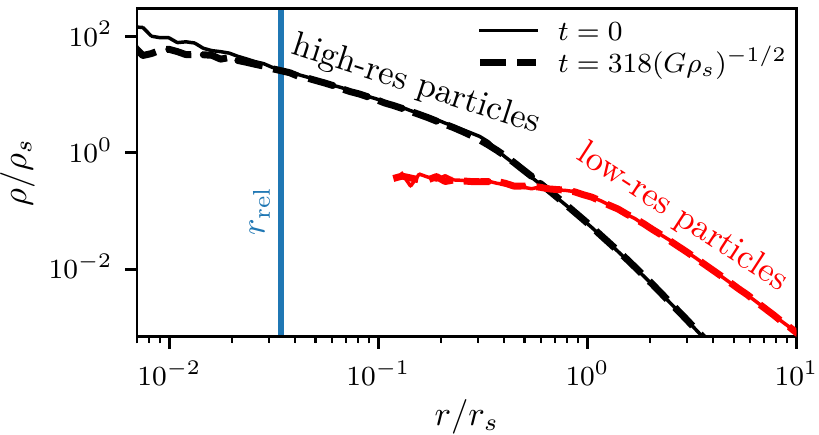}
	\caption{\label{fig:particlemass} Absence of relaxation effects associated with the use of different particle masses.  This figure shows separately the density profiles of light (high-resolution) and heavy (low-resolution) particles inside the same halo; this halo is not exposed to tidal forces.  There is no visible tendency for the heavy particles to sink to lower radii even after duration $t=318(G\rho_s)^{-1/2}$ (dashed lines).  The profile of low-resolution particles is plotted down to the radius containing 100 such particles.  As a separate effect, the density profile of light particles shallows at small radii due to two-body relaxation (between light particles alone); the resolution limit $r_\mathrm{rel}$ imposed by this effect, described in Appendix~\ref{sec:profiles}, is shown (vertical line).}
\end{figure}

As Sec.~\ref{sec:sim} notes, we sample the subhalo's central region at increased resolution such that particles whose orbital pericenters are below $r_s/3$ have $1/64$ the mass and $64$ times the number density of the other particles.  When simulation particles have different masses, it is possible for two-body interactions to artificially transfer energy from the heavy to the light particles.  To verify that this effect is not significant in our simulations, we show in Fig.~\ref{fig:particlemass} the density profiles of light and heavy particles in a subhalo not exposed to tidal forces.  Even after duration $t=318(G\rho_s)^{-1/2}$, where $\rho_s$ is the subhalo's scale density, there is no visible tendency for the heavy particles to sink to smaller radii.

\subsection{Density profiles and $J$ factors}\label{sec:profiles}

We obtain each subhalo's density profile by binning it in factors of $1.1$ in the radius.  At small radii, there is a resolution limit driven by three effects: force softening, Poisson noise, and artificial relaxation.  Each effect is associated with a minimum resolved radius below which the density profile artificially flattens.  For force softening, that radius is the distance $r_\mathrm{soft}=2.8\epsilon$, where $\epsilon$ is \textsc{Gadget}-2's force-softening parameter, at which forces become non-Newtonian.  For Poisson noise, we take it to be the radius $r_{100}$ enclosing 100 particles.  To estimate the radius $r_\mathrm{rel}$ at which artificial relaxation becomes significant, we compute the relaxation time \cite{binney1987galactic}
\begin{equation}\label{trel}
t_\mathrm{relax}= \frac{N}{8\ln \Lambda} \frac{r}{\sqrt{GM/r}}
\end{equation}
at each radius $r$, where $M$ and $N$ are the mass and particle count interior to $r$, and $\Lambda=\max\{N,r/\epsilon\}$.  If $\alpha t_\mathrm{relax}$ at radius $r$, with an appropriate proportionality constant $\alpha$, is shorter than the system age, then $r<r_\mathrm{rel}$.  The proportionality constant $\alpha$ is tuned to predict the correct $r_\mathrm{rel}$ in a simulation of the subhalo without a host; in that case, any change to the density profile is artificial since the halo was built from an equilibrium distribution.  From this calibration we use $\alpha=5$.  Figure~\ref{fig:particlemass} shows how $r_\mathrm{rel}$ marks where the density profile begins to shallow due to relaxation effects.

The minimum resolved radius of the density profile is $r_\mathrm{min}=\max\{r_\mathrm{soft},r_{100},r_\mathrm{rel}\}$.  For the purpose of accurately computing $J$ factors we extrapolate the density profile below $r_\mathrm{min}$ as $\rho=A r^{-1}$, where $A$ is the average of $\rho r$ in the three smallest radial bins above $r_\mathrm{min}$, so that
\begin{equation}\label{Jint}
J= 4\pi A^2 r_\mathrm{min}+\int_{r_\mathrm{min}}^\infty \rho(r)^2 4\pi r^2 \mathrm{d}r.
\end{equation}
Effectively, this procedure produces a lower bound on $J$ under the assumption that larger radii are always stripped more than smaller radii.  Below $r_\mathrm{min}$, we simply assume all radii are stripped equally.  We can also compute an upper bound on $J$ by assuming that radii below $r_\mathrm{min}$ are completely unaffected (so $A=\rho_s r_s$), and this allows us to estimate the uncertainty in our $J$ factors.  We find that by the termination time of each subhalo's $J$-factor trajectory (as defined in Sec.~\ref{sec:model}), the uncertainty in the $J$ factor, taken as $J_\mathrm{upper}/J_\mathrm{lower}-1$, is 31\% for one simulation (parameters $x=31$, $y=11$, $z=0.07$; see Sec.~\ref{sec:model}), smaller than 17\% for the remaining 51 simulations, and smaller than 10\% for 44 of them.

To understand the $J$ factors in the cases where $|\mathrm{d}\ln J/\mathrm{d}\ln t| \gtrsim 1$, another step is necessary.  In this regime, the elongated tidal stream can contribute significantly to the $J$ factor, making the spherical integral Eq.~(\ref{Jint}) inaccurate.  Thus, we also compute the $J$ factor as the sum over simulation particles
\begin{equation}\label{Jsum}
J=\sum_i \rho_i m_i,
\end{equation}
where $m_i$ is the mass of particle $i$ and $\rho_i$ is its local density.  The density $\rho_i$ is estimated as
\begin{equation}\label{rhoi}
\rho_i = \sum_{j=1}^{N} m_j W(r_{ij},h_i)
\end{equation}
over the $N=50$ nearest particles $j$, where $r_{ij}$ is the distance to particle $j$, $h_i$ is the distance to the $N$th particle, and $W(r,h)$ is the cubic spline kernel defined as in Ref.~\cite{springel2005cosmological}.

Equation~(\ref{Jsum}) underestimates the $J$-factor contribution at $r<r_\mathrm{min}$ due to artificial flattening of the density profile.  To accommodate the extrapolation procedure in Eq.~(\ref{Jint}) that addresses this problem, an additional step is required.  We find the bound remnant of the subhalo using a procedure similar to that in Ref.~\cite{van2017disruption}.  Beginning with the assumption that all particles are bound, we iteratively compute the gravitational potential of each particle due to all other bound particles using a Barnes-Hut octree \cite{barnes1986hierarchical} with $\theta=0.7$ and the same softening length as the simulation.  Subsequently, we mark each particle as unbound if its total energy is positive and bound if its total energy is negative.  At each step, we find the center-of-mass position and velocity of the 100 most bound particles and recenter the full system to be relative to this center of mass.  All particles are initially marked as bound, and the procedure terminates when the count of bound particles converges.\footnote{This halting condition is stricter than the one in Ref.~\cite{van2017disruption}.}

By assuming that the bound remnant is spherically symmetric, we can estimate the $J$ factor both including spherical asymmetry and compensating for the flattening of the density profile below $r_\mathrm{min}$.  If $J_\mathrm{full}$ is the $J$ factor of the full system computed using Eq.~(\ref{Jsum}) and $J_\mathrm{bd,rad}$ and $J_\mathrm{bd}$ are the $J$ factors of the bound remnant computed using  Eqs. (\ref{Jint}) and~(\ref{Jsum}), respectively, then
\begin{equation}\label{Jfull}
J=J_\mathrm{full}-J_\mathrm{bd}+J_\mathrm{bd,rad}.
\end{equation}

\subsection{Numerical convergence}

\begin{figure}[t]
	\centering
	\includegraphics[width=\columnwidth]{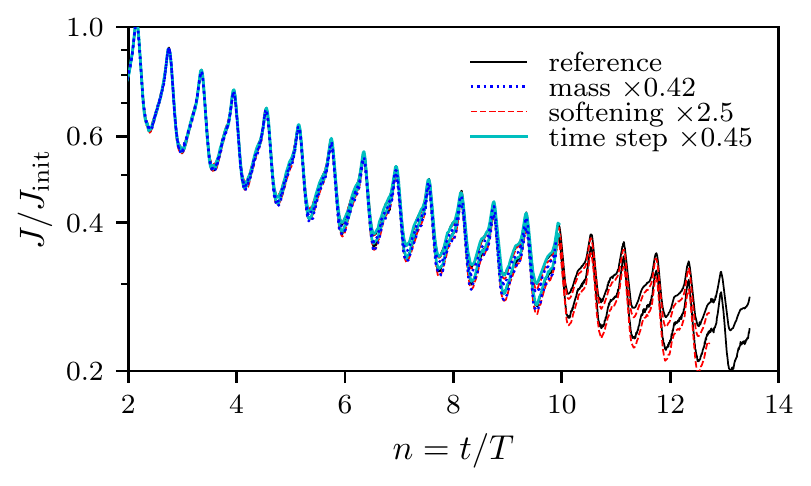}
	\caption{\label{fig:conv} Simulations of the same tidal evolution scenario carried out with different simulation parameters.  For each simulation, two $J$-factor trajectories are plotted corresponding to the lower and upper limits discussed in Appendix~\ref{sec:sim_app} (the lower limit is the value we use throughout this work).  The upper and lower limits of each simulation overlap, implying numerical convergence.}
\end{figure}

In our simulations, we set \textsc{Gadget}-2's force-softening length to be $\epsilon=0.003r_s$.  This small value is intended to evade the artificial subhalo disruption observed by Ref.~\cite{van2018dark}.  Meanwhile, the subhalo's high-resolution particles (see Sec.~\ref{sec:sim}) have mass $\num{4.3e-7} \rho_s r_s^3$.  To check that numerical artifacts in our simulations are under control, we test the impact of changing the softening length and the particle resolution.  Additionally, we test the impact of altering the (adaptive) integration time steps in order to ensure there are no artifacts arising from the application of the host's tidal field over these discrete intervals.  In Fig.~\ref{fig:conv}, we plot the $J$-factor trajectory in a reference simulation (with system parameters $x=34$, $y=0.018$, and $z=0.15$; see Sec.~\ref{sec:model}) along with three simulations of the same system with different particle resolution, force softening, and integration time steps.  We plot the upper and lower limits of the $J$-factor trajectory as discussed above.  These limits overlap for all simulation parameters, suggesting that the simulation is converged.


\section{Subhalo size}\label{sec:size}

In our simulations we applied the host halo's tidal forces using the linearized expression given by Eq.~(\ref{tidalforce}), which is valid in the limit that the subhalo is much smaller than its orbital radius.  Thus, our results are applicable in the $r_s\ll R_c$ limit.  In this appendix, we explore precisely how far the applicability of our results can be taken.  For this purpose we executed several simulations using the exact tidal force $\vec F_\mathrm{tidal}(\vec r) = \vec F(\vec R + \vec r) - \vec F(\vec R)$ instead of the linearized version, and we compare the results of these simulations to those of a simulation that employed the linearized force.  The subhalos in all of these simulations are cut off at radius $5r_s$ to avoid excessive overlap with the host's center; this change does not affect the comparison since it applies equally to every simulation.

\begin{figure}[t]
	\centering
	\includegraphics[width=\columnwidth]{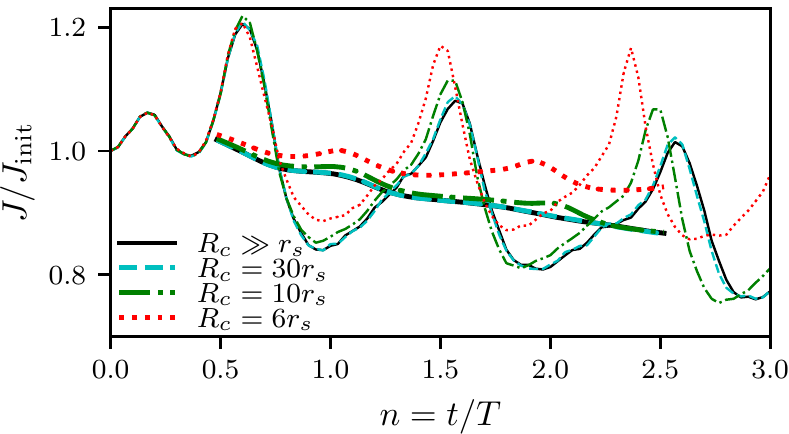}
	\caption{\label{fig:size} Influence of a subhalo's size on its tidal evolution.  This figure shows the trajectory of a subhalo's $J$ factor for several different values of the ratio $r_s/R_c$ between the subhalo's scale radius and the radius of its orbit about the host; the relative orbital radius $R_c/R_s$ is held fixed along with all other parameters.  Thin lines show full trajectories while thick lines show averages over each orbital period.  When $r_s\gtrsim 0.1 R_c$, the tidal evolution begins to diverge from the evolution in the $r_s\ll R_c$ limit (solid curve).  The double peak in the first orbit arises because of the subhalo's truncation radius in these simulations and is not relevant to the comparison.}
\end{figure}

We subjected subhalos of different sizes to the same tidal evolution scenario with parameters $x=21$, $y=0.056$, and $z=0.18$ in the parametrization given in Sec.~\ref{sec:model}.  Figure~\ref{fig:size} shows the tidal evolution of the subhalo's $J$ factor for different values of the ratio $r_s/R_c$ between the subhalo's scale radius and the radius of its orbit about the host.  We find that the tidal evolution begins to diverge markedly from that induced by the linearized tidal force when $r_s\gtrsim 0.1 R_c$.  Note that this analysis still neglects dynamical friction (including self-friction, due to the limited duration of these simulations); the influence of this effect is also sensitive to the subhalo's size.  Thus, our results are applicable if both $r_s\lesssim 0.1 R_c$ and dynamical friction can be neglected (see Sec.~\ref{sec:sim}).

\section{Computational details}\label{sec:fits}

In this appendix, we present practical ways to compute the reduced variables $x$, $y$, and $z$ along with the orbital period $T$.  For convenience, we include fitting formulas to approximate the necessary integrals.  In what follows, the host is assumed to possess an NFW profile with scale radius $R_s$ and scale density $P_s$; its mass profile is
\begin{equation}\label{nfwmass}
M(R)=4\pi P_s R_s^3\left[\ln\left(1+\frac{R}{R_s}\right)-\frac{R/R_s}{1+R/R_s}\right],
\end{equation}
its force profile is $F(R)=G M(R)/R^2$, and its potential profile is
\begin{equation}\label{nfwpot}
\Phi(R)=-4\pi G P_s R_s^2 \frac{\ln(1+R/R_s)}{R/R_s}.
\end{equation}
Meanwhile, the subhalo's orbit about the host is parametrized by the circular orbit radius $R_c$ and circularity $\eta$, and as shorthand, we define $y_c \equiv R_c/R_s$.

\subsection{Computing $x=|E_b|/\Delta E_\mathrm{imp}$}

The binding energy $E_b$ of a particle at the subhalo's scale radius $r_s$ is given by Eq.~(\ref{Eb}).  Meanwhile, the total energy $\Delta E_\mathrm{imp}$ injected into a particle at radius $r$ by tidal forces over the course of a subhalo orbit is computed using the impulse approximation, as described in Ref.~\cite{gnedin1999tidal}.  This energy depends on the particle's full three-dimensional position within the subhalo, but we simplify the picture by averaging this energy over the sphere at radius $r$.  Dimensionally, ${\Delta E_\mathrm{imp}/r^2 \sim F(R_c)/R_c}$, and we can approximate
\begin{equation}\label{impulse_fit}
\frac{\Delta E_\mathrm{imp}}{r^2}=P_1(y_c) \exp\left\{P_2(y_c) \left[1-\eta^{P_3(y_c)}\right]\right\} \frac{F(R_c)}{R_c},
\end{equation}
where $P_1(y_c)$ is defined
\begin{align}
&
P_1(y_c) = \frac{A(1+B\ln(1+y_c)-C y_c/(D+y_c))}{1+E(\ln(1+y_c)-2y_c/(2+y_c))},
\nonumber\\&
A=3.327,\ 
B=0.6463,\ 
C=0.8837,\ 
D=0.8809,
\nonumber\\&
E=0.2156,
\end{align}
$P_2(y_c)$ is defined
\begin{align}
&
P_2(y_c) = A(1+(y_c/c)^a)^b,
\nonumber\\&
A=3.005,\ 
a=3.641,\ 
b=0.08513,\ 
c=0.5703,
\end{align}
and $P_3(y_c)$ is defined
\begin{align}
&
P_3(y_c) = \frac{A(1+(y_c/c_1)^{a_1})^{b_1}}{(1+(y_c/c_2)^{a_2})^{b_2}(1+(y_c/c_3)^{a_3})^{b_3}},
\nonumber\\&
A=0.2150,\ 
a_1=1.017,\ 
b_1=0.8650,\ 
c_1=0.5057,
\nonumber\\&
a_2=2.774,\ 
b_2=0.2426,\ 
c_2=0.6415,
\nonumber\\&
a_3=0.7663,\ 
b_3=0.6508,\ 
c_3=18.84.
\end{align}
For $\eta>0.04$, this expression is accurate to within 3\% for $y_c<10$ and within 14\% for $y_c<10^3$.

\begin{figure}[b]
	\centering
	\includegraphics[width=\columnwidth]{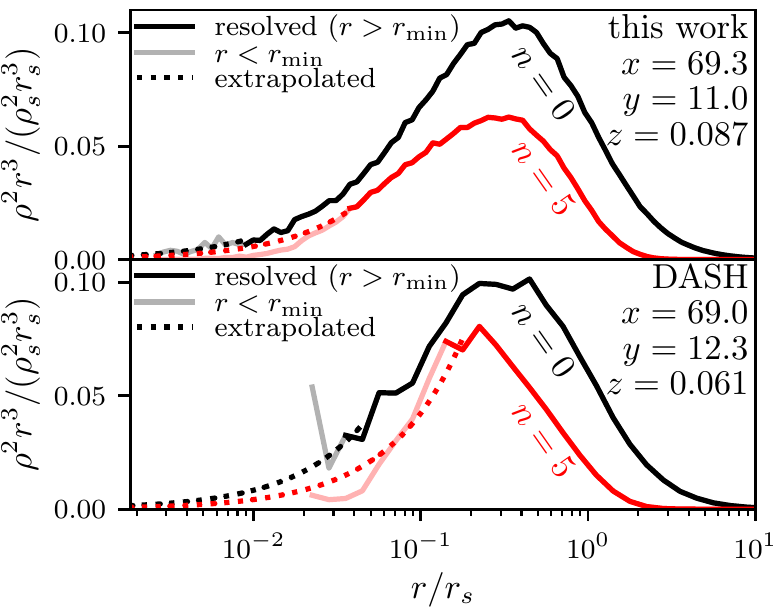}
	\caption{\label{fig:integrand} A resolution comparison between our simulations (top) and those of the DASH library (bottom).  The (log-space) integrand for the $J$ factor, $\rho^2 r^3$, is plotted for an example subhalo from each catalogue with similar system parameters ($x$, $y$, and $z$; see Sec.~\ref{sec:model}) at $n=0$ and $n=5$ orbits.  The $J$ factor is the area under the curve.  Below the resolution limit $r_\mathrm{min}$, we plot a pessimistic extrapolation of the density profile; see Appendix~\ref{sec:sim_app}.}
\end{figure}

\begin{figure*}[t]
	\centering
	\begin{tabular}{cc}
		\includegraphics[width=\columnwidth]{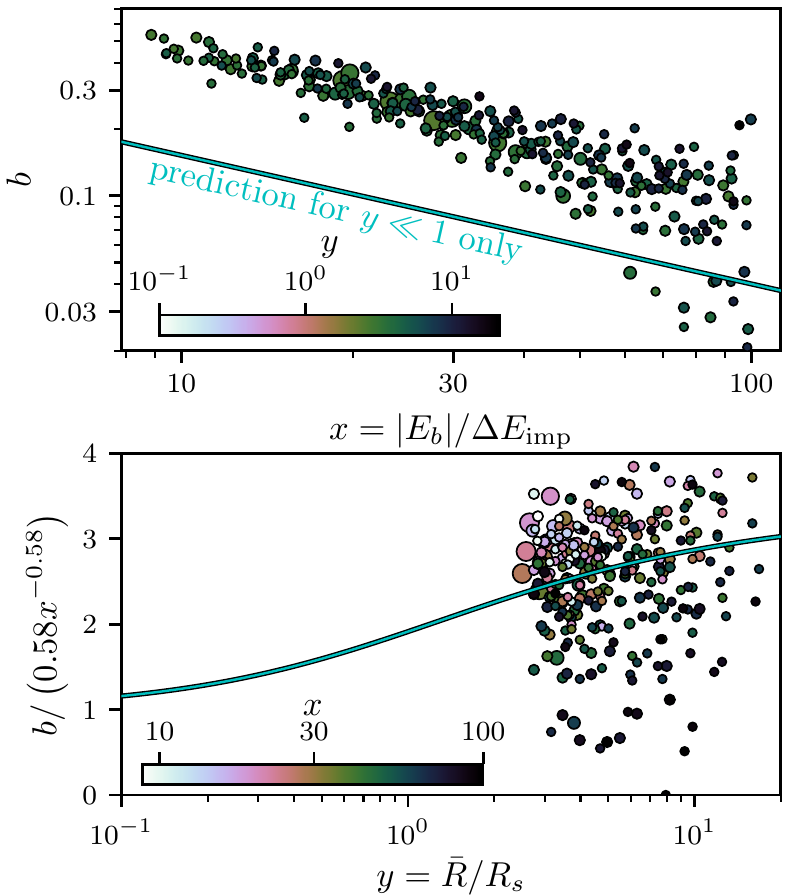}
		&
		\shortstack{
			\includegraphics[width=.945\columnwidth]{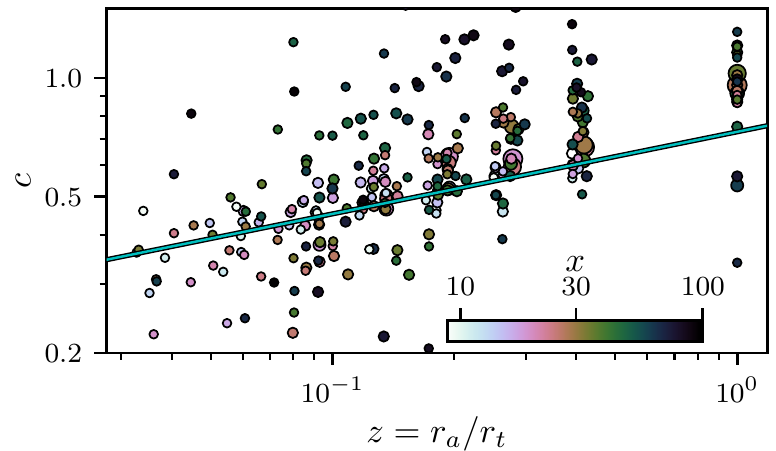}
			\\
			\includegraphics[width=.945\columnwidth]{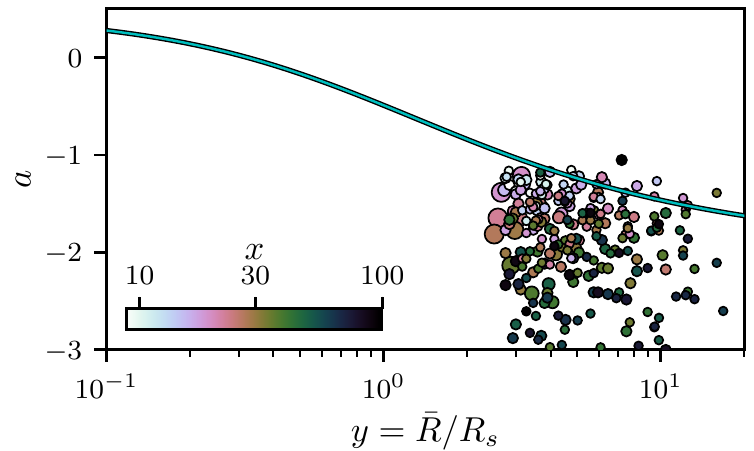}
		}
	\end{tabular}
	\caption{\label{fig:dash} A test of our model against the DASH simulations.  This figure plots the $J$-factor trajectory parameters $a$, $b$, and $c$ for the DASH simulations against the system parameters $x$, $y$, and $z$.  The solid curves are our model predictions; they are the same curves shown in Figs. \ref{fig:b}, \ref{fig:a}, and~\ref{fig:c}.  Note that the offset between the solid line and the simulations in the upper-left panel is not a discrepancy, for the solid line is only valid for $y\ll 1$.  The DASH simulations exhibit significant scatter but broadly support our model with some systematic discrepancies discussed in Appendix~\ref{sec:dash}.  The radius of each marker is proportional to the number of orbital periods, which ranges from 5 to 11.}
\end{figure*}

\subsection{Computing $y=\bar R/R_s$}

The time-averaged radius $\bar R$ of the orbit is approximately $R_c$, and, in fact, $\bar R/ R_c\to 1$ as $R_c/R_s\to 0$.  More broadly, the expression
\begin{align}
&
\frac{\bar R}{R_c} = \frac{1\!+\!B(1\!-\!F\eta^G)\ln(1\!+\!y_c)\!-\!C y_c/[D(1\!-\!H\eta^I)\!+\!y_c]}{1+E(\ln(1+y_c)-2y_c/(2+y_c))},
\nonumber\\&
B=0.3777,\ 
C=0.4892,\ 
D=2.412,\ 
E=0.2426,
\nonumber\\ &
F=0.3556,\ 
G=1.860,\ 
H=0.1665,
\end{align}
is accurate to within 0.3\% for $\eta>0.04$ and $y_c < 10^3$.

\subsection{Computing $z=r_a/r_t$}

We define the adiabatic shielding radius $r_a$ and the tidal radius $r_t$ as the solutions to $m(r_a)/r_a^3 = \rho_a$ and $m(r_t)/r_t^3 = \rho_t$, respectively, where $m(r)$ is the subhalo's initial NFW mass profile [see Eq.~(\ref{nfwmass})].  Here, $\rho_a$ and $\rho_t$ are functions of the subhalo's orbit; in particular,
\begin{equation}\label{rhoa_rhot}
\rho_a\equiv \frac{V_p^2}{G R_p^2} = \eta^2\frac{M(R_c)R_c}{R_p^4}
\ \ \text{and} \ \ 
\rho_t\equiv \frac{M(R_a)}{R_a^3},
\end{equation}
where $R_p$ and $R_a$ are the orbital pericenter and apocenter radii, which may be obtained as the two solutions $R$ to
\begin{equation}\label{ra_rp}
\Phi(R_c)-\Phi(R)+\left(1-\eta^2\frac{R_c^2}{R^2}\right)\frac{GM(R_c)}{2 R_c}=0.
\end{equation}

\subsection{Computing $T$}
Dimensionally, the radial orbit period $T \sim t_0$, where ${t_0\equiv \sqrt{R_c/F(R_c)}}$.  More precisely, the expression
\begin{align}
&
\frac{T}{t_0} = \frac{A(1+F\eta^G)[1+B\ln(1+y_c)-C y_c/(D+y_c)]}{1+E(1+H\eta^I)(\ln(1+y_c)-2y_c/(2+y_c))},
\nonumber\\&
A=3.460,\ 
B=0.6076,\ 
C=0.8831,\ 
D=2.312,
\nonumber\\ &
E=0.3325,\ 
F=0.04827,\ 
G=1.261,
\nonumber\\ &
H=0.03606,\ 
I=1.288,\ 
\end{align}
is accurate to within 0.2\% for $\eta>0.04$ and $y_c < 10^3$.

\section{Comparison to the DASH library}\label{sec:dash}

Reference~\cite{ogiya2019dash} published a library called Dynamical Aspects of SubHaloes (DASH) of idealized subhalo simulations.  This library includes the results of 2177 simulations, with different system parameters, of an $N$-body subhalo orbiting an analytic host potential.  These simulations resolve significantly less of the subhalo density profile than do ours; as shown in Fig.~\ref{fig:integrand}, they can leave large fractions of the $J$ factor unresolved.  Also, the DASH library covers a smaller parameter range in $x$, $y$, and $z$.  Nevertheless, because of the extraordinary volume of this library, it can serve as a test for our model.

We use the procedure in Appendix~\ref{sec:sim_app} to find the $J$-factor trajectory of each DASH simulation,\footnote{We use a larger $\alpha=20$ to find $r_\mathrm{rel}$ for the DASH simulations, obtained by recalibrating for these simulations.  Note that larger $\alpha$ implies more optimism about simulation resolution.} imposing an additional constraint that the trajectory halt when the maximum uncertainty in the $J$ factor is larger than a factor of 3.  Next, we fit the parameters $a$, $b$, and $c$ to this trajectory as in Sec.~\ref{sec:model}.  For the DASH simulations, Fig.~\ref{fig:dash} plots (in the same way as Figs. \ref{fig:b}, \ref{fig:a}, and~\ref{fig:c}) the trajectory parameters $a$, $b$, and $c$ against the system parameters $x$, $y$, and $z$.  Superposed are our model predictions, as solid lines, using the parameters obtained in Sec.~\ref{sec:model}.

We first remark that all DASH simulations have $y>2$, so we cannot directly test Eq.~(\ref{b_}) describing the behavior of $b$ in the $y\ll 1$ self-similar regime.  Nevertheless, the upper-left panel of Fig.~\ref{fig:dash} shows that the DASH simulations exhibit roughly the same power-law behavior $b\propto x^{-0.58}$ predicted by Eq.~(\ref{b}) (the offset between our curve and the simulations here is not a discrepancy).  The lower panels show the sensitivity of $a$ and $b$ to $y$.  Because the DASH simulations only cover a small range of $y$, we cannot verify the functional form of each parameter in $y$.  Also, there is substantial scatter, especially at large $x$.  Nevertheless, our model predicts roughly the correct values of $a$ and $b$ for these simulations, although there is a tendency for the simulations to have smaller values of $a$ and larger values of $b$.  Finally, although the scatter in $c$ is quite large, the relationship between $c$ and $z$ is approximately borne out in the DASH simulations.

The tendency for the DASH simulations to yield small $a$ and large $b$ can be understood as a resolution artifact.  Below the resolution limit, we extrapolate the density profile in a way that always underestimates the $J$ factor (see Appendix~\ref{sec:sim_app}).  This underestimation both increases the immediate loss of the $J$ factor, reducing $a$, and increases the rate at which $J$ decays (since artificial relaxation worsens the resolution over time), raising $b$.

Also, there is a tendency for systems at the large-$x$ end to exhibit large scatter in $a$, $b$, and $c$ as well as a precipitous drop in $b$ (sometimes even to $b<0$).  This trend is also an unphysical artifact.  In our simulations, we observed the same trend when $x\gtrsim 200$, which is why our simulation sample in Sec.~\ref{sec:model} only includes $x<200$.  For the lower-resolution DASH simulations, the trend begins at $x\gtrsim 50$.  The numerical difficulty with large $x$ is unclear, but it is likely connected to the fact that large $x$ implies the subhalo's internal forces are much stronger than the external tides.  The vast difference in the scales of these forces could lead to issues in numerical precision when the tiny tidal forces are added to the large internal forces.

\section{The broader density profile}\label{sec:rmax}

\begin{figure}[t]
	\centering
	\includegraphics[width=\columnwidth]{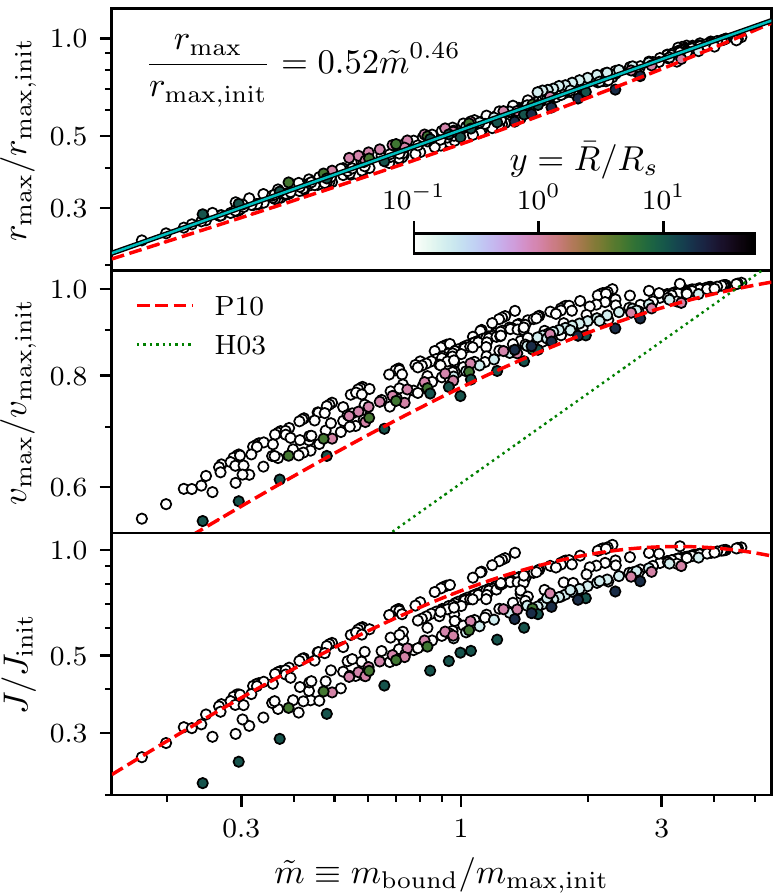}
	\caption{\label{fig:Mbd} The relationship of the subhalo properties $r_\mathrm{max}$ (top), $v_\mathrm{max}$ (middle), and $J$ (bottom) to its bound mass $m_\mathrm{bound}$ after tidal stripping.  $r_\mathrm{max}$ is cleanly related to $m_\mathrm{bound}$, but the scatter is larger for $v_\mathrm{max}$ and still larger for $J$.  Each point represents the average over a single orbit in our simulations, and the solid lines represent the displayed fitting functions.  The dashed and dotted lines correspond to the predictions of P10 and H03, respectively, assuming that the initial mass is $4.5m_\mathrm{max,init}$.  The P10 prediction in the last panel additionally assumes an NFW profile.}
\end{figure}

\begin{figure}[t]
\centering
\includegraphics[width=\columnwidth]{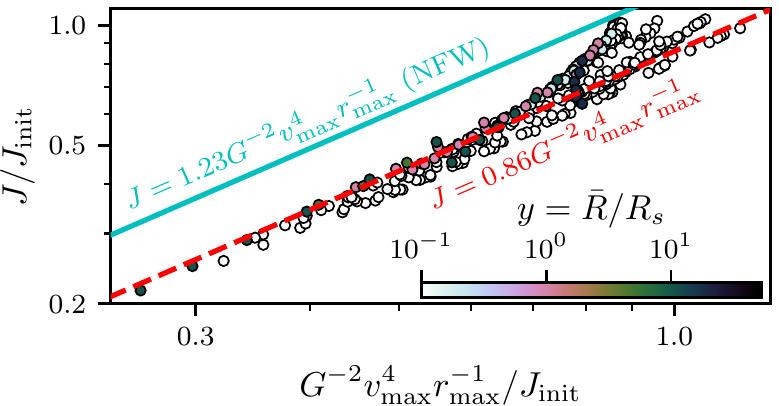}
\caption{\label{fig:Jvmax} The relationship between a subhalo's $J$ factor and its structural parameters $r_\mathrm{max}$ and $v_\mathrm{max}$; for an NFW profile, $J=1.23G^{-2}v_\mathrm{max}^4 r_\mathrm{max}^{-1}$ (solid line).  In our simulations, the $J$ factor lies consistently about 30\% below the value that would be expected assuming an NFW profile, as illustrated by the dashed line.  Each point represents the average over a single orbit in our simulations.}
\end{figure}

References \cite{hayashi2003structural} (hereafter H03) and~\cite{penarrubia2010impact} (hereafter P10) studied the tidal evolution of a subhalo's density profile, focusing on the structural parameters $v_\mathrm{max}$, the maximum circular velocity within the subhalo, and $r_\mathrm{max}$, the radius at which this velocity is attained.  Prior treatments of the annihilation rate in subhalos (e.g., Refs.~\cite{bartels2015boosting,hiroshima2018modeling}) have employed these works' predictions of $r_\mathrm{max}$ and $v_\mathrm{max}$, along with the assumption that subhalos retain NFW profiles, to predict subhalo $J$ factors.  To understand the connection between our work and these prior works, this appendix investigates the evolution of $r_\mathrm{max}$ and $v_\mathrm{max}$ in our simulations.

At each snapshot of our simulations, we find $r_\mathrm{max}$ as the radius $r<r_t$ that maximizes $v_\mathrm{circ}=\sqrt{Gm(r)/r}$, and $v_\mathrm{max}$ is the corresponding maximum.  We only consider snapshots up to the point where $|\mathrm{d}\ln J/\mathrm{d}\ln t|=1$, as discussed in Sec.~\ref{sec:model}.  Additionally, we halt the $r_\mathrm{max}$ and $v_\mathrm{max}$ trajectory when $r_\mathrm{max}$ becomes smaller than the resolution limit (see Appendix~\ref{sec:sim_app}), and we only include simulations whose trajectories cover at least five orbits about the host.  This restriction reduces our simulation count to $41$, $33$ of which are in the self-similar regime ($R_c/R_s<0.3$).

\subsection{Relations between structural parameters}

Following H03 and P10, we first explore the relationship between a subhalo's structural parameters and its total mass loss.  These prior works parametrize the mass loss using the ratio $m_\mathrm{bound}/m_\mathrm{acc}$, where $m_\mathrm{bound}$ is the mass that remains bound to the subhalo and $m_\mathrm{acc}$ is its virial mass at accretion.  However, this parametrization implies that the impact of tides is strongly sensitive to the subhalo's initial concentration, and we propose that this sensitivity is unphysical since the outer layers may be stripped almost immediately upon accretion onto the host.  To evade this problem, we instead parametrize the mass loss using the ratio $\tilde m \equiv m_\mathrm{bound}/m_\mathrm{max,init}$ of $m_\mathrm{bound}$ to the mass initially enclosed within $r_\mathrm{max}$; this ratio is initially larger than unity.  We compute $m_\mathrm{bound}$ using the procedure in Appendix~\ref{sec:sim_app}, and Fig.~\ref{fig:Mbd} shows these relationships.  For comparison we also plot the predictions of H03 and P10 assuming $m_\mathrm{acc}=4.5m_\mathrm{max,init}$, which corresponds to subhalo concentration $c_\mathrm{sub}\simeq 20$ at accretion.

We find that $r_\mathrm{max}$ is cleanly related to $m_\mathrm{bound}$ by a power law.  Additionally, for both $r_\mathrm{max}$ and $v_\mathrm{max}$, the predictions of P10 (with $c_\mathrm{sub}\simeq 20$) work reasonably well as long as $\bar R>R_s$.  However, the bottom panel of Fig.~\ref{fig:Mbd} shows that there is substantial scatter in the relationship between $J$ and $m_\mathrm{bound}$, and P10 does not accurately predict the $J$ factor if an NFW profile is assumed.  The scatter partially results from the modest scatter in $v_\mathrm{max}$, since $J\propto v_\mathrm{max}^4$, but it also reflects that tidally altered density profiles differ significantly from NFW.  Figure~\ref{fig:Jvmax} investigates this effect further and shows that a subhalo's $J$ factor is roughly 30\% smaller than what would be predicted from $r_\mathrm{max}$ and $v_\mathrm{max}$ assuming an NFW profile.

\subsection{Time evolution of structural parameters}

\begin{figure*}[p]
	\centering
	\begin{tabular}{cc}
		\includegraphics[width=.9\columnwidth]{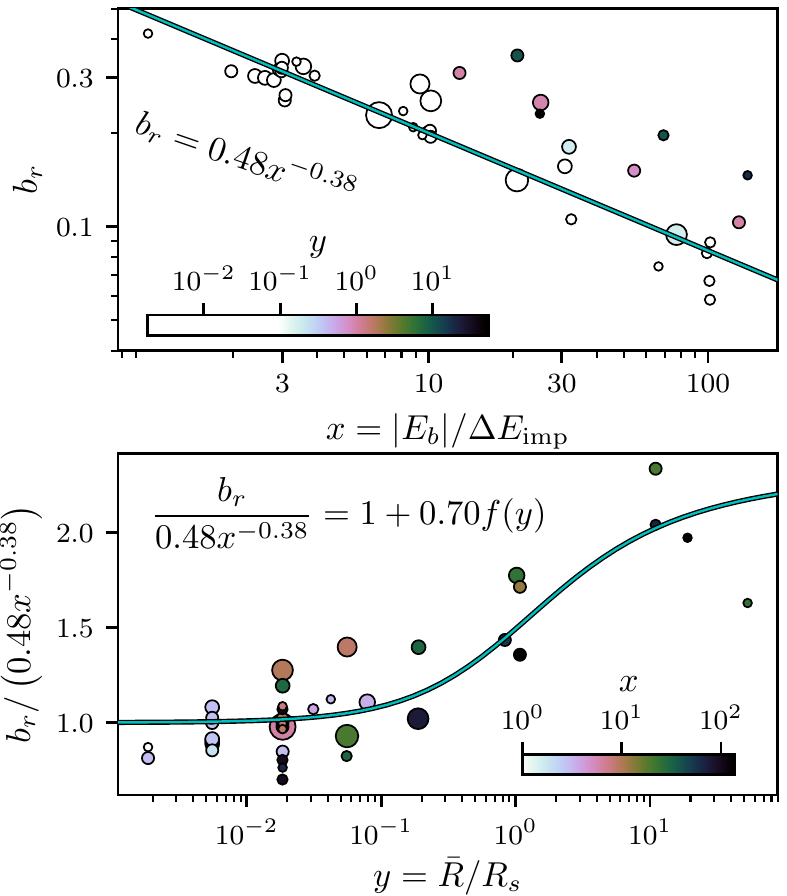}
		&
		\includegraphics[width=.9\columnwidth]{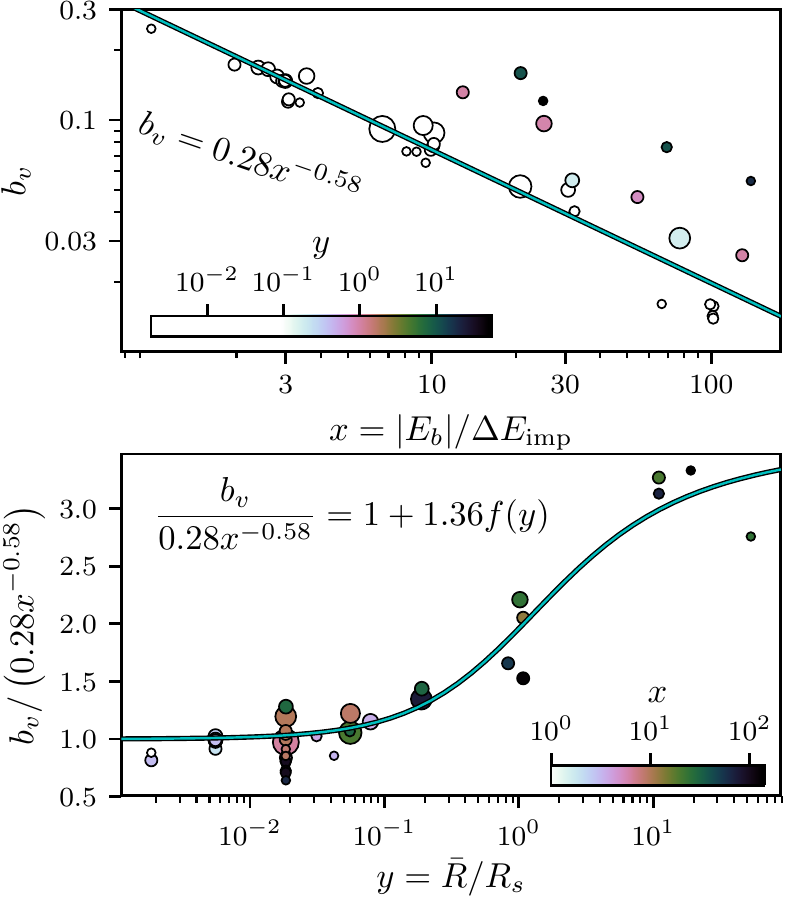}
		\\
		\includegraphics[width=.9\columnwidth]{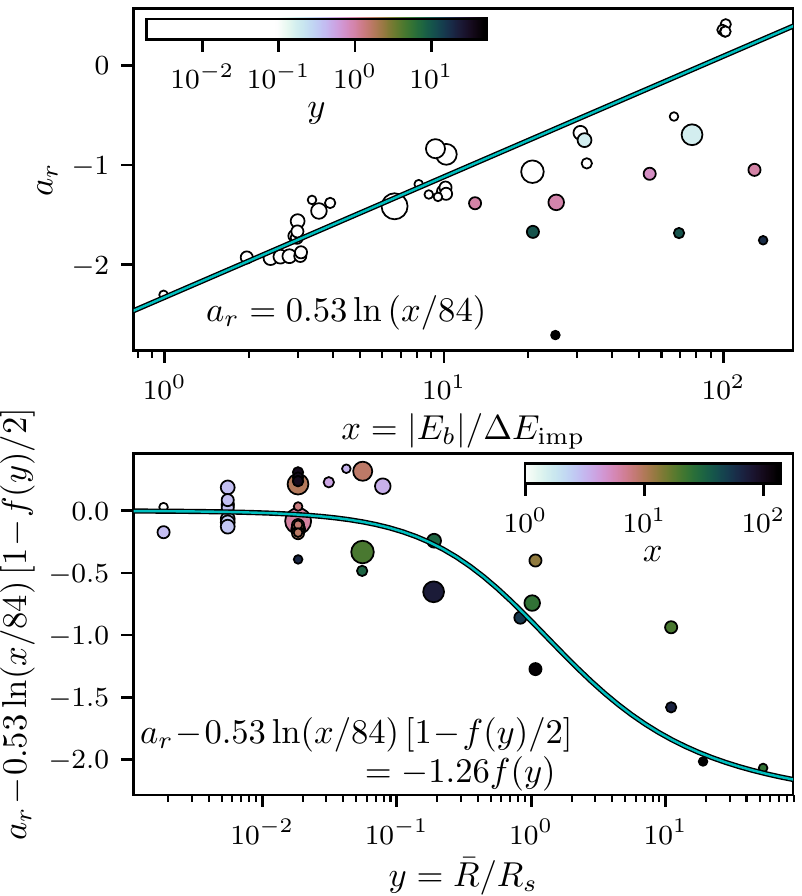}
		&
		\includegraphics[width=.9\columnwidth]{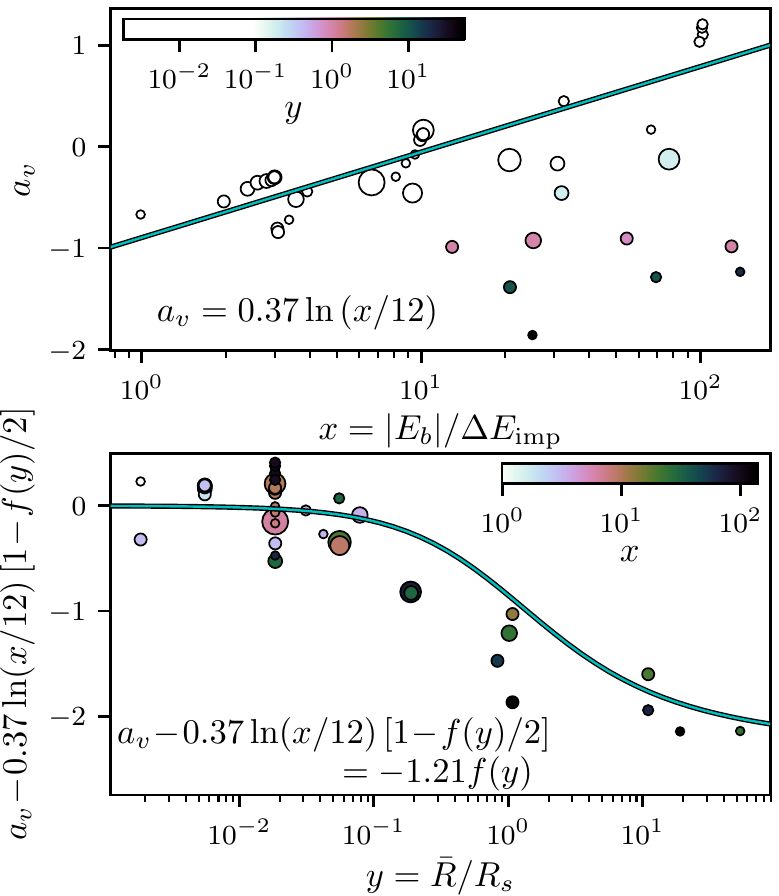}
		\\
		\includegraphics[width=.9\columnwidth]{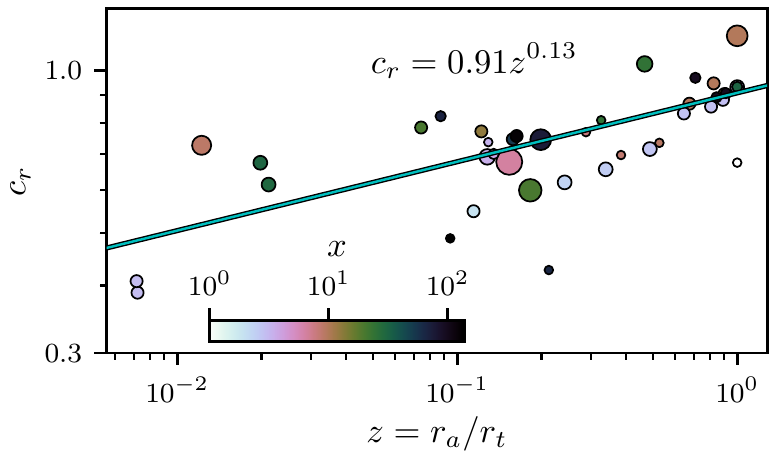}
		&
		\includegraphics[width=.9\columnwidth]{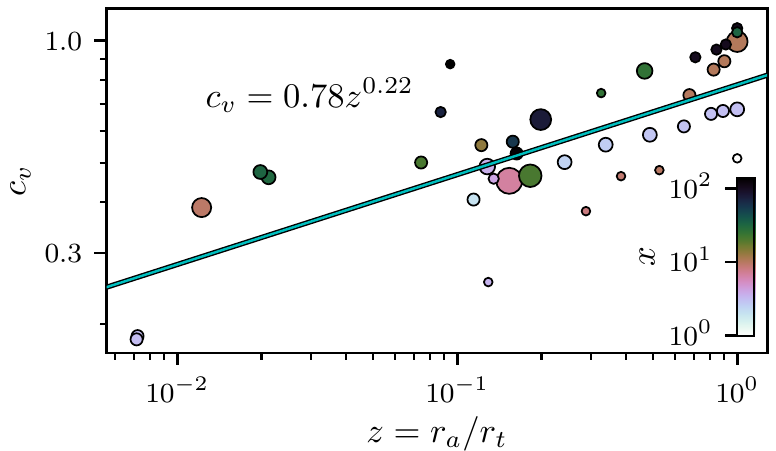}
	\end{tabular}
	\caption{\label{fig:rmax} The $r_\mathrm{max}$ trajectory parameters $a_r$, $b_r$, and $c_r$ (left panels) and the $v_\mathrm{max}$ trajectory parameters $a_v$, $b_v$, and $c_v$ (right panels) plotted against the system parameters $x$, $y$, and $z$ (cf. Figs. \ref{fig:b}, \ref{fig:a}, and~\ref{fig:c}); see Eqs. (\ref{rmodel}) and~(\ref{mmodel}).  The radius of each marker is proportional to the number of orbital periods, which ranges from 5 to 15 for this sample.}
\end{figure*}

We can also predict the evolution of the subhalo's structural parameters more explicitly.  Subjected to tidal forces, $r_\mathrm{max}$ and $v_\mathrm{max}$ follow qualitatively similar trajectories to the $J$ factor:
\begin{align}\label{rmodel}
\ln \frac{r_\mathrm{max}}{r_\mathrm{max,init}} &= b_r\left[a_r-\frac{1}{1-c_r}\left(n^{1-c_r}-1\right)\right],
\\\label{mmodel}
\ln \frac{v_\mathrm{max}}{v_\mathrm{max,init}} &= b_v\left[a_v-\frac{1}{1-c_v}\left(n^{1-c_v}-1\right)\right],
\end{align}
where $n=t/T$ is the number of orbits [compare Eq.~(\ref{model})].  Note that the total bound mass, $m_\mathrm{bd}$, behaves similarly; its trajectory follows from Eq.~(\ref{rmodel}) by inverting the equation in the top panel of Fig.~\ref{fig:Mbd}.  As shown in Fig.~\ref{fig:rmax}, $b_r$, $c_r$, $b_v$, and $c_v$ appear to depend on the system parameters $x$, $y$, and $z$ in the same way that $b$ and $c$ did:
\begin{align}
b_r &= 0.48 x^{-0.38}\left[1+0.70 f(y)\right],
\\
c_r &= 0.91 z^{0.13},
\\
b_v &= 0.28 x^{-0.58}\left[1+1.36 f(y)\right],
\\
c_v &= 0.78 z^{0.22}.
\end{align}
However, the middle panels of Fig.~\ref{fig:rmax} show that unlike $a$, the parameters $a_r$ and $a_v$ depend not only on $y$ but also on $x$.  Moreover, they are only sensitive to $x$ in the self-similar regime ($R_c/R_s<0.3$).  We fit the equation $a_r=a_{r0}\ln(x/a_{r1})$, and likewise for $a_v$, in the self-similar regime.  Next, we fit $a_r-a_{r0}\ln(x/a_{r1})[1-f(y)/2]=-a_{r2} f(y)$, and likewise for $a_v$, using all simulations.  The function $f(y)$ asymptotes at 2 for large $y$, so the combination $[1-f(y)/2]$ suppresses the $x$-dependent part of $a_r$ and $a_v$ at large $r$.  Hence, we obtain
\begin{align}
a_r &= 0.53 \ln(x/84)\left[1-f(y)/2\right]-1.26 f(y)
\\
a_v &= 0.37 \ln(x/12)\left[1-f(y)/2\right]-1.21 f(y),
\end{align}
as depicted in Fig.~\ref{fig:rmax}.

Broadly, there is more scatter in the trajectory parameters $a_r$, $b_r$, $c_r$, $a_v$, $b_v$, and $c_v$ of $r_\mathrm{max}$ and $v_\mathrm{max}$ than in the parameters $a$, $b$, and $c$ of the $J$ factor, when plotted against the system parameters $x$, $y$, and $z$ (cf. Figs. \ref{fig:b}, \ref{fig:a}, and~\ref{fig:c}).  The source of this scatter is unclear, but it is likely that $r_\mathrm{max}$ and $v_\mathrm{max}$ are more sensitive than $J$ to additional effects beyond those accounted for by the parameters $x$, $y$, and $z$.  Such heightened sensitivity is plausible for two reasons.  First, $r_\mathrm{max}$ and $v_\mathrm{max}$ are more sensitive than $J$ to the density profile at large radii, which could depend on details of the tidal forces to which the inner profile is insensitive.  Second, $r_\mathrm{max}$ and $v_\mathrm{max}$, being defined using the condition $\mathrm{d}(m(r)/r)/\mathrm{d}r=4\pi r \rho(r) - m(r)/r^2=0$, can be sensitive to fine details in the density profile $\rho(r)$.  As an integrated quantity, the $J$ factor does not exhibit this sensitivity.


\bibliography{references}

 \newcommand{\noop}[1]{}
\begin{thebibliography}{125}%
\makeatletter
\providecommand \@ifxundefined [1]{%
 \@ifx{#1\undefined}
}%
\providecommand \@ifnum [1]{%
 \ifnum #1\expandafter \@firstoftwo
 \else \expandafter \@secondoftwo
 \fi
}%
\providecommand \@ifx [1]{%
 \ifx #1\expandafter \@firstoftwo
 \else \expandafter \@secondoftwo
 \fi
}%
\providecommand \natexlab [1]{#1}%
\providecommand \enquote  [1]{``#1''}%
\providecommand \bibnamefont  [1]{#1}%
\providecommand \bibfnamefont [1]{#1}%
\providecommand \citenamefont [1]{#1}%
\providecommand \href@noop [0]{\@secondoftwo}%
\providecommand \href [0]{\begingroup \@sanitize@url \@href}%
\providecommand \@href[1]{\@@startlink{#1}\@@href}%
\providecommand \@@href[1]{\endgroup#1\@@endlink}%
\providecommand \@sanitize@url [0]{\catcode `\\12\catcode `\$12\catcode
  `\&12\catcode `\#12\catcode `\^12\catcode `\_12\catcode `\%12\relax}%
\providecommand \@@startlink[1]{}%
\providecommand \@@endlink[0]{}%
\providecommand \url  [0]{\begingroup\@sanitize@url \@url }%
\providecommand \@url [1]{\endgroup\@href {#1}{\urlprefix }}%
\providecommand \urlprefix  [0]{URL }%
\providecommand \Eprint [0]{\href }%
\providecommand \doibase [0]{http://dx.doi.org/}%
\providecommand \selectlanguage [0]{\@gobble}%
\providecommand \bibinfo  [0]{\@secondoftwo}%
\providecommand \bibfield  [0]{\@secondoftwo}%
\providecommand \translation [1]{[#1]}%
\providecommand \BibitemOpen [0]{}%
\providecommand \bibitemStop [0]{}%
\providecommand \bibitemNoStop [0]{.\EOS\space}%
\providecommand \EOS [0]{\spacefactor3000\relax}%
\providecommand \BibitemShut  [1]{\csname bibitem#1\endcsname}%
\let\auto@bib@innerbib\@empty
\bibitem [{\citenamefont {Zwicky}(1933)}]{zwicky1933rotverschiebung}%
  \BibitemOpen
  \bibfield  {author} {\bibinfo {author} {\bibfnamefont {F.}~\bibnamefont
  {Zwicky}},\ }\href@noop {} {\bibfield  {journal} {\bibinfo  {journal} {Helv.
  Phys. Acta}\ }\textbf {\bibinfo {volume} {6}},\ \bibinfo {pages} {110}
  (\bibinfo {year} {1933})}\BibitemShut {NoStop}%
\bibitem [{\citenamefont {Clowe}\ \emph {et~al.}(2004)\citenamefont {Clowe},
  \citenamefont {Gonzalez},\ and\ \citenamefont {Markevitch}}]{clowe2004weak}%
  \BibitemOpen
  \bibfield  {author} {\bibinfo {author} {\bibfnamefont {D.}~\bibnamefont
  {Clowe}}, \bibinfo {author} {\bibfnamefont {A.}~\bibnamefont {Gonzalez}}, \
  and\ \bibinfo {author} {\bibfnamefont {M.}~\bibnamefont {Markevitch}},\
  }\href@noop {} {\bibfield  {journal} {\bibinfo  {journal} {Astrophys. J.}\
  }\textbf {\bibinfo {volume} {604}},\ \bibinfo {pages} {596} (\bibinfo {year}
  {2004})},\ \Eprint {http://arxiv.org/abs/astro-ph/0312273} {astro-ph/0312273}
  \BibitemShut {NoStop}%
\bibitem [{\citenamefont {Ade}\ \emph {et~al.}(2016)\citenamefont {Ade} \emph
  {et~al.}}]{2016planck}%
  \BibitemOpen
  \bibfield  {author} {\bibinfo {author} {\bibfnamefont {P.~A.~R.}\
  \bibnamefont {Ade}} \emph {et~al.} (\bibinfo {collaboration} {Planck
  Collaboration}),\ }\href@noop {} {\bibfield  {journal} {\bibinfo  {journal}
  {Astron. Astrophys.}\ }\textbf {\bibinfo {volume} {594}},\ \bibinfo {pages}
  {A13} (\bibinfo {year} {2016})},\ \Eprint {http://arxiv.org/abs/1502.01589}
  {1502.01589} \BibitemShut {NoStop}%
\bibitem [{\citenamefont {Bertone}\ and\ \citenamefont
  {Tait}(2018)}]{bertone2018new}%
  \BibitemOpen
  \bibfield  {author} {\bibinfo {author} {\bibfnamefont {G.}~\bibnamefont
  {Bertone}}\ and\ \bibinfo {author} {\bibfnamefont {T.~M.}\ \bibnamefont
  {Tait}},\ }\href@noop {} {\bibfield  {journal} {\bibinfo  {journal} {Nature
  (London)}\ }\textbf {\bibinfo {volume} {562}},\ \bibinfo {pages} {51}
  (\bibinfo {year} {2018})},\ \Eprint {http://arxiv.org/abs/1810.01668}
  {1810.01668} \BibitemShut {NoStop}%
\bibitem [{\citenamefont {Feng}(2010)}]{feng2010dark}%
  \BibitemOpen
  \bibfield  {author} {\bibinfo {author} {\bibfnamefont {J.~L.}\ \bibnamefont
  {Feng}},\ }\href@noop {} {\bibfield  {journal} {\bibinfo  {journal} {Annu.
  Rev. Astron. Astrophys.}\ }\textbf {\bibinfo {volume} {48}},\ \bibinfo
  {pages} {495} (\bibinfo {year} {2010})},\ \Eprint
  {http://arxiv.org/abs/1003.0904} {1003.0904} \BibitemShut {NoStop}%
\bibitem [{\citenamefont {Bertone}(2010)}]{bertone2010particle}%
  \BibitemOpen
  \bibfield  {author} {\bibinfo {author} {\bibfnamefont {G.}~\bibnamefont
  {Bertone}},\ }\href@noop {} {\emph {\bibinfo {title} {Particle Dark Matter:
  Observations, Models and Searches}}}\ (\bibinfo  {publisher} {Cambridge
  University Press, Cambridge, England},\ \bibinfo {year} {2010})\BibitemShut
  {NoStop}%
\bibitem [{\citenamefont {Bertone}\ \emph {et~al.}(2005)\citenamefont
  {Bertone}, \citenamefont {Hooper},\ and\ \citenamefont
  {Silk}}]{bertone2005particle}%
  \BibitemOpen
  \bibfield  {author} {\bibinfo {author} {\bibfnamefont {G.}~\bibnamefont
  {Bertone}}, \bibinfo {author} {\bibfnamefont {D.}~\bibnamefont {Hooper}}, \
  and\ \bibinfo {author} {\bibfnamefont {J.}~\bibnamefont {Silk}},\ }\href@noop
  {} {\bibfield  {journal} {\bibinfo  {journal} {Phys. Rep.}\ }\textbf
  {\bibinfo {volume} {405}},\ \bibinfo {pages} {279} (\bibinfo {year}
  {2005})},\ \Eprint {http://arxiv.org/abs/hep-ph/0404175} {hep-ph/0404175}
  \BibitemShut {NoStop}%
\bibitem [{\citenamefont {Bergstr{\"o}m}(2000)}]{bergstrom2000non}%
  \BibitemOpen
  \bibfield  {author} {\bibinfo {author} {\bibfnamefont {L.}~\bibnamefont
  {Bergstr{\"o}m}},\ }\href@noop {} {\bibfield  {journal} {\bibinfo  {journal}
  {Rep. Prog. Phys.}\ }\textbf {\bibinfo {volume} {63}},\ \bibinfo {pages}
  {793} (\bibinfo {year} {2000})},\ \Eprint
  {http://arxiv.org/abs/hep-ph/0002126} {hep-ph/0002126} \BibitemShut {NoStop}%
\bibitem [{\citenamefont {Jungman}\ \emph {et~al.}(1996)\citenamefont
  {Jungman}, \citenamefont {Kamionkowski},\ and\ \citenamefont
  {Griest}}]{jungman1996supersymmetric}%
  \BibitemOpen
  \bibfield  {author} {\bibinfo {author} {\bibfnamefont {G.}~\bibnamefont
  {Jungman}}, \bibinfo {author} {\bibfnamefont {M.}~\bibnamefont
  {Kamionkowski}}, \ and\ \bibinfo {author} {\bibfnamefont {K.}~\bibnamefont
  {Griest}},\ }\href@noop {} {\bibfield  {journal} {\bibinfo  {journal} {Phys.
  Rep.}\ }\textbf {\bibinfo {volume} {267}},\ \bibinfo {pages} {195} (\bibinfo
  {year} {1996})},\ \Eprint {http://arxiv.org/abs/hep-ph/9506380}
  {hep-ph/9506380} \BibitemShut {NoStop}%
\bibitem [{\citenamefont {Strigari}\ \emph {et~al.}(2007)\citenamefont
  {Strigari}, \citenamefont {Koushiappas}, \citenamefont {Bullock},\ and\
  \citenamefont {Kaplinghat}}]{strigari2007precise}%
  \BibitemOpen
  \bibfield  {author} {\bibinfo {author} {\bibfnamefont {L.~E.}\ \bibnamefont
  {Strigari}}, \bibinfo {author} {\bibfnamefont {S.~M.}\ \bibnamefont
  {Koushiappas}}, \bibinfo {author} {\bibfnamefont {J.~S.}\ \bibnamefont
  {Bullock}}, \ and\ \bibinfo {author} {\bibfnamefont {M.}~\bibnamefont
  {Kaplinghat}},\ }\href@noop {} {\bibfield  {journal} {\bibinfo  {journal}
  {Phys. Rev. D}\ }\textbf {\bibinfo {volume} {75}},\ \bibinfo {pages} {083526}
  (\bibinfo {year} {2007})},\ \Eprint {http://arxiv.org/abs/astro-ph/0611925}
  {astro-ph/0611925} \BibitemShut {NoStop}%
\bibitem [{\citenamefont {Navarro}\ \emph {et~al.}(1996)\citenamefont
  {Navarro}, \citenamefont {Frenk},\ and\ \citenamefont
  {White}}]{navarro1996structure}%
  \BibitemOpen
  \bibfield  {author} {\bibinfo {author} {\bibfnamefont {J.~F.}\ \bibnamefont
  {Navarro}}, \bibinfo {author} {\bibfnamefont {C.~S.}\ \bibnamefont {Frenk}},
  \ and\ \bibinfo {author} {\bibfnamefont {S.~D.~M.}\ \bibnamefont {White}},\
  }\href@noop {} {\bibfield  {journal} {\bibinfo  {journal} {Astrophys. J.}\
  }\textbf {\bibinfo {volume} {462}},\ \bibinfo {pages} {563} (\bibinfo {year}
  {1996})},\ \Eprint {http://arxiv.org/abs/astro-ph/9508025} {astro-ph/9508025}
  \BibitemShut {NoStop}%
\bibitem [{\citenamefont {Navarro}\ \emph {et~al.}(1997)\citenamefont
  {Navarro}, \citenamefont {Frenk},\ and\ \citenamefont
  {White}}]{navarro1997universal}%
  \BibitemOpen
  \bibfield  {author} {\bibinfo {author} {\bibfnamefont {J.~F.}\ \bibnamefont
  {Navarro}}, \bibinfo {author} {\bibfnamefont {C.~S.}\ \bibnamefont {Frenk}},
  \ and\ \bibinfo {author} {\bibfnamefont {S.~D.~M.}\ \bibnamefont {White}},\
  }\href@noop {} {\bibfield  {journal} {\bibinfo  {journal} {Astrophys. J.}\
  }\textbf {\bibinfo {volume} {490}},\ \bibinfo {pages} {493} (\bibinfo {year}
  {1997})},\ \Eprint {http://arxiv.org/abs/astro-ph/9611107} {astro-ph/9611107}
  \BibitemShut {NoStop}%
\bibitem [{\citenamefont {Brooks}(2014)}]{brooks2014re}%
  \BibitemOpen
  \bibfield  {author} {\bibinfo {author} {\bibfnamefont {A.}~\bibnamefont
  {Brooks}},\ }\href@noop {} {\bibfield  {journal} {\bibinfo  {journal} {Ann.
  Phys. (Amsterdam)}\ }\textbf {\bibinfo {volume} {526}},\ \bibinfo {pages}
  {294} (\bibinfo {year} {2014})},\ \Eprint {http://arxiv.org/abs/1407.7544}
  {1407.7544} \BibitemShut {NoStop}%
\bibitem [{\citenamefont {Binney}\ and\ \citenamefont
  {Tremaine}(1987)}]{binney1987galactic}%
  \BibitemOpen
  \bibfield  {author} {\bibinfo {author} {\bibfnamefont {J.}~\bibnamefont
  {Binney}}\ and\ \bibinfo {author} {\bibfnamefont {S.}~\bibnamefont
  {Tremaine}},\ }\href@noop {} {\emph {\bibinfo {title} {Galactic Dynamics}}}\
  (\bibinfo  {publisher} {Princeton University Press},\ \bibinfo {address}
  {Princeton, NJ},\ \bibinfo {year} {1987})\BibitemShut {NoStop}%
\bibitem [{\citenamefont {Mo}\ \emph {et~al.}(2010)\citenamefont {Mo},
  \citenamefont {Van~den Bosch},\ and\ \citenamefont {White}}]{mo2010galaxy}%
  \BibitemOpen
  \bibfield  {author} {\bibinfo {author} {\bibfnamefont {H.}~\bibnamefont
  {Mo}}, \bibinfo {author} {\bibfnamefont {F.}~\bibnamefont {Van~den Bosch}}, \
  and\ \bibinfo {author} {\bibfnamefont {S.}~\bibnamefont {White}},\
  }\href@noop {} {\emph {\bibinfo {title} {Galaxy Formation and Evolution}}}\
  (\bibinfo  {publisher} {Cambridge University Press, Cambridge, England},\
  \bibinfo {year} {2010})\BibitemShut {NoStop}%
\bibitem [{\citenamefont {Chandrasekhar}(1943)}]{chandrasekhar1943dynamical}%
  \BibitemOpen
  \bibfield  {author} {\bibinfo {author} {\bibfnamefont {S.}~\bibnamefont
  {Chandrasekhar}},\ }\href@noop {} {\bibfield  {journal} {\bibinfo  {journal}
  {Astrophys. J.}\ }\textbf {\bibinfo {volume} {97}},\ \bibinfo {pages} {255}
  (\bibinfo {year} {1943})}\BibitemShut {NoStop}%
\bibitem [{\citenamefont {Goerdt}\ \emph {et~al.}(2007)\citenamefont {Goerdt},
  \citenamefont {Gnedin}, \citenamefont {Moore}, \citenamefont {Diemand},\ and\
  \citenamefont {Stadel}}]{goerdt2007survival}%
  \BibitemOpen
  \bibfield  {author} {\bibinfo {author} {\bibfnamefont {T.}~\bibnamefont
  {Goerdt}}, \bibinfo {author} {\bibfnamefont {O.~Y.}\ \bibnamefont {Gnedin}},
  \bibinfo {author} {\bibfnamefont {B.}~\bibnamefont {Moore}}, \bibinfo
  {author} {\bibfnamefont {J.}~\bibnamefont {Diemand}}, \ and\ \bibinfo
  {author} {\bibfnamefont {J.}~\bibnamefont {Stadel}},\ }\href@noop {}
  {\bibfield  {journal} {\bibinfo  {journal} {Mon. Not. R. Astron. Soc.}\
  }\textbf {\bibinfo {volume} {375}},\ \bibinfo {pages} {191} (\bibinfo {year}
  {2007})},\ \Eprint {http://arxiv.org/abs/astro-ph/0608495} {astro-ph/0608495}
  \BibitemShut {NoStop}%
\bibitem [{\citenamefont {Penarrubia}\ \emph {et~al.}(2010)\citenamefont
  {Penarrubia}, \citenamefont {Benson}, \citenamefont {Walker}, \citenamefont
  {Gilmore}, \citenamefont {McConnachie},\ and\ \citenamefont
  {Mayer}}]{penarrubia2010impact}%
  \BibitemOpen
  \bibfield  {author} {\bibinfo {author} {\bibfnamefont {J.}~\bibnamefont
  {Penarrubia}}, \bibinfo {author} {\bibfnamefont {A.~J.}\ \bibnamefont
  {Benson}}, \bibinfo {author} {\bibfnamefont {M.~G.}\ \bibnamefont {Walker}},
  \bibinfo {author} {\bibfnamefont {G.}~\bibnamefont {Gilmore}}, \bibinfo
  {author} {\bibfnamefont {A.~W.}\ \bibnamefont {McConnachie}}, \ and\ \bibinfo
  {author} {\bibfnamefont {L.}~\bibnamefont {Mayer}},\ }\href@noop {}
  {\bibfield  {journal} {\bibinfo  {journal} {Mon. Not. R. Astron. Soc.}\
  }\textbf {\bibinfo {volume} {406}},\ \bibinfo {pages} {1290} (\bibinfo {year}
  {2010})},\ \Eprint {http://arxiv.org/abs/1002.3376} {1002.3376} \BibitemShut
  {NoStop}%
\bibitem [{\citenamefont {Errani}\ and\ \citenamefont
  {Pe{\~n}arrubia}()}]{errani2019can}%
  \BibitemOpen
  \bibfield  {author} {\bibinfo {author} {\bibfnamefont {R.}~\bibnamefont
  {Errani}}\ and\ \bibinfo {author} {\bibfnamefont {J.}~\bibnamefont
  {Pe{\~n}arrubia}},\ }\href@noop {} {\ }\Eprint
  {http://arxiv.org/abs/1906.01642} {1906.01642} \BibitemShut {NoStop}%
\bibitem [{\citenamefont {Berezinsky}\ \emph {et~al.}(2008)\citenamefont
  {Berezinsky}, \citenamefont {Dokuchaev},\ and\ \citenamefont
  {Eroshenko}}]{berezinsky2008remnants}%
  \BibitemOpen
  \bibfield  {author} {\bibinfo {author} {\bibfnamefont {V.}~\bibnamefont
  {Berezinsky}}, \bibinfo {author} {\bibfnamefont {V.}~\bibnamefont
  {Dokuchaev}}, \ and\ \bibinfo {author} {\bibfnamefont {Y.}~\bibnamefont
  {Eroshenko}},\ }\href@noop {} {\bibfield  {journal} {\bibinfo  {journal}
  {Phys. Rev. D}\ }\textbf {\bibinfo {volume} {77}},\ \bibinfo {pages} {083519}
  (\bibinfo {year} {2008})},\ \Eprint {http://arxiv.org/abs/0712.3499}
  {0712.3499} \BibitemShut {NoStop}%
\bibitem [{\citenamefont {Berezinsky}\ \emph {et~al.}(2003)\citenamefont
  {Berezinsky}, \citenamefont {Dokuchaev},\ and\ \citenamefont
  {Eroshenko}}]{berezinsky2003small}%
  \BibitemOpen
  \bibfield  {author} {\bibinfo {author} {\bibfnamefont {V.}~\bibnamefont
  {Berezinsky}}, \bibinfo {author} {\bibfnamefont {V.}~\bibnamefont
  {Dokuchaev}}, \ and\ \bibinfo {author} {\bibfnamefont {Y.}~\bibnamefont
  {Eroshenko}},\ }\href@noop {} {\bibfield  {journal} {\bibinfo  {journal}
  {Phys. Rev. D}\ }\textbf {\bibinfo {volume} {68}},\ \bibinfo {pages} {103003}
  (\bibinfo {year} {2003})},\ \Eprint {http://arxiv.org/abs/astro-ph/0301551}
  {astro-ph/0301551} \BibitemShut {NoStop}%
\bibitem [{\citenamefont {Diemand}\ \emph {et~al.}(2005)\citenamefont
  {Diemand}, \citenamefont {Moore},\ and\ \citenamefont
  {Stadel}}]{diemand2005earth}%
  \BibitemOpen
  \bibfield  {author} {\bibinfo {author} {\bibfnamefont {J.}~\bibnamefont
  {Diemand}}, \bibinfo {author} {\bibfnamefont {B.}~\bibnamefont {Moore}}, \
  and\ \bibinfo {author} {\bibfnamefont {J.}~\bibnamefont {Stadel}},\
  }\href@noop {} {\bibfield  {journal} {\bibinfo  {journal} {Nature (London)}\
  }\textbf {\bibinfo {volume} {433}},\ \bibinfo {pages} {389} (\bibinfo {year}
  {2005})},\ \Eprint {http://arxiv.org/abs/astro-ph/0501589} {astro-ph/0501589}
  \BibitemShut {NoStop}%
\bibitem [{\citenamefont {Pieri}\ \emph {et~al.}(2008)\citenamefont {Pieri},
  \citenamefont {Bertone},\ and\ \citenamefont {Branchini}}]{pieri2008dark}%
  \BibitemOpen
  \bibfield  {author} {\bibinfo {author} {\bibfnamefont {L.}~\bibnamefont
  {Pieri}}, \bibinfo {author} {\bibfnamefont {G.}~\bibnamefont {Bertone}}, \
  and\ \bibinfo {author} {\bibfnamefont {E.}~\bibnamefont {Branchini}},\
  }\href@noop {} {\bibfield  {journal} {\bibinfo  {journal} {Mon. Not. R.
  Astron. Soc.}\ }\textbf {\bibinfo {volume} {384}},\ \bibinfo {pages} {1627}
  (\bibinfo {year} {2008})},\ \Eprint {http://arxiv.org/abs/0706.2101}
  {0706.2101} \BibitemShut {NoStop}%
\bibitem [{\citenamefont {Ishiyama}\ \emph {et~al.}(2010)\citenamefont
  {Ishiyama}, \citenamefont {Makino},\ and\ \citenamefont
  {Ebisuzaki}}]{ishiyama2010gamma}%
  \BibitemOpen
  \bibfield  {author} {\bibinfo {author} {\bibfnamefont {T.}~\bibnamefont
  {Ishiyama}}, \bibinfo {author} {\bibfnamefont {J.}~\bibnamefont {Makino}}, \
  and\ \bibinfo {author} {\bibfnamefont {T.}~\bibnamefont {Ebisuzaki}},\
  }\href@noop {} {\bibfield  {journal} {\bibinfo  {journal} {Astrophys. J.
  Lett.}\ }\textbf {\bibinfo {volume} {723}},\ \bibinfo {pages} {L195}
  (\bibinfo {year} {2010})},\ \Eprint {http://arxiv.org/abs/1006.3392}
  {1006.3392} \BibitemShut {NoStop}%
\bibitem [{\citenamefont {Anderhalden}\ and\ \citenamefont
  {Diemand}(2013{\natexlab{a}})}]{anderhalden2013density}%
  \BibitemOpen
  \bibfield  {author} {\bibinfo {author} {\bibfnamefont {D.}~\bibnamefont
  {Anderhalden}}\ and\ \bibinfo {author} {\bibfnamefont {J.}~\bibnamefont
  {Diemand}},\ }\href@noop {} {\bibfield  {journal} {\bibinfo  {journal} {J.
  Cosmol. Astropart. Phys.}\ }\textbf {\bibinfo {volume} {04}},\ \bibinfo
  {pages} {009} (\bibinfo {year} {2013}{\natexlab{a}})},\ \Eprint
  {http://arxiv.org/abs/1302.0003} {1302.0003} \BibitemShut {NoStop}%
\bibitem [{\citenamefont {Anderhalden}\ and\ \citenamefont
  {Diemand}(2013{\natexlab{b}})}]{anderhalden2013erratum}%
  \BibitemOpen
  \bibfield  {author} {\bibinfo {author} {\bibfnamefont {D.}~\bibnamefont
  {Anderhalden}}\ and\ \bibinfo {author} {\bibfnamefont {J.}~\bibnamefont
  {Diemand}},\ }\href@noop {} {\bibfield  {journal} {\bibinfo  {journal} {J.
  Cosmol. Astropart. Phys.}\ }\textbf {\bibinfo {volume} {08}},\ \bibinfo
  {pages} {E02} (\bibinfo {year} {2013}{\natexlab{b}})}\BibitemShut {NoStop}%
\bibitem [{\citenamefont {Ishiyama}(2014)}]{ishiyama2014hierarchical}%
  \BibitemOpen
  \bibfield  {author} {\bibinfo {author} {\bibfnamefont {T.}~\bibnamefont
  {Ishiyama}},\ }\href@noop {} {\bibfield  {journal} {\bibinfo  {journal}
  {Astrophys. J.}\ }\textbf {\bibinfo {volume} {788}},\ \bibinfo {pages} {27}
  (\bibinfo {year} {2014})},\ \Eprint {http://arxiv.org/abs/1404.1650}
  {1404.1650} \BibitemShut {NoStop}%
\bibitem [{\citenamefont {S{\'a}nchez-Conde}\ and\ \citenamefont
  {Prada}(2014)}]{sanchez2014flattening}%
  \BibitemOpen
  \bibfield  {author} {\bibinfo {author} {\bibfnamefont {M.~A.}\ \bibnamefont
  {S{\'a}nchez-Conde}}\ and\ \bibinfo {author} {\bibfnamefont {F.}~\bibnamefont
  {Prada}},\ }\href@noop {} {\bibfield  {journal} {\bibinfo  {journal} {Mon.
  Not. R. Astron. Soc.}\ }\textbf {\bibinfo {volume} {442}},\ \bibinfo {pages}
  {2271} (\bibinfo {year} {2014})},\ \Eprint {http://arxiv.org/abs/1312.1729}
  {1312.1729} \BibitemShut {NoStop}%
\bibitem [{\citenamefont {Anderson}\ \emph {et~al.}(2016)\citenamefont
  {Anderson}, \citenamefont {Zimmer}, \citenamefont {Conrad}, \citenamefont
  {Gustafsson}, \citenamefont {S{\'a}nchez-Conde},\ and\ \citenamefont
  {Caputo}}]{anderson2016search}%
  \BibitemOpen
  \bibfield  {author} {\bibinfo {author} {\bibfnamefont {B.}~\bibnamefont
  {Anderson}}, \bibinfo {author} {\bibfnamefont {S.}~\bibnamefont {Zimmer}},
  \bibinfo {author} {\bibfnamefont {J.}~\bibnamefont {Conrad}}, \bibinfo
  {author} {\bibfnamefont {M.}~\bibnamefont {Gustafsson}}, \bibinfo {author}
  {\bibfnamefont {M.}~\bibnamefont {S{\'a}nchez-Conde}}, \ and\ \bibinfo
  {author} {\bibfnamefont {R.}~\bibnamefont {Caputo}},\ }\href@noop {}
  {\bibfield  {journal} {\bibinfo  {journal} {J. Cosmol. Astropart. Phys.}\
  }\textbf {\bibinfo {volume} {02}},\ \bibinfo {pages} {026} (\bibinfo {year}
  {2016})},\ \Eprint {http://arxiv.org/abs/1511.00014} {1511.00014}
  \BibitemShut {NoStop}%
\bibitem [{\citenamefont {Gao}\ \emph {et~al.}(2012)\citenamefont {Gao},
  \citenamefont {Frenk}, \citenamefont {Jenkins}, \citenamefont {Springel},\
  and\ \citenamefont {White}}]{gao2011will}%
  \BibitemOpen
  \bibfield  {author} {\bibinfo {author} {\bibfnamefont {L.}~\bibnamefont
  {Gao}}, \bibinfo {author} {\bibfnamefont {C.}~\bibnamefont {Frenk}}, \bibinfo
  {author} {\bibfnamefont {A.}~\bibnamefont {Jenkins}}, \bibinfo {author}
  {\bibfnamefont {V.}~\bibnamefont {Springel}}, \ and\ \bibinfo {author}
  {\bibfnamefont {S.}~\bibnamefont {White}},\ }\href@noop {} {\bibfield
  {journal} {\bibinfo  {journal} {Mon. Not. R. Astron. Soc.}\ }\textbf
  {\bibinfo {volume} {419}},\ \bibinfo {pages} {1721} (\bibinfo {year}
  {2012})},\ \Eprint {http://arxiv.org/abs/1107.1916} {1107.1916} \BibitemShut
  {NoStop}%
\bibitem [{\citenamefont {Springel}\ \emph {et~al.}(2008)\citenamefont
  {Springel}, \citenamefont {White}, \citenamefont {Frenk}, \citenamefont
  {Navarro}, \citenamefont {Jenkins}, \citenamefont {Vogelsberger},
  \citenamefont {Wang}, \citenamefont {Ludlow},\ and\ \citenamefont
  {Helmi}}]{springel2008prospects}%
  \BibitemOpen
  \bibfield  {author} {\bibinfo {author} {\bibfnamefont {V.}~\bibnamefont
  {Springel}}, \bibinfo {author} {\bibfnamefont {S.~D.}\ \bibnamefont {White}},
  \bibinfo {author} {\bibfnamefont {C.~S.}\ \bibnamefont {Frenk}}, \bibinfo
  {author} {\bibfnamefont {J.~F.}\ \bibnamefont {Navarro}}, \bibinfo {author}
  {\bibfnamefont {A.}~\bibnamefont {Jenkins}}, \bibinfo {author} {\bibfnamefont
  {M.}~\bibnamefont {Vogelsberger}}, \bibinfo {author} {\bibfnamefont
  {J.}~\bibnamefont {Wang}}, \bibinfo {author} {\bibfnamefont {A.}~\bibnamefont
  {Ludlow}}, \ and\ \bibinfo {author} {\bibfnamefont {A.}~\bibnamefont
  {Helmi}},\ }\href@noop {} {\bibfield  {journal} {\bibinfo  {journal} {Nature
  (London)}\ }\textbf {\bibinfo {volume} {456}},\ \bibinfo {pages} {73}
  (\bibinfo {year} {2008})},\ \Eprint {http://arxiv.org/abs/0809.0894}
  {0809.0894} \BibitemShut {NoStop}%
\bibitem [{\citenamefont {Stref}\ and\ \citenamefont
  {Lavalle}(2017)}]{stref2017modeling}%
  \BibitemOpen
  \bibfield  {author} {\bibinfo {author} {\bibfnamefont {M.}~\bibnamefont
  {Stref}}\ and\ \bibinfo {author} {\bibfnamefont {J.}~\bibnamefont
  {Lavalle}},\ }\href@noop {} {\bibfield  {journal} {\bibinfo  {journal} {Phys.
  Rev. D}\ }\textbf {\bibinfo {volume} {95}},\ \bibinfo {pages} {063003}
  (\bibinfo {year} {2017})},\ \Eprint {http://arxiv.org/abs/1610.02233}
  {1610.02233} \BibitemShut {NoStop}%
\bibitem [{\citenamefont {Stref}\ \emph {et~al.}(2019)\citenamefont {Stref},
  \citenamefont {Lacroix},\ and\ \citenamefont {Lavalle}}]{stref2019remnants}%
  \BibitemOpen
  \bibfield  {author} {\bibinfo {author} {\bibfnamefont {M.}~\bibnamefont
  {Stref}}, \bibinfo {author} {\bibfnamefont {T.}~\bibnamefont {Lacroix}}, \
  and\ \bibinfo {author} {\bibfnamefont {J.}~\bibnamefont {Lavalle}},\
  }\href@noop {} {\bibfield  {journal} {\bibinfo  {journal} {Galaxies}\
  }\textbf {\bibinfo {volume} {7}},\ \bibinfo {pages} {65} (\bibinfo {year}
  {2019})},\ \Eprint {http://arxiv.org/abs/1905.02008} {1905.02008}
  \BibitemShut {NoStop}%
\bibitem [{\citenamefont {Bartels}\ and\ \citenamefont
  {Ando}(2015)}]{bartels2015boosting}%
  \BibitemOpen
  \bibfield  {author} {\bibinfo {author} {\bibfnamefont {R.}~\bibnamefont
  {Bartels}}\ and\ \bibinfo {author} {\bibfnamefont {S.}~\bibnamefont {Ando}},\
  }\href@noop {} {\bibfield  {journal} {\bibinfo  {journal} {Phys. Rev. D}\
  }\textbf {\bibinfo {volume} {92}},\ \bibinfo {pages} {123508} (\bibinfo
  {year} {2015})},\ \Eprint {http://arxiv.org/abs/1507.08656} {1507.08656}
  \BibitemShut {NoStop}%
\bibitem [{\citenamefont {Hiroshima}\ \emph {et~al.}(2018)\citenamefont
  {Hiroshima}, \citenamefont {Ando},\ and\ \citenamefont
  {Ishiyama}}]{hiroshima2018modeling}%
  \BibitemOpen
  \bibfield  {author} {\bibinfo {author} {\bibfnamefont {N.}~\bibnamefont
  {Hiroshima}}, \bibinfo {author} {\bibfnamefont {S.}~\bibnamefont {Ando}}, \
  and\ \bibinfo {author} {\bibfnamefont {T.}~\bibnamefont {Ishiyama}},\
  }\href@noop {} {\bibfield  {journal} {\bibinfo  {journal} {Phys. Rev. D}\
  }\textbf {\bibinfo {volume} {97}},\ \bibinfo {pages} {123002} (\bibinfo
  {year} {2018})},\ \Eprint {http://arxiv.org/abs/1803.07691} {1803.07691}
  \BibitemShut {NoStop}%
\bibitem [{\citenamefont {Ando}\ \emph {et~al.}(2019)\citenamefont {Ando},
  \citenamefont {Ishiyama},\ and\ \citenamefont {Hiroshima}}]{ando2019halo}%
  \BibitemOpen
  \bibfield  {author} {\bibinfo {author} {\bibfnamefont {S.}~\bibnamefont
  {Ando}}, \bibinfo {author} {\bibfnamefont {T.}~\bibnamefont {Ishiyama}}, \
  and\ \bibinfo {author} {\bibfnamefont {N.}~\bibnamefont {Hiroshima}},\
  }\href@noop {} {\bibfield  {journal} {\bibinfo  {journal} {Galaxies}\
  }\textbf {\bibinfo {volume} {7}},\ \bibinfo {pages} {68} (\bibinfo {year}
  {2019})},\ \Eprint {http://arxiv.org/abs/1903.11427} {1903.11427}
  \BibitemShut {NoStop}%
\bibitem [{\citenamefont {Erickcek}\ and\ \citenamefont
  {Sigurdson}(2011)}]{erickcek2011reheating}%
  \BibitemOpen
  \bibfield  {author} {\bibinfo {author} {\bibfnamefont {A.~L.}\ \bibnamefont
  {Erickcek}}\ and\ \bibinfo {author} {\bibfnamefont {K.}~\bibnamefont
  {Sigurdson}},\ }\href@noop {} {\bibfield  {journal} {\bibinfo  {journal}
  {Phys. Rev. D}\ }\textbf {\bibinfo {volume} {84}},\ \bibinfo {pages} {083503}
  (\bibinfo {year} {2011})},\ \Eprint {http://arxiv.org/abs/1106.0536}
  {1106.0536} \BibitemShut {NoStop}%
\bibitem [{\citenamefont {Barenboim}\ and\ \citenamefont
  {Rasero}(2014)}]{barenboim2014structure}%
  \BibitemOpen
  \bibfield  {author} {\bibinfo {author} {\bibfnamefont {G.}~\bibnamefont
  {Barenboim}}\ and\ \bibinfo {author} {\bibfnamefont {J.}~\bibnamefont
  {Rasero}},\ }\href@noop {} {\bibfield  {journal} {\bibinfo  {journal} {J.
  High Energy Phys.}\ }\textbf {\bibinfo {volume} {4}},\ \bibinfo {pages} {138}
  (\bibinfo {year} {2014})},\ \Eprint {http://arxiv.org/abs/1311.4034}
  {1311.4034} \BibitemShut {NoStop}%
\bibitem [{\citenamefont {Fan}\ \emph {et~al.}(2014)\citenamefont {Fan},
  \citenamefont {{\"O}zsoy},\ and\ \citenamefont {Watson}}]{fan2014nonthermal}%
  \BibitemOpen
  \bibfield  {author} {\bibinfo {author} {\bibfnamefont {J.~J.}\ \bibnamefont
  {Fan}}, \bibinfo {author} {\bibfnamefont {O.}~\bibnamefont {{\"O}zsoy}}, \
  and\ \bibinfo {author} {\bibfnamefont {S.}~\bibnamefont {Watson}},\
  }\href@noop {} {\bibfield  {journal} {\bibinfo  {journal} {Phys. Rev. D}\
  }\textbf {\bibinfo {volume} {90}},\ \bibinfo {pages} {043536} (\bibinfo
  {year} {2014})},\ \Eprint {http://arxiv.org/abs/1405.7373} {1405.7373}
  \BibitemShut {NoStop}%
\bibitem [{\citenamefont {Erickcek}(2015)}]{erickcek2015dark}%
  \BibitemOpen
  \bibfield  {author} {\bibinfo {author} {\bibfnamefont {A.~L.}\ \bibnamefont
  {Erickcek}},\ }\href@noop {} {\bibfield  {journal} {\bibinfo  {journal}
  {Phys. Rev. D}\ }\textbf {\bibinfo {volume} {92}},\ \bibinfo {pages} {103505}
  (\bibinfo {year} {2015})},\ \Eprint {http://arxiv.org/abs/1504.03335}
  {1504.03335} \BibitemShut {NoStop}%
\bibitem [{\citenamefont {Redmond}\ \emph {et~al.}(2018)\citenamefont
  {Redmond}, \citenamefont {Trezza},\ and\ \citenamefont
  {Erickcek}}]{redmond2018growth}%
  \BibitemOpen
  \bibfield  {author} {\bibinfo {author} {\bibfnamefont {K.}~\bibnamefont
  {Redmond}}, \bibinfo {author} {\bibfnamefont {A.}~\bibnamefont {Trezza}}, \
  and\ \bibinfo {author} {\bibfnamefont {A.~L.}\ \bibnamefont {Erickcek}},\
  }\href@noop {} {\bibfield  {journal} {\bibinfo  {journal} {Phys. Rev. D}\
  }\textbf {\bibinfo {volume} {98}},\ \bibinfo {pages} {063504} (\bibinfo
  {year} {2018})},\ \Eprint {http://arxiv.org/abs/1807.01327} {1807.01327}
  \BibitemShut {NoStop}%
\bibitem [{\citenamefont {Silk}\ and\ \citenamefont
  {Turner}(1987)}]{silk1987double}%
  \BibitemOpen
  \bibfield  {author} {\bibinfo {author} {\bibfnamefont {J.}~\bibnamefont
  {Silk}}\ and\ \bibinfo {author} {\bibfnamefont {M.~S.}\ \bibnamefont
  {Turner}},\ }\href@noop {} {\bibfield  {journal} {\bibinfo  {journal} {Phys.
  Rev. D}\ }\textbf {\bibinfo {volume} {35}},\ \bibinfo {pages} {419} (\bibinfo
  {year} {1987})}\BibitemShut {NoStop}%
\bibitem [{\citenamefont {Salopek}\ \emph {et~al.}(1989)\citenamefont
  {Salopek}, \citenamefont {Bond},\ and\ \citenamefont
  {Bardeen}}]{salopek1989designing}%
  \BibitemOpen
  \bibfield  {author} {\bibinfo {author} {\bibfnamefont {D.~S.}\ \bibnamefont
  {Salopek}}, \bibinfo {author} {\bibfnamefont {J.~R.}\ \bibnamefont {Bond}}, \
  and\ \bibinfo {author} {\bibfnamefont {J.~M.}\ \bibnamefont {Bardeen}},\
  }\href@noop {} {\bibfield  {journal} {\bibinfo  {journal} {Phys. Rev. D}\
  }\textbf {\bibinfo {volume} {40}},\ \bibinfo {pages} {1753} (\bibinfo {year}
  {1989})}\BibitemShut {NoStop}%
\bibitem [{\citenamefont {Starobinskij}(1992)}]{starobinsky1992}%
  \BibitemOpen
  \bibfield  {author} {\bibinfo {author} {\bibfnamefont {A.~A.}\ \bibnamefont
  {Starobinskij}},\ }\href@noop {} {\bibfield  {journal} {\bibinfo  {journal}
  {JETP Lett.}\ }\textbf {\bibinfo {volume} {55}},\ \bibinfo {pages} {489}
  (\bibinfo {year} {1992})}\BibitemShut {NoStop}%
\bibitem [{\citenamefont {Ivanov}\ \emph {et~al.}(1994)\citenamefont {Ivanov},
  \citenamefont {Naselsky},\ and\ \citenamefont
  {Novikov}}]{ivanov1994inflation}%
  \BibitemOpen
  \bibfield  {author} {\bibinfo {author} {\bibfnamefont {P.}~\bibnamefont
  {Ivanov}}, \bibinfo {author} {\bibfnamefont {P.}~\bibnamefont {Naselsky}}, \
  and\ \bibinfo {author} {\bibfnamefont {I.}~\bibnamefont {Novikov}},\
  }\href@noop {} {\bibfield  {journal} {\bibinfo  {journal} {Phys. Rev. D}\
  }\textbf {\bibinfo {volume} {50}},\ \bibinfo {pages} {7173} (\bibinfo {year}
  {1994})}\BibitemShut {NoStop}%
\bibitem [{\citenamefont {Randall}\ \emph {et~al.}(1996)\citenamefont
  {Randall}, \citenamefont {Solja{\v{c}}i{\'c}},\ and\ \citenamefont
  {Guth}}]{randall1996supernatural}%
  \BibitemOpen
  \bibfield  {author} {\bibinfo {author} {\bibfnamefont {L.}~\bibnamefont
  {Randall}}, \bibinfo {author} {\bibfnamefont {M.}~\bibnamefont
  {Solja{\v{c}}i{\'c}}}, \ and\ \bibinfo {author} {\bibfnamefont {A.~H.}\
  \bibnamefont {Guth}},\ }\href@noop {} {\bibfield  {journal} {\bibinfo
  {journal} {Nucl. Phys.}\ }\textbf {\bibinfo {volume} {B472}},\ \bibinfo
  {pages} {377} (\bibinfo {year} {1996})},\ \Eprint
  {http://arxiv.org/abs/hep-ph/9512439} {hep-ph/9512439} \BibitemShut {NoStop}%
\bibitem [{\citenamefont {Stewart}(1997)}]{stewart1997flattening}%
  \BibitemOpen
  \bibfield  {author} {\bibinfo {author} {\bibfnamefont {E.~D.}\ \bibnamefont
  {Stewart}},\ }\href@noop {} {\bibfield  {journal} {\bibinfo  {journal} {Phys.
  Rev. D}\ }\textbf {\bibinfo {volume} {56}},\ \bibinfo {pages} {2019}
  (\bibinfo {year} {1997})},\ \Eprint {http://arxiv.org/abs/hep-ph/9703232}
  {hep-ph/9703232} \BibitemShut {NoStop}%
\bibitem [{\citenamefont {Adams}\ \emph {et~al.}(1997)\citenamefont {Adams},
  \citenamefont {Ross},\ and\ \citenamefont {Sarkar}}]{adams1997multiple}%
  \BibitemOpen
  \bibfield  {author} {\bibinfo {author} {\bibfnamefont {J.~A.}\ \bibnamefont
  {Adams}}, \bibinfo {author} {\bibfnamefont {G.~G.}\ \bibnamefont {Ross}}, \
  and\ \bibinfo {author} {\bibfnamefont {S.}~\bibnamefont {Sarkar}},\
  }\href@noop {} {\bibfield  {journal} {\bibinfo  {journal} {Nucl. Phys.}\
  }\textbf {\bibinfo {volume} {B503}},\ \bibinfo {pages} {405} (\bibinfo {year}
  {1997})},\ \Eprint {http://arxiv.org/abs/hep-ph/9704286} {hep-ph/9704286}
  \BibitemShut {NoStop}%
\bibitem [{\citenamefont {Starobinsky}(1998)}]{starobinsky1998beyond}%
  \BibitemOpen
  \bibfield  {author} {\bibinfo {author} {\bibfnamefont {A.~A.}\ \bibnamefont
  {Starobinsky}},\ }\href@noop {} {\bibfield  {journal} {\bibinfo  {journal}
  {Gravit. Cosmol.}\ }\textbf {\bibinfo {volume} {4}},\ \bibinfo {pages} {489}
  (\bibinfo {year} {1998})},\ \Eprint {http://arxiv.org/abs/astro-ph/9811360}
  {astro-ph/9811360} \BibitemShut {NoStop}%
\bibitem [{\citenamefont {Covi}\ and\ \citenamefont
  {Lyth}(1999)}]{covi1999running}%
  \BibitemOpen
  \bibfield  {author} {\bibinfo {author} {\bibfnamefont {L.}~\bibnamefont
  {Covi}}\ and\ \bibinfo {author} {\bibfnamefont {D.~H.}\ \bibnamefont
  {Lyth}},\ }\href@noop {} {\bibfield  {journal} {\bibinfo  {journal} {Phys.
  Rev. D}\ }\textbf {\bibinfo {volume} {59}},\ \bibinfo {pages} {063515}
  (\bibinfo {year} {1999})},\ \Eprint {http://arxiv.org/abs/hep-ph/9809562}
  {hep-ph/9809562} \BibitemShut {NoStop}%
\bibitem [{\citenamefont {Martin}\ \emph {et~al.}(2000)\citenamefont {Martin},
  \citenamefont {Riazuelo},\ and\ \citenamefont
  {Sakellariadou}}]{martin2000nonvacuum}%
  \BibitemOpen
  \bibfield  {author} {\bibinfo {author} {\bibfnamefont {J.}~\bibnamefont
  {Martin}}, \bibinfo {author} {\bibfnamefont {A.}~\bibnamefont {Riazuelo}}, \
  and\ \bibinfo {author} {\bibfnamefont {M.}~\bibnamefont {Sakellariadou}},\
  }\href@noop {} {\bibfield  {journal} {\bibinfo  {journal} {Phys. Rev. D}\
  }\textbf {\bibinfo {volume} {61}},\ \bibinfo {pages} {083518} (\bibinfo
  {year} {2000})},\ \Eprint {http://arxiv.org/abs/astro-ph/9904167}
  {astro-ph/9904167} \BibitemShut {NoStop}%
\bibitem [{\citenamefont {Chung}\ \emph {et~al.}(2000)\citenamefont {Chung},
  \citenamefont {Kolb}, \citenamefont {Riotto},\ and\ \citenamefont
  {Tkachev}}]{chung2000probing}%
  \BibitemOpen
  \bibfield  {author} {\bibinfo {author} {\bibfnamefont {D.~J.~H.}\
  \bibnamefont {Chung}}, \bibinfo {author} {\bibfnamefont {E.~W.}\ \bibnamefont
  {Kolb}}, \bibinfo {author} {\bibfnamefont {A.}~\bibnamefont {Riotto}}, \ and\
  \bibinfo {author} {\bibfnamefont {I.~I.}\ \bibnamefont {Tkachev}},\
  }\href@noop {} {\bibfield  {journal} {\bibinfo  {journal} {Phys. Rev. D}\
  }\textbf {\bibinfo {volume} {62}},\ \bibinfo {pages} {043508} (\bibinfo
  {year} {2000})},\ \Eprint {http://arxiv.org/abs/hep-ph/9910437}
  {hep-ph/9910437} \BibitemShut {NoStop}%
\bibitem [{\citenamefont {Martin}\ and\ \citenamefont
  {Brandenberger}(2001)}]{martin2001trans}%
  \BibitemOpen
  \bibfield  {author} {\bibinfo {author} {\bibfnamefont {J.}~\bibnamefont
  {Martin}}\ and\ \bibinfo {author} {\bibfnamefont {R.~H.}\ \bibnamefont
  {Brandenberger}},\ }\href@noop {} {\bibfield  {journal} {\bibinfo  {journal}
  {Phys. Rev. D}\ }\textbf {\bibinfo {volume} {63}},\ \bibinfo {pages} {123501}
  (\bibinfo {year} {2001})},\ \Eprint {http://arxiv.org/abs/hep-th/0005209}
  {hep-th/0005209} \BibitemShut {NoStop}%
\bibitem [{\citenamefont {Joy}\ \emph {et~al.}(2008)\citenamefont {Joy},
  \citenamefont {Sahni},\ and\ \citenamefont {Starobinsky}}]{joy2008new}%
  \BibitemOpen
  \bibfield  {author} {\bibinfo {author} {\bibfnamefont {M.}~\bibnamefont
  {Joy}}, \bibinfo {author} {\bibfnamefont {V.}~\bibnamefont {Sahni}}, \ and\
  \bibinfo {author} {\bibfnamefont {A.~A.}\ \bibnamefont {Starobinsky}},\
  }\href@noop {} {\bibfield  {journal} {\bibinfo  {journal} {Phys. Rev. D}\
  }\textbf {\bibinfo {volume} {77}},\ \bibinfo {pages} {023514} (\bibinfo
  {year} {2008})},\ \Eprint {http://arxiv.org/abs/0711.1585} {0711.1585}
  \BibitemShut {NoStop}%
\bibitem [{\citenamefont {Barnaby}\ and\ \citenamefont
  {Huang}(2009)}]{barnaby2009particle}%
  \BibitemOpen
  \bibfield  {author} {\bibinfo {author} {\bibfnamefont {N.}~\bibnamefont
  {Barnaby}}\ and\ \bibinfo {author} {\bibfnamefont {Z.}~\bibnamefont
  {Huang}},\ }\href@noop {} {\bibfield  {journal} {\bibinfo  {journal} {Phys.
  Rev. D}\ }\textbf {\bibinfo {volume} {80}},\ \bibinfo {pages} {126018}
  (\bibinfo {year} {2009})},\ \Eprint {http://arxiv.org/abs/0909.0751}
  {0909.0751} \BibitemShut {NoStop}%
\bibitem [{\citenamefont {Barnaby}(2010)}]{barnaby2010features}%
  \BibitemOpen
  \bibfield  {author} {\bibinfo {author} {\bibfnamefont {N.}~\bibnamefont
  {Barnaby}},\ }\href@noop {} {\bibfield  {journal} {\bibinfo  {journal} {Phys.
  Rev. D}\ }\textbf {\bibinfo {volume} {82}},\ \bibinfo {pages} {106009}
  (\bibinfo {year} {2010})},\ \Eprint {http://arxiv.org/abs/1006.4615}
  {1006.4615} \BibitemShut {NoStop}%
\bibitem [{\citenamefont {Ben-Dayan}\ and\ \citenamefont
  {Brustein}(2010)}]{ben2010cosmic}%
  \BibitemOpen
  \bibfield  {author} {\bibinfo {author} {\bibfnamefont {I.}~\bibnamefont
  {Ben-Dayan}}\ and\ \bibinfo {author} {\bibfnamefont {R.}~\bibnamefont
  {Brustein}},\ }\href@noop {} {\bibfield  {journal} {\bibinfo  {journal} {J.
  Cosmol. Astropart. Phys.}\ }\textbf {\bibinfo {volume} {09}},\ \bibinfo
  {pages} {007} (\bibinfo {year} {2010})},\ \Eprint
  {http://arxiv.org/abs/0907.2384} {0907.2384} \BibitemShut {NoStop}%
\bibitem [{\citenamefont {Gong}\ and\ \citenamefont
  {Sasaki}(2011)}]{gong2011waterfall}%
  \BibitemOpen
  \bibfield  {author} {\bibinfo {author} {\bibfnamefont {J.-O.}\ \bibnamefont
  {Gong}}\ and\ \bibinfo {author} {\bibfnamefont {M.}~\bibnamefont {Sasaki}},\
  }\href@noop {} {\bibfield  {journal} {\bibinfo  {journal} {J. Cosmol.
  Astropart. Phys.}\ }\textbf {\bibinfo {volume} {03}},\ \bibinfo {pages} {028}
  (\bibinfo {year} {2011})},\ \Eprint {http://arxiv.org/abs/1010.3405}
  {1010.3405} \BibitemShut {NoStop}%
\bibitem [{\citenamefont {Lyth}(2011)}]{lyth2011contribution}%
  \BibitemOpen
  \bibfield  {author} {\bibinfo {author} {\bibfnamefont {D.~H.}\ \bibnamefont
  {Lyth}},\ }\href@noop {} {\bibfield  {journal} {\bibinfo  {journal} {J.
  Cosmol. Astropart. Phys.}\ }\textbf {\bibinfo {volume} {07}},\ \bibinfo
  {pages} {035} (\bibinfo {year} {2011})},\ \Eprint
  {http://arxiv.org/abs/1012.4617} {1012.4617} \BibitemShut {NoStop}%
\bibitem [{\citenamefont {Bugaev}\ and\ \citenamefont
  {Klimai}(2011)}]{bugaev2011curvature}%
  \BibitemOpen
  \bibfield  {author} {\bibinfo {author} {\bibfnamefont {E.}~\bibnamefont
  {Bugaev}}\ and\ \bibinfo {author} {\bibfnamefont {P.}~\bibnamefont
  {Klimai}},\ }\href@noop {} {\bibfield  {journal} {\bibinfo  {journal} {J.
  Cosmol. Astropart. Phys.}\ }\textbf {\bibinfo {volume} {11}},\ \bibinfo
  {pages} {028} (\bibinfo {year} {2011})},\ \Eprint
  {http://arxiv.org/abs/1107.3754} {1107.3754} \BibitemShut {NoStop}%
\bibitem [{\citenamefont {Barnaby}\ and\ \citenamefont
  {Peloso}(2011)}]{barnaby2011large}%
  \BibitemOpen
  \bibfield  {author} {\bibinfo {author} {\bibfnamefont {N.}~\bibnamefont
  {Barnaby}}\ and\ \bibinfo {author} {\bibfnamefont {M.}~\bibnamefont
  {Peloso}},\ }\href@noop {} {\bibfield  {journal} {\bibinfo  {journal} {Phys.
  Rev. Lett.}\ }\textbf {\bibinfo {volume} {106}},\ \bibinfo {pages} {181301}
  (\bibinfo {year} {2011})},\ \Eprint {http://arxiv.org/abs/1011.1500}
  {1011.1500} \BibitemShut {NoStop}%
\bibitem [{\citenamefont {Ach{\'u}carro}\ \emph {et~al.}(2011)\citenamefont
  {Ach{\'u}carro}, \citenamefont {Gong}, \citenamefont {Hardeman},
  \citenamefont {Palma},\ and\ \citenamefont {Patil}}]{achucarro2011features}%
  \BibitemOpen
  \bibfield  {author} {\bibinfo {author} {\bibfnamefont {A.}~\bibnamefont
  {Ach{\'u}carro}}, \bibinfo {author} {\bibfnamefont {J.-O.}\ \bibnamefont
  {Gong}}, \bibinfo {author} {\bibfnamefont {S.}~\bibnamefont {Hardeman}},
  \bibinfo {author} {\bibfnamefont {G.~A.}\ \bibnamefont {Palma}}, \ and\
  \bibinfo {author} {\bibfnamefont {S.~P.}\ \bibnamefont {Patil}},\ }\href@noop
  {} {\bibfield  {journal} {\bibinfo  {journal} {J. Cosmol. Astropart. Phys.}\
  }\textbf {\bibinfo {volume} {01}},\ \bibinfo {pages} {030} (\bibinfo {year}
  {2011})},\ \Eprint {http://arxiv.org/abs/1010.3693} {1010.3693} \BibitemShut
  {NoStop}%
\bibitem [{\citenamefont {Cespedes}\ \emph {et~al.}(2012)\citenamefont
  {Cespedes}, \citenamefont {Atal},\ and\ \citenamefont
  {Palma}}]{cespedes2012importance}%
  \BibitemOpen
  \bibfield  {author} {\bibinfo {author} {\bibfnamefont {S.}~\bibnamefont
  {Cespedes}}, \bibinfo {author} {\bibfnamefont {V.}~\bibnamefont {Atal}}, \
  and\ \bibinfo {author} {\bibfnamefont {G.~A.}\ \bibnamefont {Palma}},\
  }\href@noop {} {\bibfield  {journal} {\bibinfo  {journal} {J. Cosmol.
  Astropart. Phys.}\ }\textbf {\bibinfo {volume} {05}},\ \bibinfo {pages} {008}
  (\bibinfo {year} {2012})},\ \Eprint {http://arxiv.org/abs/1201.4848}
  {1201.4848} \BibitemShut {NoStop}%
\bibitem [{\citenamefont {Barnaby}\ \emph {et~al.}(2012)\citenamefont
  {Barnaby}, \citenamefont {Pajer},\ and\ \citenamefont
  {Peloso}}]{barnaby2012gauge}%
  \BibitemOpen
  \bibfield  {author} {\bibinfo {author} {\bibfnamefont {N.}~\bibnamefont
  {Barnaby}}, \bibinfo {author} {\bibfnamefont {E.}~\bibnamefont {Pajer}}, \
  and\ \bibinfo {author} {\bibfnamefont {M.}~\bibnamefont {Peloso}},\
  }\href@noop {} {\bibfield  {journal} {\bibinfo  {journal} {Phys. Rev. D}\
  }\textbf {\bibinfo {volume} {85}},\ \bibinfo {pages} {023525} (\bibinfo
  {year} {2012})},\ \Eprint {http://arxiv.org/abs/1110.3327} {1110.3327}
  \BibitemShut {NoStop}%
\bibitem [{\citenamefont {Bringmann}\ \emph {et~al.}(2012)\citenamefont
  {Bringmann}, \citenamefont {Scott},\ and\ \citenamefont
  {Akrami}}]{bringmann2012improved}%
  \BibitemOpen
  \bibfield  {author} {\bibinfo {author} {\bibfnamefont {T.}~\bibnamefont
  {Bringmann}}, \bibinfo {author} {\bibfnamefont {P.}~\bibnamefont {Scott}}, \
  and\ \bibinfo {author} {\bibfnamefont {Y.}~\bibnamefont {Akrami}},\
  }\href@noop {} {\bibfield  {journal} {\bibinfo  {journal} {Phys. Rev. D}\
  }\textbf {\bibinfo {volume} {85}},\ \bibinfo {pages} {125027} (\bibinfo
  {year} {2012})},\ \Eprint {http://arxiv.org/abs/1110.2484} {1110.2484}
  \BibitemShut {NoStop}%
\bibitem [{\citenamefont {Delos}\ \emph
  {et~al.}(2018{\natexlab{a}})\citenamefont {Delos}, \citenamefont {Erickcek},
  \citenamefont {Bailey},\ and\ \citenamefont {Alvarez}}]{delos2018density}%
  \BibitemOpen
  \bibfield  {author} {\bibinfo {author} {\bibfnamefont {M.~S.}\ \bibnamefont
  {Delos}}, \bibinfo {author} {\bibfnamefont {A.~L.}\ \bibnamefont {Erickcek}},
  \bibinfo {author} {\bibfnamefont {A.~P.}\ \bibnamefont {Bailey}}, \ and\
  \bibinfo {author} {\bibfnamefont {M.~A.}\ \bibnamefont {Alvarez}},\
  }\href@noop {} {\bibfield  {journal} {\bibinfo  {journal} {Phys. Rev. D}\
  }\textbf {\bibinfo {volume} {98}},\ \bibinfo {pages} {063527} (\bibinfo
  {year} {2018}{\natexlab{a}})},\ \Eprint {http://arxiv.org/abs/1806.07389}
  {1806.07389} \BibitemShut {NoStop}%
\bibitem [{\citenamefont {Blanco}\ \emph {et~al.}()\citenamefont {Blanco},
  \citenamefont {Delos}, \citenamefont {Erickcek},\ and\ \citenamefont
  {Hooper}}]{blanco2019annihilation}%
  \BibitemOpen
  \bibfield  {author} {\bibinfo {author} {\bibfnamefont {C.}~\bibnamefont
  {Blanco}}, \bibinfo {author} {\bibfnamefont {M.~S.}\ \bibnamefont {Delos}},
  \bibinfo {author} {\bibfnamefont {A.~L.}\ \bibnamefont {Erickcek}}, \ and\
  \bibinfo {author} {\bibfnamefont {D.}~\bibnamefont {Hooper}},\ }\href@noop {}
  {\ }\Eprint {http://arxiv.org/abs/1906.00010} {1906.00010} \BibitemShut
  {NoStop}%
\bibitem [{\citenamefont {van~den Bosch}(2017)}]{van2017dissecting}%
  \BibitemOpen
  \bibfield  {author} {\bibinfo {author} {\bibfnamefont {F.~C.}\ \bibnamefont
  {van~den Bosch}},\ }\href@noop {} {\bibfield  {journal} {\bibinfo  {journal}
  {Mon. Not. R. Astron. Soc.}\ }\textbf {\bibinfo {volume} {468}},\ \bibinfo
  {pages} {885} (\bibinfo {year} {2017})},\ \Eprint
  {http://arxiv.org/abs/1611.02657} {1611.02657} \BibitemShut {NoStop}%
\bibitem [{\citenamefont {van~den Bosch}\ \emph {et~al.}(2018)\citenamefont
  {van~den Bosch}, \citenamefont {Ogiya}, \citenamefont {Hahn},\ and\
  \citenamefont {Burkert}}]{van2017disruption}%
  \BibitemOpen
  \bibfield  {author} {\bibinfo {author} {\bibfnamefont {F.~C.}\ \bibnamefont
  {van~den Bosch}}, \bibinfo {author} {\bibfnamefont {G.}~\bibnamefont
  {Ogiya}}, \bibinfo {author} {\bibfnamefont {O.}~\bibnamefont {Hahn}}, \ and\
  \bibinfo {author} {\bibfnamefont {A.}~\bibnamefont {Burkert}},\ }\href@noop
  {} {\bibfield  {journal} {\bibinfo  {journal} {Mon. Not. R. Astron. Soc.}\
  }\textbf {\bibinfo {volume} {474}},\ \bibinfo {pages} {3043} (\bibinfo {year}
  {2018})},\ \Eprint {http://arxiv.org/abs/1711.05276} {1711.05276}
  \BibitemShut {NoStop}%
\bibitem [{\citenamefont {van~den Bosch}\ and\ \citenamefont
  {Ogiya}(2018)}]{van2018dark}%
  \BibitemOpen
  \bibfield  {author} {\bibinfo {author} {\bibfnamefont {F.~C.}\ \bibnamefont
  {van~den Bosch}}\ and\ \bibinfo {author} {\bibfnamefont {G.}~\bibnamefont
  {Ogiya}},\ }\href@noop {} {\bibfield  {journal} {\bibinfo  {journal} {Mon.
  Not. R. Astron. Soc.}\ }\textbf {\bibinfo {volume} {475}},\ \bibinfo {pages}
  {4066} (\bibinfo {year} {2018})},\ \Eprint {http://arxiv.org/abs/1801.05427}
  {1801.05427} \BibitemShut {NoStop}%
\bibitem [{\citenamefont {Taylor}\ and\ \citenamefont
  {Babul}(2001)}]{taylor2001dynamics}%
  \BibitemOpen
  \bibfield  {author} {\bibinfo {author} {\bibfnamefont {J.~E.}\ \bibnamefont
  {Taylor}}\ and\ \bibinfo {author} {\bibfnamefont {A.}~\bibnamefont {Babul}},\
  }\href@noop {} {\bibfield  {journal} {\bibinfo  {journal} {Astrophys. J.}\
  }\textbf {\bibinfo {volume} {559}},\ \bibinfo {pages} {716} (\bibinfo {year}
  {2001})},\ \Eprint {http://arxiv.org/abs/astro-ph/0012305} {astro-ph/0012305}
  \BibitemShut {NoStop}%
\bibitem [{\citenamefont {Pe{\~n}arrubia}\ and\ \citenamefont
  {Benson}(2005)}]{penarrubia2005effects}%
  \BibitemOpen
  \bibfield  {author} {\bibinfo {author} {\bibfnamefont {J.}~\bibnamefont
  {Pe{\~n}arrubia}}\ and\ \bibinfo {author} {\bibfnamefont {A.~J.}\
  \bibnamefont {Benson}},\ }\href@noop {} {\bibfield  {journal} {\bibinfo
  {journal} {Mon. Not. R. Astron. Soc.}\ }\textbf {\bibinfo {volume} {364}},\
  \bibinfo {pages} {977} (\bibinfo {year} {2005})},\ \Eprint
  {http://arxiv.org/abs/astro-ph/0412370} {astro-ph/0412370} \BibitemShut
  {NoStop}%
\bibitem [{\citenamefont {van~den Bosch}\ \emph {et~al.}(2005)\citenamefont
  {van~den Bosch}, \citenamefont {Tormen},\ and\ \citenamefont
  {Giocoli}}]{van2005mass}%
  \BibitemOpen
  \bibfield  {author} {\bibinfo {author} {\bibfnamefont {F.~C.}\ \bibnamefont
  {van~den Bosch}}, \bibinfo {author} {\bibfnamefont {G.}~\bibnamefont
  {Tormen}}, \ and\ \bibinfo {author} {\bibfnamefont {C.}~\bibnamefont
  {Giocoli}},\ }\href@noop {} {\bibfield  {journal} {\bibinfo  {journal} {Mon.
  Not. R. Astron. Soc.}\ }\textbf {\bibinfo {volume} {359}},\ \bibinfo {pages}
  {1029} (\bibinfo {year} {2005})},\ \Eprint
  {http://arxiv.org/abs/astro-ph/0409201} {astro-ph/0409201} \BibitemShut
  {NoStop}%
\bibitem [{\citenamefont {Zentner}\ \emph {et~al.}(2005)\citenamefont
  {Zentner}, \citenamefont {Berlind}, \citenamefont {Bullock}, \citenamefont
  {Kravtsov},\ and\ \citenamefont {Wechsler}}]{zentner2005physics}%
  \BibitemOpen
  \bibfield  {author} {\bibinfo {author} {\bibfnamefont {A.~R.}\ \bibnamefont
  {Zentner}}, \bibinfo {author} {\bibfnamefont {A.~A.}\ \bibnamefont
  {Berlind}}, \bibinfo {author} {\bibfnamefont {J.~S.}\ \bibnamefont
  {Bullock}}, \bibinfo {author} {\bibfnamefont {A.~V.}\ \bibnamefont
  {Kravtsov}}, \ and\ \bibinfo {author} {\bibfnamefont {R.~H.}\ \bibnamefont
  {Wechsler}},\ }\href@noop {} {\bibfield  {journal} {\bibinfo  {journal}
  {Astrophys. J.}\ }\textbf {\bibinfo {volume} {624}},\ \bibinfo {pages} {505}
  (\bibinfo {year} {2005})},\ \Eprint {http://arxiv.org/abs/astro-ph/0411586}
  {astro-ph/0411586} \BibitemShut {NoStop}%
\bibitem [{\citenamefont {Kampakoglou}\ and\ \citenamefont
  {Benson}(2007)}]{kampakoglou2006tidal}%
  \BibitemOpen
  \bibfield  {author} {\bibinfo {author} {\bibfnamefont {M.}~\bibnamefont
  {Kampakoglou}}\ and\ \bibinfo {author} {\bibfnamefont {A.~J.}\ \bibnamefont
  {Benson}},\ }\href@noop {} {\bibfield  {journal} {\bibinfo  {journal} {Mon.
  Not. R. Astron. Soc.}\ }\textbf {\bibinfo {volume} {374}},\ \bibinfo {pages}
  {775} (\bibinfo {year} {2007})},\ \Eprint
  {http://arxiv.org/abs/astro-ph/0607024} {astro-ph/0607024} \BibitemShut
  {NoStop}%
\bibitem [{\citenamefont {Gan}\ \emph {et~al.}(2010)\citenamefont {Gan},
  \citenamefont {Kang}, \citenamefont {van~den Bosch},\ and\ \citenamefont
  {Hou}}]{gan2010improved}%
  \BibitemOpen
  \bibfield  {author} {\bibinfo {author} {\bibfnamefont {J.}~\bibnamefont
  {Gan}}, \bibinfo {author} {\bibfnamefont {X.}~\bibnamefont {Kang}}, \bibinfo
  {author} {\bibfnamefont {F.~C.}\ \bibnamefont {van~den Bosch}}, \ and\
  \bibinfo {author} {\bibfnamefont {J.}~\bibnamefont {Hou}},\ }\href@noop {}
  {\bibfield  {journal} {\bibinfo  {journal} {Mon. Not. R. Astron. Soc.}\
  }\textbf {\bibinfo {volume} {408}},\ \bibinfo {pages} {2201} (\bibinfo {year}
  {2010})},\ \Eprint {http://arxiv.org/abs/1007.0023} {1007.0023} \BibitemShut
  {NoStop}%
\bibitem [{\citenamefont {Pullen}\ \emph {et~al.}(2014)\citenamefont {Pullen},
  \citenamefont {Benson},\ and\ \citenamefont
  {Moustakas}}]{pullen2014nonlinear}%
  \BibitemOpen
  \bibfield  {author} {\bibinfo {author} {\bibfnamefont {A.~R.}\ \bibnamefont
  {Pullen}}, \bibinfo {author} {\bibfnamefont {A.~J.}\ \bibnamefont {Benson}},
  \ and\ \bibinfo {author} {\bibfnamefont {L.~A.}\ \bibnamefont {Moustakas}},\
  }\href@noop {} {\bibfield  {journal} {\bibinfo  {journal} {Astrophys. J.}\
  }\textbf {\bibinfo {volume} {792}},\ \bibinfo {pages} {24} (\bibinfo {year}
  {2014})},\ \Eprint {http://arxiv.org/abs/arXiv:1407.8189} {arXiv:1407.8189}
  \BibitemShut {NoStop}%
\bibitem [{\citenamefont {Jiang}\ and\ \citenamefont {van~den
  Bosch}(2016)}]{jiang2016statistics}%
  \BibitemOpen
  \bibfield  {author} {\bibinfo {author} {\bibfnamefont {F.}~\bibnamefont
  {Jiang}}\ and\ \bibinfo {author} {\bibfnamefont {F.~C.}\ \bibnamefont
  {van~den Bosch}},\ }\href@noop {} {\bibfield  {journal} {\bibinfo  {journal}
  {Mon. Not. R. Astron. Soc.}\ }\textbf {\bibinfo {volume} {458}},\ \bibinfo
  {pages} {2848} (\bibinfo {year} {2016})},\ \Eprint
  {http://arxiv.org/abs/1403.6827} {1403.6827} \BibitemShut {NoStop}%
\bibitem [{\citenamefont {Hayashi}\ \emph {et~al.}(2003)\citenamefont
  {Hayashi}, \citenamefont {Navarro}, \citenamefont {Taylor}, \citenamefont
  {Stadel},\ and\ \citenamefont {Quinn}}]{hayashi2003structural}%
  \BibitemOpen
  \bibfield  {author} {\bibinfo {author} {\bibfnamefont {E.}~\bibnamefont
  {Hayashi}}, \bibinfo {author} {\bibfnamefont {J.~F.}\ \bibnamefont
  {Navarro}}, \bibinfo {author} {\bibfnamefont {J.~E.}\ \bibnamefont {Taylor}},
  \bibinfo {author} {\bibfnamefont {J.}~\bibnamefont {Stadel}}, \ and\ \bibinfo
  {author} {\bibfnamefont {T.}~\bibnamefont {Quinn}},\ }\href@noop {}
  {\bibfield  {journal} {\bibinfo  {journal} {Astrophys. J.}\ }\textbf
  {\bibinfo {volume} {584}},\ \bibinfo {pages} {541} (\bibinfo {year}
  {2003})},\ \Eprint {http://arxiv.org/abs/astro-ph/0203004} {astro-ph/0203004}
  \BibitemShut {NoStop}%
\bibitem [{\citenamefont {Ogiya}\ \emph {et~al.}(2019)\citenamefont {Ogiya},
  \citenamefont {Van~den Bosch}, \citenamefont {Hahn}, \citenamefont {Green},
  \citenamefont {Miller},\ and\ \citenamefont {Burkert}}]{ogiya2019dash}%
  \BibitemOpen
  \bibfield  {author} {\bibinfo {author} {\bibfnamefont {G.}~\bibnamefont
  {Ogiya}}, \bibinfo {author} {\bibfnamefont {F.~C.}\ \bibnamefont {Van~den
  Bosch}}, \bibinfo {author} {\bibfnamefont {O.}~\bibnamefont {Hahn}}, \bibinfo
  {author} {\bibfnamefont {S.~B.}\ \bibnamefont {Green}}, \bibinfo {author}
  {\bibfnamefont {T.~B.}\ \bibnamefont {Miller}}, \ and\ \bibinfo {author}
  {\bibfnamefont {A.}~\bibnamefont {Burkert}},\ }\href@noop {} {\bibfield
  {journal} {\bibinfo  {journal} {Mon. Not. R. Astron. Soc.}\ }\textbf
  {\bibinfo {volume} {485}},\ \bibinfo {pages} {189} (\bibinfo {year}
  {2019})},\ \Eprint {http://arxiv.org/abs/1901.08601} {1901.08601}
  \BibitemShut {NoStop}%
\bibitem [{\citenamefont {Kazantzidis}\ \emph {et~al.}(2004)\citenamefont
  {Kazantzidis}, \citenamefont {Mayer}, \citenamefont {Mastropietro},
  \citenamefont {Diemand}, \citenamefont {Stadel},\ and\ \citenamefont
  {Moore}}]{kazantzidis2004density}%
  \BibitemOpen
  \bibfield  {author} {\bibinfo {author} {\bibfnamefont {S.}~\bibnamefont
  {Kazantzidis}}, \bibinfo {author} {\bibfnamefont {L.}~\bibnamefont {Mayer}},
  \bibinfo {author} {\bibfnamefont {C.}~\bibnamefont {Mastropietro}}, \bibinfo
  {author} {\bibfnamefont {J.}~\bibnamefont {Diemand}}, \bibinfo {author}
  {\bibfnamefont {J.}~\bibnamefont {Stadel}}, \ and\ \bibinfo {author}
  {\bibfnamefont {B.}~\bibnamefont {Moore}},\ }\href@noop {} {\bibfield
  {journal} {\bibinfo  {journal} {Astrophys. J.}\ }\textbf {\bibinfo {volume}
  {608}},\ \bibinfo {pages} {663} (\bibinfo {year} {2004})},\ \Eprint
  {http://arxiv.org/abs/astro-ph/0312194} {astro-ph/0312194} \BibitemShut
  {NoStop}%
\bibitem [{\citenamefont {Read}\ \emph {et~al.}(2006)\citenamefont {Read},
  \citenamefont {Wilkinson}, \citenamefont {Evans}, \citenamefont {Gilmore},\
  and\ \citenamefont {Kleyna}}]{read2006tidal}%
  \BibitemOpen
  \bibfield  {author} {\bibinfo {author} {\bibfnamefont {J.~I.}\ \bibnamefont
  {Read}}, \bibinfo {author} {\bibfnamefont {M.}~\bibnamefont {Wilkinson}},
  \bibinfo {author} {\bibfnamefont {N.}~\bibnamefont {Evans}}, \bibinfo
  {author} {\bibfnamefont {G.}~\bibnamefont {Gilmore}}, \ and\ \bibinfo
  {author} {\bibfnamefont {J.~T.}\ \bibnamefont {Kleyna}},\ }\href@noop {}
  {\bibfield  {journal} {\bibinfo  {journal} {Mon. Not. R. Astron. Soc.}\
  }\textbf {\bibinfo {volume} {366}},\ \bibinfo {pages} {429} (\bibinfo {year}
  {2006})},\ \Eprint {http://arxiv.org/abs/astro-ph/0506687} {astro-ph/0506687}
  \BibitemShut {NoStop}%
\bibitem [{\citenamefont {Moore}(1994)}]{moore1994evidence}%
  \BibitemOpen
  \bibfield  {author} {\bibinfo {author} {\bibfnamefont {B.}~\bibnamefont
  {Moore}},\ }\href@noop {} {\bibfield  {journal} {\bibinfo  {journal} {Nature
  (London)}\ }\textbf {\bibinfo {volume} {370}},\ \bibinfo {pages} {629}
  (\bibinfo {year} {1994})}\BibitemShut {NoStop}%
\bibitem [{\citenamefont {Read}\ \emph {et~al.}(2018)\citenamefont {Read},
  \citenamefont {Walker},\ and\ \citenamefont {Steger}}]{read2018case}%
  \BibitemOpen
  \bibfield  {author} {\bibinfo {author} {\bibfnamefont {J.~I.}\ \bibnamefont
  {Read}}, \bibinfo {author} {\bibfnamefont {M.~G.}\ \bibnamefont {Walker}}, \
  and\ \bibinfo {author} {\bibfnamefont {P.}~\bibnamefont {Steger}},\
  }\href@noop {} {\bibfield  {journal} {\bibinfo  {journal} {Mon. Not. R.
  Astron. Soc.}\ }\textbf {\bibinfo {volume} {481}},\ \bibinfo {pages} {860}
  (\bibinfo {year} {2018})},\ \Eprint {http://arxiv.org/abs/1805.06934}
  {1805.06934} \BibitemShut {NoStop}%
\bibitem [{\citenamefont {Polisensky}\ and\ \citenamefont
  {Ricotti}(2015)}]{polisensky2015fingerprints}%
  \BibitemOpen
  \bibfield  {author} {\bibinfo {author} {\bibfnamefont {E.}~\bibnamefont
  {Polisensky}}\ and\ \bibinfo {author} {\bibfnamefont {M.}~\bibnamefont
  {Ricotti}},\ }\href@noop {} {\bibfield  {journal} {\bibinfo  {journal} {Mon.
  Not. R. Astron. Soc.}\ }\textbf {\bibinfo {volume} {450}},\ \bibinfo {pages}
  {2172} (\bibinfo {year} {2015})},\ \Eprint {http://arxiv.org/abs/1504.02126}
  {1504.02126} \BibitemShut {NoStop}%
\bibitem [{\citenamefont {Ogiya}\ and\ \citenamefont
  {Hahn}(2018)}]{ogiya2017sets}%
  \BibitemOpen
  \bibfield  {author} {\bibinfo {author} {\bibfnamefont {G.}~\bibnamefont
  {Ogiya}}\ and\ \bibinfo {author} {\bibfnamefont {O.}~\bibnamefont {Hahn}},\
  }\href@noop {} {\bibfield  {journal} {\bibinfo  {journal} {Mon. Not. R.
  Astron. Soc.}\ }\textbf {\bibinfo {volume} {473}},\ \bibinfo {pages} {4339}
  (\bibinfo {year} {2018})},\ \Eprint {http://arxiv.org/abs/1707.07693}
  {1707.07693} \BibitemShut {NoStop}%
\bibitem [{\citenamefont {Delos}\ \emph
  {et~al.}(2018{\natexlab{b}})\citenamefont {Delos}, \citenamefont {Erickcek},
  \citenamefont {Bailey},\ and\ \citenamefont
  {Alvarez}}]{delos2018ultracompact}%
  \BibitemOpen
  \bibfield  {author} {\bibinfo {author} {\bibfnamefont {M.~S.}\ \bibnamefont
  {Delos}}, \bibinfo {author} {\bibfnamefont {A.~L.}\ \bibnamefont {Erickcek}},
  \bibinfo {author} {\bibfnamefont {A.~P.}\ \bibnamefont {Bailey}}, \ and\
  \bibinfo {author} {\bibfnamefont {M.~A.}\ \bibnamefont {Alvarez}},\
  }\href@noop {} {\bibfield  {journal} {\bibinfo  {journal} {Phys. Rev. D}\
  }\textbf {\bibinfo {volume} {97}},\ \bibinfo {pages} {041303(R)} (\bibinfo
  {year} {2018}{\natexlab{b}})},\ \Eprint {http://arxiv.org/abs/1712.05421}
  {1712.05421} \BibitemShut {NoStop}%
\bibitem [{\citenamefont {Angulo}\ \emph {et~al.}(2017)\citenamefont {Angulo},
  \citenamefont {Hahn}, \citenamefont {Ludlow},\ and\ \citenamefont
  {Bonoli}}]{angulo2017earth}%
  \BibitemOpen
  \bibfield  {author} {\bibinfo {author} {\bibfnamefont {R.~E.}\ \bibnamefont
  {Angulo}}, \bibinfo {author} {\bibfnamefont {O.}~\bibnamefont {Hahn}},
  \bibinfo {author} {\bibfnamefont {A.~D.}\ \bibnamefont {Ludlow}}, \ and\
  \bibinfo {author} {\bibfnamefont {S.}~\bibnamefont {Bonoli}},\ }\href@noop {}
  {\bibfield  {journal} {\bibinfo  {journal} {Mon. Not. R. Astron. Soc.}\
  }\textbf {\bibinfo {volume} {471}},\ \bibinfo {pages} {4687} (\bibinfo {year}
  {2017})},\ \Eprint {http://arxiv.org/abs/1604.03131} {1604.03131}
  \BibitemShut {NoStop}%
\bibitem [{\citenamefont {Delos}\ \emph {et~al.}(2019)\citenamefont {Delos},
  \citenamefont {Bruff},\ and\ \citenamefont {Erickcek}}]{delos2019predicting}%
  \BibitemOpen
  \bibfield  {author} {\bibinfo {author} {\bibfnamefont {M.~S.}\ \bibnamefont
  {Delos}}, \bibinfo {author} {\bibfnamefont {M.}~\bibnamefont {Bruff}}, \ and\
  \bibinfo {author} {\bibfnamefont {A.~L.}\ \bibnamefont {Erickcek}},\
  }\href@noop {} {\bibfield  {journal} {\bibinfo  {journal} {Phys. Rev. D}\
  }\textbf {\bibinfo {volume} {100}},\ \bibinfo {pages} {023523} (\bibinfo
  {year} {2019})}\BibitemShut {NoStop}%
\bibitem [{\citenamefont {Ogiya}\ \emph {et~al.}(2016)\citenamefont {Ogiya},
  \citenamefont {Nagai},\ and\ \citenamefont {Ishiyama}}]{ogiya2016dynamical}%
  \BibitemOpen
  \bibfield  {author} {\bibinfo {author} {\bibfnamefont {G.}~\bibnamefont
  {Ogiya}}, \bibinfo {author} {\bibfnamefont {D.}~\bibnamefont {Nagai}}, \ and\
  \bibinfo {author} {\bibfnamefont {T.}~\bibnamefont {Ishiyama}},\ }\href@noop
  {} {\bibfield  {journal} {\bibinfo  {journal} {Mon. Not. R. Astron. Soc.}\
  }\textbf {\bibinfo {volume} {461}},\ \bibinfo {pages} {3385} (\bibinfo {year}
  {2016})},\ \Eprint {http://arxiv.org/abs/1604.02866} {1604.02866}
  \BibitemShut {NoStop}%
\bibitem [{\citenamefont {Springel}\ \emph {et~al.}(2001)\citenamefont
  {Springel}, \citenamefont {Yoshida},\ and\ \citenamefont
  {White}}]{springel2001gadget}%
  \BibitemOpen
  \bibfield  {author} {\bibinfo {author} {\bibfnamefont {V.}~\bibnamefont
  {Springel}}, \bibinfo {author} {\bibfnamefont {N.}~\bibnamefont {Yoshida}}, \
  and\ \bibinfo {author} {\bibfnamefont {S.~D.~M.}\ \bibnamefont {White}},\
  }\href@noop {} {\bibfield  {journal} {\bibinfo  {journal} {New Astron.}\
  }\textbf {\bibinfo {volume} {6}},\ \bibinfo {pages} {79} (\bibinfo {year}
  {2001})},\ \Eprint {http://arxiv.org/abs/astro-ph/0003162} {astro-ph/0003162}
  \BibitemShut {NoStop}%
\bibitem [{\citenamefont {Springel}(2005)}]{springel2005cosmological}%
  \BibitemOpen
  \bibfield  {author} {\bibinfo {author} {\bibfnamefont {V.}~\bibnamefont
  {Springel}},\ }\href@noop {} {\bibfield  {journal} {\bibinfo  {journal} {Mon.
  Not. R. Astron. Soc.}\ }\textbf {\bibinfo {volume} {364}},\ \bibinfo {pages}
  {1105} (\bibinfo {year} {2005})},\ \Eprint
  {http://arxiv.org/abs/astro-ph/0505010} {astro-ph/0505010} \BibitemShut
  {NoStop}%
\bibitem [{\citenamefont {Widrow}(2000)}]{widrow2000distribution}%
  \BibitemOpen
  \bibfield  {author} {\bibinfo {author} {\bibfnamefont {L.~M.}\ \bibnamefont
  {Widrow}},\ }\href@noop {} {\bibfield  {journal} {\bibinfo  {journal}
  {Astrophys. J. Suppl. Ser.}\ }\textbf {\bibinfo {volume} {131}},\ \bibinfo
  {pages} {39} (\bibinfo {year} {2000})}\BibitemShut {NoStop}%
\bibitem [{\citenamefont {Fujii}\ \emph {et~al.}(2006)\citenamefont {Fujii},
  \citenamefont {Funato},\ and\ \citenamefont {Makino}}]{fujii2006dynamical}%
  \BibitemOpen
  \bibfield  {author} {\bibinfo {author} {\bibfnamefont {M.}~\bibnamefont
  {Fujii}}, \bibinfo {author} {\bibfnamefont {Y.}~\bibnamefont {Funato}}, \
  and\ \bibinfo {author} {\bibfnamefont {J.}~\bibnamefont {Makino}},\
  }\href@noop {} {\bibfield  {journal} {\bibinfo  {journal} {Publ. Astron. Soc.
  Jpn.}\ }\textbf {\bibinfo {volume} {58}},\ \bibinfo {pages} {743} (\bibinfo
  {year} {2006})},\ \Eprint {http://arxiv.org/abs/astro-ph/0511651}
  {astro-ph/0511651} \BibitemShut {NoStop}%
\bibitem [{\citenamefont {Fellhauer}\ and\ \citenamefont
  {Lin}(2007)}]{fellhauer2007influence}%
  \BibitemOpen
  \bibfield  {author} {\bibinfo {author} {\bibfnamefont {M.}~\bibnamefont
  {Fellhauer}}\ and\ \bibinfo {author} {\bibfnamefont {D.}~\bibnamefont
  {Lin}},\ }\href@noop {} {\bibfield  {journal} {\bibinfo  {journal} {Mon. Not.
  R. Astron. Soc.}\ }\textbf {\bibinfo {volume} {375}},\ \bibinfo {pages} {604}
  (\bibinfo {year} {2007})},\ \Eprint {http://arxiv.org/abs/astro-ph/0611557}
  {astro-ph/0611557} \BibitemShut {NoStop}%
\bibitem [{\citenamefont {Gnedin}\ \emph {et~al.}(1999)\citenamefont {Gnedin},
  \citenamefont {Hernquist},\ and\ \citenamefont {Ostriker}}]{gnedin1999tidal}%
  \BibitemOpen
  \bibfield  {author} {\bibinfo {author} {\bibfnamefont {O.~Y.}\ \bibnamefont
  {Gnedin}}, \bibinfo {author} {\bibfnamefont {L.}~\bibnamefont {Hernquist}}, \
  and\ \bibinfo {author} {\bibfnamefont {J.~P.}\ \bibnamefont {Ostriker}},\
  }\href@noop {} {\bibfield  {journal} {\bibinfo  {journal} {Astrophys. J.}\
  }\textbf {\bibinfo {volume} {514}},\ \bibinfo {pages} {109} (\bibinfo {year}
  {1999})},\ \Eprint {http://arxiv.org/abs/astro-ph/9709161} {astro-ph/9709161}
  \BibitemShut {NoStop}%
\bibitem [{\citenamefont {Spitzer}(1987)}]{spitzer1987dynamical}%
  \BibitemOpen
  \bibfield  {author} {\bibinfo {author} {\bibfnamefont {L.~S.}\ \bibnamefont
  {Spitzer}, \bibfnamefont {Jr}},\ }\href@noop {} {\emph {\bibinfo {title}
  {Dynamical Evolution of Globular Clusters}}}\ (\bibinfo  {publisher}
  {Princeton University Press, Princeton, NJ},\ \bibinfo {year}
  {1987})\BibitemShut {NoStop}%
\bibitem [{\citenamefont
  {Weinberg}(1994{\natexlab{a}})}]{weinberg1994adiabatic}%
  \BibitemOpen
  \bibfield  {author} {\bibinfo {author} {\bibfnamefont {M.~D.}\ \bibnamefont
  {Weinberg}},\ }\href@noop {} {\bibfield  {journal} {\bibinfo  {journal}
  {Astron. J.}\ }\textbf {\bibinfo {volume} {108}},\ \bibinfo {pages} {1398}
  (\bibinfo {year} {1994}{\natexlab{a}})},\ \Eprint
  {http://arxiv.org/abs/astro-ph/9404015} {astro-ph/9404015} \BibitemShut
  {NoStop}%
\bibitem [{\citenamefont
  {Weinberg}(1994{\natexlab{b}})}]{weinberg1994adiabatic2}%
  \BibitemOpen
  \bibfield  {author} {\bibinfo {author} {\bibfnamefont {M.~D.}\ \bibnamefont
  {Weinberg}},\ }\href@noop {} {\bibfield  {journal} {\bibinfo  {journal}
  {Astron. J.}\ }\textbf {\bibinfo {volume} {108}},\ \bibinfo {pages} {1403}
  (\bibinfo {year} {1994}{\natexlab{b}})},\ \Eprint
  {http://arxiv.org/abs/astro-ph/9404016} {astro-ph/9404016} \BibitemShut
  {NoStop}%
\bibitem [{\citenamefont {Gnedin}\ and\ \citenamefont
  {Ostriker}(1999)}]{gnedin1999self}%
  \BibitemOpen
  \bibfield  {author} {\bibinfo {author} {\bibfnamefont {O.~Y.}\ \bibnamefont
  {Gnedin}}\ and\ \bibinfo {author} {\bibfnamefont {J.~P.}\ \bibnamefont
  {Ostriker}},\ }\href@noop {} {\bibfield  {journal} {\bibinfo  {journal}
  {Astrophys. J.}\ }\textbf {\bibinfo {volume} {513}},\ \bibinfo {pages} {626}
  (\bibinfo {year} {1999})},\ \Eprint {http://arxiv.org/abs/astro-ph/9902326}
  {astro-ph/9902326} \BibitemShut {NoStop}%
\bibitem [{\citenamefont {Klypin}\ \emph {et~al.}(1999)\citenamefont {Klypin},
  \citenamefont {Gottl{\"o}ber}, \citenamefont {Kravtsov},\ and\ \citenamefont
  {Khokhlov}}]{klypin1999galaxies}%
  \BibitemOpen
  \bibfield  {author} {\bibinfo {author} {\bibfnamefont {A.}~\bibnamefont
  {Klypin}}, \bibinfo {author} {\bibfnamefont {S.}~\bibnamefont
  {Gottl{\"o}ber}}, \bibinfo {author} {\bibfnamefont {A.~V.}\ \bibnamefont
  {Kravtsov}}, \ and\ \bibinfo {author} {\bibfnamefont {A.~M.}\ \bibnamefont
  {Khokhlov}},\ }\href@noop {} {\bibfield  {journal} {\bibinfo  {journal}
  {Astrophys. J.}\ }\textbf {\bibinfo {volume} {516}},\ \bibinfo {pages} {530}
  (\bibinfo {year} {1999})},\ \Eprint {http://arxiv.org/abs/astro-ph/9708191}
  {astro-ph/9708191} \BibitemShut {NoStop}%
\bibitem [{\citenamefont {Klypin}\ \emph {et~al.}(2015)\citenamefont {Klypin},
  \citenamefont {Prada}, \citenamefont {Yepes}, \citenamefont {He{\ss}},\ and\
  \citenamefont {Gottl{\"o}ber}}]{klypin2015halo}%
  \BibitemOpen
  \bibfield  {author} {\bibinfo {author} {\bibfnamefont {A.}~\bibnamefont
  {Klypin}}, \bibinfo {author} {\bibfnamefont {F.}~\bibnamefont {Prada}},
  \bibinfo {author} {\bibfnamefont {G.}~\bibnamefont {Yepes}}, \bibinfo
  {author} {\bibfnamefont {S.}~\bibnamefont {He{\ss}}}, \ and\ \bibinfo
  {author} {\bibfnamefont {S.}~\bibnamefont {Gottl{\"o}ber}},\ }\href@noop {}
  {\bibfield  {journal} {\bibinfo  {journal} {Mon. Not. R. Astron. Soc.}\
  }\textbf {\bibinfo {volume} {447}},\ \bibinfo {pages} {3693} (\bibinfo {year}
  {2015})},\ \Eprint {http://arxiv.org/abs/1310.3740} {1310.3740} \BibitemShut
  {NoStop}%
\bibitem [{\citenamefont {Delos}(2019)}]{delos2019evolution}%
  \BibitemOpen
  \bibfield  {author} {\bibinfo {author} {\bibfnamefont {M.~S.}\ \bibnamefont
  {Delos}},\ }\href@noop {} {\  (\bibinfo {year} {2019})},\ \Eprint
  {http://arxiv.org/abs/1907.13133} {1907.13133} \BibitemShut {NoStop}%
\bibitem [{\citenamefont {Tormen}(1997)}]{tormen1997rise}%
  \BibitemOpen
  \bibfield  {author} {\bibinfo {author} {\bibfnamefont {G.}~\bibnamefont
  {Tormen}},\ }\href@noop {} {\bibfield  {journal} {\bibinfo  {journal} {Mon.
  Not. R. Astron. Soc.}\ }\textbf {\bibinfo {volume} {290}},\ \bibinfo {pages}
  {411} (\bibinfo {year} {1997})},\ \Eprint
  {http://arxiv.org/abs/astro-ph/9611078} {astro-ph/9611078} \BibitemShut
  {NoStop}%
\bibitem [{\citenamefont {Khochfar}\ and\ \citenamefont
  {Burkert}(2006)}]{khochfar2006orbital}%
  \BibitemOpen
  \bibfield  {author} {\bibinfo {author} {\bibfnamefont {S.}~\bibnamefont
  {Khochfar}}\ and\ \bibinfo {author} {\bibfnamefont {A.}~\bibnamefont
  {Burkert}},\ }\href@noop {} {\bibfield  {journal} {\bibinfo  {journal}
  {Astron. Astrophys.}\ }\textbf {\bibinfo {volume} {445}},\ \bibinfo {pages}
  {403} (\bibinfo {year} {2006})},\ \Eprint
  {http://arxiv.org/abs/astro-ph/0309611} {astro-ph/0309611} \BibitemShut
  {NoStop}%
\bibitem [{\citenamefont {Wetzel}(2011)}]{wetzel2011orbits}%
  \BibitemOpen
  \bibfield  {author} {\bibinfo {author} {\bibfnamefont {A.~R.}\ \bibnamefont
  {Wetzel}},\ }\href@noop {} {\bibfield  {journal} {\bibinfo  {journal} {Mon.
  Not. R. Astron. Soc.}\ }\textbf {\bibinfo {volume} {412}},\ \bibinfo {pages}
  {49} (\bibinfo {year} {2011})},\ \Eprint {http://arxiv.org/abs/1001.4792}
  {1001.4792} \BibitemShut {NoStop}%
\bibitem [{\citenamefont {Jiang}\ \emph {et~al.}(2015)\citenamefont {Jiang},
  \citenamefont {Cole}, \citenamefont {Sawala},\ and\ \citenamefont
  {Frenk}}]{jiang2015orbital}%
  \BibitemOpen
  \bibfield  {author} {\bibinfo {author} {\bibfnamefont {L.}~\bibnamefont
  {Jiang}}, \bibinfo {author} {\bibfnamefont {S.}~\bibnamefont {Cole}},
  \bibinfo {author} {\bibfnamefont {T.}~\bibnamefont {Sawala}}, \ and\ \bibinfo
  {author} {\bibfnamefont {C.~S.}\ \bibnamefont {Frenk}},\ }\href@noop {}
  {\bibfield  {journal} {\bibinfo  {journal} {Mon. Not. R. Astron. Soc.}\
  }\textbf {\bibinfo {volume} {448}},\ \bibinfo {pages} {1674} (\bibinfo {year}
  {2015})},\ \Eprint {http://arxiv.org/abs/1409.1179} {1409.1179} \BibitemShut
  {NoStop}%
\bibitem [{\citenamefont {Delos}\ \emph {et~al.}(shed)\citenamefont {Delos},
  \citenamefont {Erickcek},\ and\ \citenamefont {Linden}}]{delos2019gamma}%
  \BibitemOpen
  \bibfield  {author} {\bibinfo {author} {\bibfnamefont {M.~S.}\ \bibnamefont
  {Delos}}, \bibinfo {author} {\bibfnamefont {A.~L.}\ \bibnamefont {Erickcek}},
  \ and\ \bibinfo {author} {\bibfnamefont {T.}~\bibnamefont {Linden}},\
  }\href@noop {} {\  (\bibinfo {year} {\noop{2019}to be
  published})}\BibitemShut {NoStop}%
\bibitem [{\citenamefont {Navarro}\ \emph {et~al.}(2010)\citenamefont
  {Navarro}, \citenamefont {Ludlow}, \citenamefont {Springel}, \citenamefont
  {Wang}, \citenamefont {Vogelsberger}, \citenamefont {White}, \citenamefont
  {Jenkins}, \citenamefont {Frenk},\ and\ \citenamefont
  {Helmi}}]{navarro2010diversity}%
  \BibitemOpen
  \bibfield  {author} {\bibinfo {author} {\bibfnamefont {J.~F.}\ \bibnamefont
  {Navarro}}, \bibinfo {author} {\bibfnamefont {A.}~\bibnamefont {Ludlow}},
  \bibinfo {author} {\bibfnamefont {V.}~\bibnamefont {Springel}}, \bibinfo
  {author} {\bibfnamefont {J.}~\bibnamefont {Wang}}, \bibinfo {author}
  {\bibfnamefont {M.}~\bibnamefont {Vogelsberger}}, \bibinfo {author}
  {\bibfnamefont {S.~D.~M.}\ \bibnamefont {White}}, \bibinfo {author}
  {\bibfnamefont {A.}~\bibnamefont {Jenkins}}, \bibinfo {author} {\bibfnamefont
  {C.~S.}\ \bibnamefont {Frenk}}, \ and\ \bibinfo {author} {\bibfnamefont
  {A.}~\bibnamefont {Helmi}},\ }\href@noop {} {\bibfield  {journal} {\bibinfo
  {journal} {Mon. Not. R. Astron. Soc.}\ }\textbf {\bibinfo {volume} {402}},\
  \bibinfo {pages} {21} (\bibinfo {year} {2010})}\BibitemShut {NoStop}%
\bibitem [{\citenamefont {Bullock}\ and\ \citenamefont
  {Boylan-Kolchin}(2017)}]{bullock2017small}%
  \BibitemOpen
  \bibfield  {author} {\bibinfo {author} {\bibfnamefont {J.}~\bibnamefont
  {Bullock}}\ and\ \bibinfo {author} {\bibfnamefont {M.}~\bibnamefont
  {Boylan-Kolchin}},\ }\href@noop {} {\bibfield  {journal} {\bibinfo  {journal}
  {Annu. Rev. Astron. Astrophys.}\ }\textbf {\bibinfo {volume} {55}},\ \bibinfo
  {pages} {343} (\bibinfo {year} {2017})},\ \Eprint
  {http://arxiv.org/abs/1707.04256} {1707.04256} \BibitemShut {NoStop}%
\bibitem [{\citenamefont {Berezinsky}\ \emph {et~al.}(2006)\citenamefont
  {Berezinsky}, \citenamefont {Dokuchaev},\ and\ \citenamefont
  {Eroshenko}}]{berezinsky2006destruction}%
  \BibitemOpen
  \bibfield  {author} {\bibinfo {author} {\bibfnamefont {V.}~\bibnamefont
  {Berezinsky}}, \bibinfo {author} {\bibfnamefont {V.}~\bibnamefont
  {Dokuchaev}}, \ and\ \bibinfo {author} {\bibfnamefont {Y.}~\bibnamefont
  {Eroshenko}},\ }\href@noop {} {\bibfield  {journal} {\bibinfo  {journal}
  {Phys. Rev. D}\ }\textbf {\bibinfo {volume} {73}},\ \bibinfo {pages} {063504}
  (\bibinfo {year} {2006})},\ \Eprint {http://arxiv.org/abs/astro-ph/0511494}
  {astro-ph/0511494} \BibitemShut {NoStop}%
\bibitem [{\citenamefont {Zhao}\ \emph {et~al.}(2007)\citenamefont {Zhao},
  \citenamefont {Hooper}, \citenamefont {Angus}, \citenamefont {Taylor},\ and\
  \citenamefont {Silk}}]{zhao2007tidal}%
  \BibitemOpen
  \bibfield  {author} {\bibinfo {author} {\bibfnamefont {H.}~\bibnamefont
  {Zhao}}, \bibinfo {author} {\bibfnamefont {D.}~\bibnamefont {Hooper}},
  \bibinfo {author} {\bibfnamefont {G.~W.}\ \bibnamefont {Angus}}, \bibinfo
  {author} {\bibfnamefont {J.~E.}\ \bibnamefont {Taylor}}, \ and\ \bibinfo
  {author} {\bibfnamefont {J.}~\bibnamefont {Silk}},\ }\href@noop {} {\bibfield
   {journal} {\bibinfo  {journal} {Astrophys. J.}\ }\textbf {\bibinfo {volume}
  {654}},\ \bibinfo {pages} {697} (\bibinfo {year} {2007})},\ \Eprint
  {http://arxiv.org/abs/astro-ph/0508215} {astro-ph/0508215} \BibitemShut
  {NoStop}%
\bibitem [{\citenamefont {Green}\ and\ \citenamefont
  {Goodwin}(2007)}]{green2007mini}%
  \BibitemOpen
  \bibfield  {author} {\bibinfo {author} {\bibfnamefont {A.~M.}\ \bibnamefont
  {Green}}\ and\ \bibinfo {author} {\bibfnamefont {S.~P.}\ \bibnamefont
  {Goodwin}},\ }\href@noop {} {\bibfield  {journal} {\bibinfo  {journal} {Mon.
  Not. R. Astron. Soc.}\ }\textbf {\bibinfo {volume} {375}},\ \bibinfo {pages}
  {1111} (\bibinfo {year} {2007})},\ \Eprint
  {http://arxiv.org/abs/astro-ph/0604142} {astro-ph/0604142} \BibitemShut
  {NoStop}%
\bibitem [{\citenamefont {Schneider}\ \emph {et~al.}(2010)\citenamefont
  {Schneider}, \citenamefont {Krauss},\ and\ \citenamefont
  {Moore}}]{schneider2010impact}%
  \BibitemOpen
  \bibfield  {author} {\bibinfo {author} {\bibfnamefont {A.}~\bibnamefont
  {Schneider}}, \bibinfo {author} {\bibfnamefont {L.}~\bibnamefont {Krauss}}, \
  and\ \bibinfo {author} {\bibfnamefont {B.}~\bibnamefont {Moore}},\
  }\href@noop {} {\bibfield  {journal} {\bibinfo  {journal} {Phys. Rev. D}\
  }\textbf {\bibinfo {volume} {82}},\ \bibinfo {pages} {063525} (\bibinfo
  {year} {2010})},\ \Eprint {http://arxiv.org/abs/1004.5432} {1004.5432}
  \BibitemShut {NoStop}%
\bibitem [{\citenamefont {Angus}\ and\ \citenamefont
  {Zhao}(2007)}]{angus2007cold}%
  \BibitemOpen
  \bibfield  {author} {\bibinfo {author} {\bibfnamefont {G.}~\bibnamefont
  {Angus}}\ and\ \bibinfo {author} {\bibfnamefont {H.}~\bibnamefont {Zhao}},\
  }\href@noop {} {\bibfield  {journal} {\bibinfo  {journal} {Mon. Not. R.
  Astron. Soc.}\ }\textbf {\bibinfo {volume} {375}},\ \bibinfo {pages} {1146}
  (\bibinfo {year} {2007})},\ \Eprint {http://arxiv.org/abs/astro-ph/0608580}
  {astro-ph/0608580} \BibitemShut {NoStop}%
\bibitem [{\citenamefont {Berezinsky}\ \emph {et~al.}(2014)\citenamefont
  {Berezinsky}, \citenamefont {Dokuchaev},\ and\ \citenamefont
  {Eroshenko}}]{berezinsky2014small}%
  \BibitemOpen
  \bibfield  {author} {\bibinfo {author} {\bibfnamefont {V.~S.}\ \bibnamefont
  {Berezinsky}}, \bibinfo {author} {\bibfnamefont {V.~I.}\ \bibnamefont
  {Dokuchaev}}, \ and\ \bibinfo {author} {\bibfnamefont {Y.~N.}\ \bibnamefont
  {Eroshenko}},\ }\href@noop {} {\bibfield  {journal} {\bibinfo  {journal}
  {Phys. Usp.}\ }\textbf {\bibinfo {volume} {57}},\ \bibinfo {pages} {1}
  (\bibinfo {year} {2014})},\ \Eprint {http://arxiv.org/abs/1405.2204}
  {1405.2204} \BibitemShut {NoStop}%
\bibitem [{\citenamefont {D'Onghia}\ \emph {et~al.}(2010)\citenamefont
  {D'Onghia}, \citenamefont {Springel}, \citenamefont {Hernquist},\ and\
  \citenamefont {Keres}}]{d2010substructure}%
  \BibitemOpen
  \bibfield  {author} {\bibinfo {author} {\bibfnamefont {E.}~\bibnamefont
  {D'Onghia}}, \bibinfo {author} {\bibfnamefont {V.}~\bibnamefont {Springel}},
  \bibinfo {author} {\bibfnamefont {L.}~\bibnamefont {Hernquist}}, \ and\
  \bibinfo {author} {\bibfnamefont {D.}~\bibnamefont {Keres}},\ }\href@noop {}
  {\bibfield  {journal} {\bibinfo  {journal} {Astrophys. J.}\ }\textbf
  {\bibinfo {volume} {709}},\ \bibinfo {pages} {1138} (\bibinfo {year}
  {2010})},\ \Eprint {http://arxiv.org/abs/0907.3482} {0907.3482} \BibitemShut
  {NoStop}%
\bibitem [{\citenamefont {Zhu}\ \emph {et~al.}(2016)\citenamefont {Zhu},
  \citenamefont {Marinacci}, \citenamefont {Maji}, \citenamefont {Li},
  \citenamefont {Springel},\ and\ \citenamefont {Hernquist}}]{zhu2016baryonic}%
  \BibitemOpen
  \bibfield  {author} {\bibinfo {author} {\bibfnamefont {Q.}~\bibnamefont
  {Zhu}}, \bibinfo {author} {\bibfnamefont {F.}~\bibnamefont {Marinacci}},
  \bibinfo {author} {\bibfnamefont {M.}~\bibnamefont {Maji}}, \bibinfo {author}
  {\bibfnamefont {Y.}~\bibnamefont {Li}}, \bibinfo {author} {\bibfnamefont
  {V.}~\bibnamefont {Springel}}, \ and\ \bibinfo {author} {\bibfnamefont
  {L.}~\bibnamefont {Hernquist}},\ }\href@noop {} {\bibfield  {journal}
  {\bibinfo  {journal} {Mon. Not. R. Astron. Soc.}\ }\textbf {\bibinfo {volume}
  {458}},\ \bibinfo {pages} {1559} (\bibinfo {year} {2016})},\ \Eprint
  {http://arxiv.org/abs/1506.05537} {1506.05537} \BibitemShut {NoStop}%
\bibitem [{\citenamefont {Errani}\ \emph {et~al.}(2016)\citenamefont {Errani},
  \citenamefont {Pe{\~n}arrubia}, \citenamefont {Laporte},\ and\ \citenamefont
  {G{\'o}mez}}]{errani2016effect}%
  \BibitemOpen
  \bibfield  {author} {\bibinfo {author} {\bibfnamefont {R.}~\bibnamefont
  {Errani}}, \bibinfo {author} {\bibfnamefont {J.}~\bibnamefont
  {Pe{\~n}arrubia}}, \bibinfo {author} {\bibfnamefont {C.~F.}\ \bibnamefont
  {Laporte}}, \ and\ \bibinfo {author} {\bibfnamefont {F.~A.}\ \bibnamefont
  {G{\'o}mez}},\ }\href@noop {} {\bibfield  {journal} {\bibinfo  {journal}
  {Mon. Not. R. Astron. Soc. Lett.}\ }\textbf {\bibinfo {volume} {465}},\
  \bibinfo {pages} {L59} (\bibinfo {year} {2016})},\ \Eprint
  {http://arxiv.org/abs/1608.01849} {1608.01849} \BibitemShut {NoStop}%
\bibitem [{\citenamefont {Garrison-Kimmel}\ \emph {et~al.}(2017)\citenamefont
  {Garrison-Kimmel}, \citenamefont {Wetzel}, \citenamefont {Bullock},
  \citenamefont {Hopkins}, \citenamefont {Boylan-Kolchin}, \citenamefont
  {Faucher-Gigu{\`e}re}, \citenamefont {Kere{\v{s}}}, \citenamefont {Quataert},
  \citenamefont {Sanderson}, \citenamefont {Graus} \emph
  {et~al.}}]{garrison2017not}%
  \BibitemOpen
  \bibfield  {author} {\bibinfo {author} {\bibfnamefont {S.}~\bibnamefont
  {Garrison-Kimmel}}, \bibinfo {author} {\bibfnamefont {A.}~\bibnamefont
  {Wetzel}}, \bibinfo {author} {\bibfnamefont {J.~S.}\ \bibnamefont {Bullock}},
  \bibinfo {author} {\bibfnamefont {P.~F.}\ \bibnamefont {Hopkins}}, \bibinfo
  {author} {\bibfnamefont {M.}~\bibnamefont {Boylan-Kolchin}}, \bibinfo
  {author} {\bibfnamefont {C.-A.}\ \bibnamefont {Faucher-Gigu{\`e}re}},
  \bibinfo {author} {\bibfnamefont {D.}~\bibnamefont {Kere{\v{s}}}}, \bibinfo
  {author} {\bibfnamefont {E.}~\bibnamefont {Quataert}}, \bibinfo {author}
  {\bibfnamefont {R.~E.}\ \bibnamefont {Sanderson}}, \bibinfo {author}
  {\bibfnamefont {A.~S.}\ \bibnamefont {Graus}},  \emph {et~al.},\ }\href@noop
  {} {\bibfield  {journal} {\bibinfo  {journal} {Mon. Not. R. Astron. Soc.}\
  }\textbf {\bibinfo {volume} {471}},\ \bibinfo {pages} {1709} (\bibinfo {year}
  {2017})},\ \Eprint {http://arxiv.org/abs/1701.03792} {1701.03792}
  \BibitemShut {NoStop}%
\bibitem [{\citenamefont {Kelley}\ \emph {et~al.}(2019)\citenamefont {Kelley},
  \citenamefont {Bullock}, \citenamefont {Garrison-Kimmel}, \citenamefont
  {Boylan-Kolchin}, \citenamefont {Pawlowski},\ and\ \citenamefont
  {Graus}}]{kelley2018phat}%
  \BibitemOpen
  \bibfield  {author} {\bibinfo {author} {\bibfnamefont {T.}~\bibnamefont
  {Kelley}}, \bibinfo {author} {\bibfnamefont {J.~S.}\ \bibnamefont {Bullock}},
  \bibinfo {author} {\bibfnamefont {S.}~\bibnamefont {Garrison-Kimmel}},
  \bibinfo {author} {\bibfnamefont {M.}~\bibnamefont {Boylan-Kolchin}},
  \bibinfo {author} {\bibfnamefont {M.~S.}\ \bibnamefont {Pawlowski}}, \ and\
  \bibinfo {author} {\bibfnamefont {A.~S.}\ \bibnamefont {Graus}},\ }\href@noop
  {} {\bibfield  {journal} {\bibinfo  {journal} {Mon. Not. R. Astron. Soc.}\
  }\textbf {\bibinfo {volume} {487}},\ \bibinfo {pages} {4409} (\bibinfo {year}
  {2019})},\ \Eprint {http://arxiv.org/abs/1811.12413} {1811.12413}
  \BibitemShut {NoStop}%
\bibitem [{\citenamefont {H{\"u}tten}\ \emph {et~al.}(2019)\citenamefont
  {H{\"u}tten}, \citenamefont {Stref}, \citenamefont {Combet}, \citenamefont
  {Maurin},\ and\ \citenamefont {Lavalle}}]{hutten2019gamma}%
  \BibitemOpen
  \bibfield  {author} {\bibinfo {author} {\bibfnamefont {M.}~\bibnamefont
  {H{\"u}tten}}, \bibinfo {author} {\bibfnamefont {M.}~\bibnamefont {Stref}},
  \bibinfo {author} {\bibfnamefont {C.}~\bibnamefont {Combet}}, \bibinfo
  {author} {\bibfnamefont {D.}~\bibnamefont {Maurin}}, \ and\ \bibinfo {author}
  {\bibfnamefont {J.}~\bibnamefont {Lavalle}},\ }\href@noop {} {\bibfield
  {journal} {\bibinfo  {journal} {Galaxies}\ }\textbf {\bibinfo {volume} {7}},\
  \bibinfo {pages} {60} (\bibinfo {year} {2019})},\ \Eprint
  {http://arxiv.org/abs/1904.10935} {1904.10935} \BibitemShut {NoStop}%
\bibitem [{\citenamefont {McConnachie}(2012)}]{mcconnachie2012observed}%
  \BibitemOpen
  \bibfield  {author} {\bibinfo {author} {\bibfnamefont {A.~W.}\ \bibnamefont
  {McConnachie}},\ }\href@noop {} {\bibfield  {journal} {\bibinfo  {journal}
  {Astron. J.}\ }\textbf {\bibinfo {volume} {144}},\ \bibinfo {pages} {4}
  (\bibinfo {year} {2012})},\ \Eprint {http://arxiv.org/abs/1204.1562}
  {1204.1562} \BibitemShut {NoStop}%
\bibitem [{\citenamefont {Green}(2011)}]{green2011colour}%
  \BibitemOpen
  \bibfield  {author} {\bibinfo {author} {\bibfnamefont {D.~A.}\ \bibnamefont
  {Green}},\ }\href@noop {} {\bibfield  {journal} {\bibinfo  {journal} {Bull.
  Astron. Soc. India}\ }\textbf {\bibinfo {volume} {39}},\ \bibinfo {pages}
  {289} (\bibinfo {year} {2011})},\ \Eprint {http://arxiv.org/abs/1108.5083}
  {1108.5083} \BibitemShut {NoStop}%
\bibitem [{\citenamefont {Barnes}\ and\ \citenamefont
  {Hut}(1986)}]{barnes1986hierarchical}%
  \BibitemOpen
  \bibfield  {author} {\bibinfo {author} {\bibfnamefont {J.}~\bibnamefont
  {Barnes}}\ and\ \bibinfo {author} {\bibfnamefont {P.}~\bibnamefont {Hut}},\
  }\href@noop {} {\bibfield  {journal} {\bibinfo  {journal} {Nature}\ }\textbf
  {\bibinfo {volume} {324}},\ \bibinfo {pages} {446} (\bibinfo {year}
  {1986})}\BibitemShut {NoStop}%
\end{thebibliography}%

\end{document}